%% file: be3.tex
\documentclass[11pt]{article}
\usepackage{amsmath,amsthm,amsfonts,amssymb, graphicx,mathabx,epsfig}
\usepackage{bbm}
\usepackage{authblk}
\usepackage{setspace}
\usepackage[usenames]{color}
\usepackage[active]{srcltx}
\newcommand{\ie}{{\it i.e. }}
\newcommand{\eg}{{\it e.g. }}
\newcommand{\cf}{{\it cf.}}
\renewcommand{\Re}{\operatorname{\mathrm{Re}}}
\renewcommand{\Im}{\operatorname{\mathrm{Im}}}

\newcommand{\diag}{\operatorname{\mathrm{diag}}}
\renewcommand{\d}{\mathrm{d}}

\newcommand{\e}{\varepsilon}
\newcommand{\dbar}{\kern-.1em{\raise.8ex\hbox{ -}}\kern-.6em{d}}
  
\def\half{\mbox{$\frac 1 2$}}

\def\?{\marginpar{not sure}}

\newcommand{\comment}[1]{}

\newtheorem{thm}{Theorem}[section]

\newtheorem{lemma}[thm]{Lemma}
\newtheorem{prop}[thm]{Proposition}

\newtheorem{rem}[thm]{Remark}

\newtheorem{defi}[thm]{Definition}
\def \be{\begin{equation}}
\def \ee{\end{equation}}
\def \ben{\begin{equation*}}
\def \een{\end{equation*}}
\def \bea{\begin{eqnarray}}
\def \eea{\end{eqnarray}}

\def\qed{\hfill\raise1pt\hbox{\vrule height5pt width5pt depth0pt}}
\def\nn{\nonumber}
\def\sgn{\mathop\mathrm{sgn}}
\def\io{\infty}
\def\GL{\mathrm{GL}}

\def\pf{\mathrm{pf}\,}

\def\L{\Lambda}
\def\l{\lambda}
\def\r{\rho}
\def\s{\sigma}
\def\a{\alpha}
\def\b{\beta}
\def\d{\delta}
\def\D{\Delta}
\def\m{\mu}
\def\g{\gamma}
\def\G{\Gamma}
\def\e{\varepsilon}
\def\t{\tau}
\def\th{\theta}
\def\Th{\Theta}

\definecolor{light}{gray}{.75}

\begin{document}
\title{Bulk-edge correspondence for two-dimensional topological insulators}

\author{G.M. Graf and M. Porta
\\
\small{Theoretische Physik, ETH Zurich, 8093 Zurich, Switzerland} }

\maketitle

\begin{abstract}
Topological insulators can be characterized alternatively in terms of bulk or edge properties. We prove the equivalence between the two descriptions for two-dimensional solids in the single-particle picture. We give a new formulation of the $\mathbb{Z}_{2}$-invariant, which allows for a bulk index not relying on a (two-dimensional) Brillouin zone. When available though, that index is shown to agree with known formulations. The method also applies to integer quantum Hall systems. We discuss a further variant of the correspondence, based on scattering theory.
\end{abstract}
\section{Introduction}

{\it Topological insulators} are materials that behave as ordinary insulators in the bulk, in that they exhibit an excitation gap, whereas the edge has robust, gapless modes. They have been theoretically predicted as a class \cite{KM} and as a specific compound \cite{BHZ}, in which the effect was then observed \cite{K, Hs}. In analogy with quantum Hall (QH) systems, the presence of edge states is a robust property of the system, in that it is stable under moderate changes of parameters. That calls for an explanation in terms of topological invariants.

As pointed out in \cite{KM}, a key feature of topological insulators is fermionic {\it time-reversal symmetry}. It was shown that two-dimensional time-reversal symmetric insulators admit two topologically distinct phases: The phases can not be deformed into one another as long as the bulk gap and the symmetry are there. In one phase the insulator is an ordinary one; that is, it does not carry currents in the bulk nor at the edges. In the other phase edge states are present. They come in pairs with opposite velocity and spin. This phase is called the {\it quantum spin Hall} (QSH) {\it phase} and its signature is a nonzero spin current, which in contrast to the QH current is not quantized, as a rule. For integer quantum Hall systems the topological invariant is a Chern number, and can be read off from the value of the Hall conductivity; for quantum spin Hall systems the topological invariant is not expressing the value of the spin current, but the parity of the number of edge states.

In this paper we prove the {\it bulk-edge correspondence} for two-dimensional topological insulators and for independent particles. For short we introduce a $\mathbb{Z}_{2}$ bulk topological invariant (to be shown equivalent to others), and we show that it is equal to the number of pairs of edge states modulo $2$. For sure, that duality has been discussed in the literature, see \eg\cite{FK}, but we maintain that there is room for a strict mathematical approach; just as for quantum Hall systems, where Laughlin's argument has gained in precision and detail by the subsequent mathematical discussion. We should though mention \cite{QWZ,ASV}, whose results are further compared to ours below. 

The paper is organized as follows. In Section~\ref{sec1.1} we consider a general class of two-dimensional single-particle lattice Hamiltonians, with and without edge. They are symmetric under fermionic time reversal and hence potentially describe topological insulators. One noteworthy feature is that periodicity is postulated only in the direction parallel to the edge. In Section~\ref{mr} we formulate the duality in its most basic version, in a precise though still preliminary form. The purpose of Section~\ref{secexa} is to show how graphene and related model materials, which have been considered as candidates for topological insulators, fit into our scheme. As a side remark, we give a simple new proof of the absence of edge states for armchair boundary conditions. In Section~\ref{trib} we discuss some abstract vector bundles on the 2-torus. By considering their sections and transition matrices, we classify them in terms of a $\mathbb{Z}_{2}$-invariant, defined similarly but not identically to one found in \cite{FK}. So equipped, we define the bulk index in Section~\ref{sec2.1} and state in full the basic version of duality (Theorem~\ref{thm1}). It is immediately followed by the main steps of the proof, while technical details are postponed to main part of Section~\ref{secproofs}. The part of the article so far described provides a full and self-contained account of basic bulk-edge duality. It is the most general part, and yet makes up for less than half of its length.

In Section~\ref{secbloch} we formulate a bulk index for the specific and more familiar case where the lattice Hamiltonian is doubly periodic. As it is to be expected, the relevant 2-torus is now the Brillouin zone. We discuss how that index arises from the general one, leaving details to Section~\ref{BlBu}. All the results obtained up to that point have a counterpart in the case of QH systems, which we present in Section~\ref{QHI} for illustration, because they are simpler in that context. As a matter of fact, for such systems we include an independent, alternate version of the duality based on scattering theory and more precisely on Levinson's theorem (Section~\ref{scatt}), as well as a comparison between the two versions (Section~\ref{comp}). Related details are found in Section~\ref{bec}. We conjecture an analogous alternate version for QSH systems. Finally, Section~\ref{sr} contains some results about indices, including a comparison of ours with some of those found in the literature.

Let us comment on the relation to other work. In \cite{KM} an edge and a bulk characterization of a topological insulator is given, but without proof of a mathematical link. In \cite{QWZ} a correspondence between bulk and edge description is given. There however the bulk is put in correspondence with twisted boundary conditions allowing tunneling between a pair of edges; they include open boundaries as a special case. Moreover, the bulk invariant used there is physically different from a $\mathbb{Z}_{2}$-classification and may not reflect itself in open boundaries. In \cite{ASV} the bulk-edge duality for topological insulators is derived in the same sense as ours, among other results. The setting, however, is not quite as general as ours, as far as the side of the bulk is concerned. As the edge is concerned, an index related to the Maslov index is defined and shown to be equivalent to that of \cite{KM}. As for the bulk, it is assumed that the Hamiltonian is the perturbation of one commuting with spin, and hence consisting to several copies of a quantum Hall system. It is assumed that the gap remains open as the perturbation is switched on. Thus spin resolved Chern numbers can be defined by homotopy \cite{P, SWSH} and the duality inferred from that of the quantum Hall case \cite{SKR}.

Further, assorted comments are: Some of our examples are discussed in \cite{ASV} in a similar vein. Related indices for topological insulators are discussed in \cite{ASV, FK, HK, KM, MB, R}. A bulk-edge duality result for general symmetry classes is found in \cite{EG}, differing from the results presented here in various ways; for instance the index takes values in $\mathbb{Z}$, with just hints at the $\mathbb{Z}_{2}$ case, at least in two dimensions. A choice of a torus and bundle similar to ours in Sect.~\ref{mr} is found in \cite{T}. The results of Section~\ref{secbloch} depend on the analytic properties of band functions \cite{Ko}. In \cite{Hat0} bulk-edge duality for Hall systems is pinpointed at the birth of edge states of band edges. This insight is the reason for using Levinson's theorem, though the method is otherwise different and the result more general. Last but not least, the existence of a complementary approach to gapped systems, including interacting ones, should be mentioned. It is based on effective, topological field theories as a tool to explore the response of a system in the limit of low frequencies and long wavelengths; see \cite{W, FrK, FZ, Z, FrS, FrST} for early examples.  

\section{Setting and results}
We shall introduce a class of {\it bulk}, resp. {\it edge}, single-particle discrete Schr\"odinger operators, by which we describe insulators, topological or otherwise, which extend over a plane, resp. a half-plane. We conclude the section with a loose description of the results.

\subsection{The Schr\"odinger operators}\label{sec1.1}

Consider, at first and in less than final generality, a tight binding Hamiltonian with nearest neighbor hopping on the lattice $\mathbb{Z}\times\mathbb{Z}$, resp. $\mathbb{N}\times\mathbb{Z}$. The Hamiltonian is assumed to be periodic in the direction along the edge, but not necessarily across it. The period may be taken to be equal to one without loss: In fact sites within a period may be regarded as labels of internal degrees of freedom, among others like \eg spin. We may thus perform a Bloch decomposition with respect to the longitudinal quasi-momentum $k\in S^{1}:= \mathbb{R}/ 2\pi\mathbb{Z}$, which remains a good quantum number even in presence of the boundary. 

We so end up with a family of Hamiltonians $H(k)$ defined on the one-dimensional lattice $\mathbb{Z}\ni n$, resp. $\mathbb{N}$, and acting on wave-functions $\psi_n\in\mathbb{C}^{N}$, where $N$ is the number of internal degrees of freedom.

More generally, these objects are stated as follows. 

\begin{defi}{\em[Bulk Hamiltonian]}\label{bh} The Hamiltonian, acting on $\psi\in\ell^2(\mathbb{Z}; \mathbb{C}^{N})$ and parametrized by $k\in S^{1}$, is 
\be
\bigl( H(k)\psi \bigr)_{n} = A(k)\psi_{n-1} + A(k)^{*}\psi_{n+1} + V_{n}(k)\psi_{n}\;,\qquad (n\in \mathbb{Z},\;\psi_{n}\in \mathbb{C}^{N})\;.\label{1.1}
\ee
The {\em potential} $V_{n}(k)$ and the {\em hopping matrices} $A(k)$ are $N\times N$ matrices having a $C^{1}$-dependence on $k$, uniformly in $n$. We assume $V_{n}(k) = V_{n}(k)^*$, where $^{*}$ denotes the matrix-adjoint, and that $A(k)\in\GL(N)$. (Recall that $\GL(N)\subset M_{N}(\mathbb{C})$ consists of invertible matrices of order $N$.)
\end{defi}

We then consider the restriction of the Hamiltonian to $\mathbb{N}=\{1,2,\ldots\}$ (we find it convenient to omit zero), while allowing for changes within a finite distance $n_0\ge 0$ from the edge.

\begin{defi}{\em[Edge Hamiltonian]}\label{eh} The Hamiltonian, acting on $\psi\in\ell^2(\mathbb{N}; \mathbb{C}^{N})$, is 
\be
\bigl( H^{\sharp}(k)\psi \bigr)_{n} = A(k)\psi_{n-1} + A(k)^{*}\psi_{n+1} + V^{\sharp}_{n}(k)\psi_{n}\;,\qquad (n\in \mathbb{N},\;\psi_{n}\in \mathbb{C}^{N})\;,\label{1.2}
\ee
where $V^{\sharp}$ satisfies the above properties of $V$, as well as 
\be\label{1.2a}
V^{\sharp}_{n}(k) = V_{n}(k)\;,\qquad (n> n_0)\;. 
\ee
Moreover we assume the Dirichlet boundary condition, meaning that for $n=1$ Eq.~(\ref{1.2}) is to be read with $\psi_{0}=0$.
\end{defi}
\begin{rem}The Dirichlet condition is by no means special. Since
\[
(\varphi,  H^{\sharp}\psi)-(H^{\sharp}\varphi,  \psi)=
\varphi_1^*A\psi_{0}-\varphi_{0}^*A^*\psi_1\;,
\]
any boundary condition $\psi_{0}=\Lambda\psi_1$, ($\Lambda(k)\in M_{N}(\mathbb{C})$) defines a self-adjoint Hamiltonian if $(A\Lambda)^*=A\Lambda$. That amounts to the Dirichlet condition after adding $\delta_{n1}A\Lambda$ to $V^{\sharp}_{n}$.
\end{rem}

Contrary to what the above motivation might suggest, the Hamiltonians (\ref{1.1}, \ref{1.2}) are prompted by more than just the square lattice. In the next section we will show that several models based on the honeycomb lattice, which have been considered \cite{KM} in relation with the quantum spin Hall effect, fit the scheme. Moreover, one-dimensional spin pumps \cite{FK} also match the description, with $k$ playing the role of time. 

The topological classification applies to {\it insulators} that are invariant under odd (or fermionic) {\it time-reversal symmetry}. In the sequel we specify these notions.

\begin{defi}{\em[Time-reversal symmetry]}\label{trs}
The symmetry is a map $\Th: \mathbb{C}^{N}\to \mathbb{C}^{N}$ with the following properties.
\begin{itemize}
\item[i)] $\Th$ is antilinear and $\Th^{2} = -1$;
%unitary w.r.t. to the standard inner product of $\mathbb{C}^{N}$;
\item[ii)] $\Th^*\Th=1$;
\item[iii)] For all $k\in S^{1}$, 
\be
H(-k) = \Th H(k) \Th^{-1}\;,\label{1.14b}
\ee
where $\Th$ also denotes the map induced on $\ell^2(\mathbb{Z}; \mathbb{C}^{N})$. Likewise for $H^{\sharp}$.
\end{itemize}
\end{defi}
As a result, $N$ is even. In the models described in the next section, properties (i, ii) arise from the time-reversal of a spin-$\half$ particle, and (iii) from the symmetry of the Hamiltonian. 

The Bulk Hamiltonian of an insulator is supposed to have a spectral gap at Fermi energy $\m$ and for all $k$:
\be
\m\notin\s(H(k))\;,\qquad (k\in S^{1})\;.\label{1.13}
\ee
Typically $H^{\sharp}$ does not satisfy the gap condition. In fact, while for the essential spectra we have $\s_\mathrm{ess}(H^{\sharp}(k))\subset\s_\mathrm{ess}(H(k))$, the edge Hamiltonian may have discrete eigenvalues crossing $\m$ for some values of $k$.

\subsection{The main result in brief}\label{mr}

We can informally introduce two indices, $\mathcal{I},\mathcal{I}^{\sharp}\in\{\pm 1\}$, defined in terms of $H$ and $H^{\sharp}$, respectively. On the circle the involution $k\mapsto -k$ has two fixed points, $k=0,\pi$. The {\it edge index} $\mathcal{I}^{\sharp}=(-1)^n$ is the parity of the number $n$ of those $k\in[0,\pi]$ at which an eigenvalue of $H^{\sharp}(k)$ equals $\m$, at least if the eigenvalue crossings are simple. The index $\mathcal{I}$ requires more explanation. By definition $H(k)$ does not have eigenvalues $z\notin\s(H(k))$; yet we may regard the Schr\"odinger equation
\[
(H(k)-z)\psi=0
\]
as a second-order difference equation in $n\in\mathbb{Z}$. As such, it has $2N$ linearly independent solutions, but we focus attention on those which decay as $n\to +\io$. They form a linear space, $E_{z,k}$, of dimension $N$. By (i, iii) we have
\be
(H(-k) - \bar z)\Th\psi =  \Th (H(k) - z)\psi\;,\nn %\label{1.14c}
\ee
and thus $E_{\bar z,-k}=\Th E_{z,k}$. We choose a reflection symmetric complex contour, $\gamma=\bar\gamma$, encircling the part of the spectrum of $H(k)$ lying below $\m$, and we set $\mathbb{T}=\gamma\times S^{1}$. We so have: (a) an involution $(z,k)\mapsto (\bar z,-k)$ on the torus $\mathbb{T}$; (b) a vector bundle with base $\mathbb{T}$ and fibers $E_{z,k}$; which (c) are compatible with $\Th$ in the stated sense. To any vector bundle with these features, including (i), but irrespective of the concrete definition of its fibers, an index will be associated in Sect.~\ref{trib}. The {\it bulk index} $\mathcal{I}$ is that index for the particular bundle arising from $H$ as described. The main result of this work is that $\mathcal{I}=\mathcal{I}^{\sharp}$. 
\begin{figure}[hbtp]
\centering
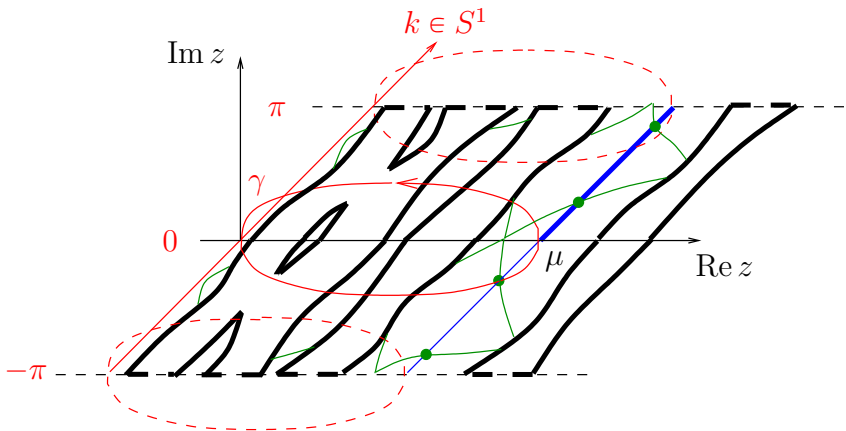
\caption{The torus $\mathbb{T}$ is the product of the loop $\g$ in the $z$-plane with the $k$-axis (both red); it is to be glued along the dashed loops. Thick lines (in black) delimit parts of the {\it bulk} spectrum $\s(H(k))$, as a function of $k$. Curves (in green) connecting them are discrete eigenvalues $\e(k)\in\s(H^{\sharp}(k))$, representing {\it edge} states. They intersect the torus only at crossing points (dots) along the Fermi line $\{z=\m\}\times S^{1}$ (in blue), one half of it (thick) being relevant for the index $\mathcal{I}^{\sharp}$. The spectra are symmetric in $k\mapsto -k$.}\label{fig2}
\end{figure}

It should be noted that the result relies on just one translational symmetry. 

\section{Examples}\label{secexa}

We show how to obtain Hamiltonians (\ref{1.1}, \ref{1.2}) from the Schr\"odinger operator on the honeycomb lattice, which models {\it graphene} in the single-particle approximation. We will consider two types of boundary conditions: {\it zigzag} and {\it armchair}. It is known that
%, in absence of a gap (which is a non-generic case), 
the spectrum of the Schr\"odinger operator depends on which boundary condition is chosen, \cite{Nak, Hat}. In particular, a zigzag boundary implies the presence of {\it zero-energy edge states}, while they are absent for the armchair boundary. In Example~\ref{ex2} we give a new proof of the last statement.

\subsection{Graphene}\label{ex2} 
\begin{figure}[hbtp]
\centering
\def\svgwidth{200pt}
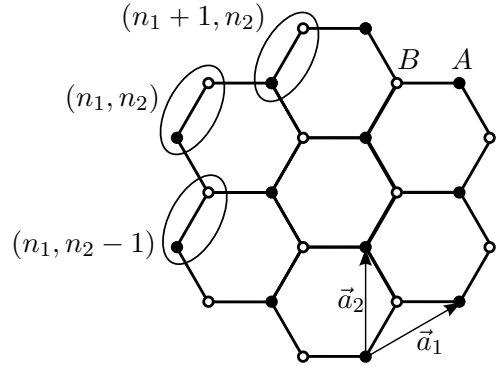
\caption{The honeycomb lattice. Coordinates of lattice sites are relative to the basis $\{\vec a_{1},\vec a_{2}\}$.}\label{fig0}
\end{figure}

Let $\Lambda = \Lambda_{A}\cup\Lambda_{B}$ be the (infinite) honeycomb lattice and its bipartite decomposition into two triangular lattices $\Lambda_{A}$, $\Lambda_{B}$. Upon fixing an origin, they are
\begin{gather}
\Lambda_{A} := \big\{ \vec n = n_1 \vec a_1 + n_{2}\vec a_2\mid (n_{1},n_{2})\in \mathbb{Z}^{2}\big\}\;,\qquad \Lambda_{B} := \Lambda_{A} + \vec \d\;,\nn\\
|\vec a_1|=|\vec a_2| \;,\qquad \angle(\vec a_1,\vec a_2)=\frac{\pi}{3}\;,
%\vec a_1 \cdot\vec a_2=\half|\vec a_1||\vec a_2| \;,
\qquad \vec \d = \frac{1}{3}(\vec a_1+\vec a_2)\;.\nn%\label{1.4a}
\end{gather}
Any site in $\Lambda_{A}$ has three nearest neighbors in $\Lambda_{B}$, shifted by $\vec \d$ or by another equivalent vector. See Fig.~\ref{fig0}. The model for graphene is simply the Schr\"odinger operator $H_0$ for a particle hopping between nearest neighbors (with hopping parameter $-t$).
\comment{Given $\vec n\in\L_{A}$, its three nearest neighbors are $\vec n+\vec \d_{i}\in\L_{B}$, $i = 1,2,3$, with $\vec \d_{1}$ defined in (\ref{1.4a}) and $\vec \d_{2} = \frac{1}{2}(-1,\,\sqrt{3})$, $\vec \d_{3} = \frac{1}{2}(-1,\,-\sqrt{3})$. }
%The dual lattice $\Lambda_{A}^{*}$ of the triangular lattice $\Lambda_A$ is given by
%
%\bea
%\Lambda_{A}^{*} := \big\{ \vec k = k_{1}\vec b_{1} + k_2\vec b_{2}: (k_1,k_2)\in \mathbb{Z}_{2}\big\}\;,\nn\\
%\vec b_{1} := \Bigl( -\frac{4\pi}{3},\,0 \Bigr)\;,\qquad \vec b_{2} := \Bigl( \frac{2\pi}{3}\,, \frac{2\pi}{\sqrt{3}} \Bigr)\;.\nn %\label{1.4b}
%\eea
%

\begin{figure}[hbtp]
\centering
\def\svgwidth{350pt}
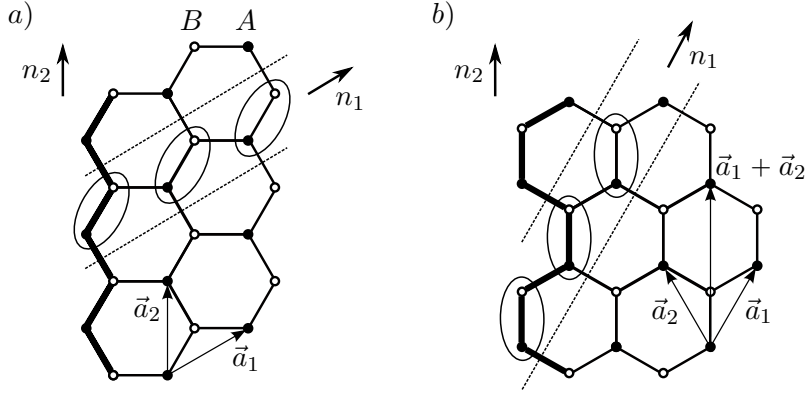
\caption{The honeycomb lattice with zigzag $a)$ and armchair $b)$ boundary conditions. The lattice has been rotated, so as to have $n_1$ fixed along the vertical.}\label{fig1}
\end{figure}

The (Bravais) lattice of translations of $\L$ is $\L_{A}$. We reduce $\L$ to $\L_{A}$ by dimerizing neighbors shifted by $\vec \d$. We retain the position $\vec n$ of the $A$-site as that of the dimer, and the values of the wave function at the two sites as pseudospin components:
\be
\psi_{\vec n} := \begin{pmatrix} \psi^{A}_{\vec n} \\ \psi^{B}_{\vec n} \end{pmatrix} \in \mathbb{C}^{2}\;.\label{1.4c}
\ee 
With these notations, the Schr\"odinger operator $H_0$ takes the form 
\be
\bigl(H_0\psi\bigr)_{n_1,n_2} = -t\begin{pmatrix} \psi^{B}_{n_1,n_2} + \psi^{B}_{n_1, n_2-1} + \psi^{B}_{n_1-1,n_2} \\ \psi^{A}_{n_1,n_2+1} + \psi^{A}_{n_1+1,n_2} + \psi^{A}_{n_1,n_2}\end{pmatrix} \;.\label{1.4d}
\ee
The expression for $H_0$ could of course have ended up differently. Let us name the choices underlying its construction: a sublattice shift vector (above: $\vec\d$), defining the dimer, and two primitive lattice vectors ($\vec a_1$ and $\vec a_2$), defining adjacency between dimers. For instance for $\vec a_1^{\,'}=\vec a_1$, $\vec a_2^{\,'}=\vec a_1+\vec a_2$, and hence $n_1'=n_1-n_2$, $n_2'=n_2$, Eq.~(\ref{1.4d}) becomes
\be
\bigl( \widetilde H_0\psi\bigr)_{n_1,n_2} = -t\begin{pmatrix} \psi^{B}_{n_1,n_2} + \psi^{B}_{n_1+1, n_2-1} + \psi^{B}_{n_1-1,n_2} \\ \psi^{A}_{n_1-1,n_2+1} + \psi^{A}_{n_1+1,n_2} + \psi^{A}_{n_1,n_2}\end{pmatrix}\;, \nn %\label{1.4e}
\ee
after dropping primes. See Fig.~\ref{fig1}.

The Bloch decomposition w.r.t. $n_2\in\mathbb Z$
\be
\psi_{\vec n} = \int_{S^{1}} \frac{dk}{2\pi}\,e^{ik n_2}\psi_{n_1}(k)\;,\qquad (\psi_{n_1}(k)\in\mathbb{C}^{2})\label{1.4f}
\ee
fibers the Hamiltonian, $(H_0\psi)_{n_1}(k)=(H_0(k)\psi(k))_{n_{1}}$. In fact we obtain from Eq.~(\ref{1.4d})
\be
\bigl(H_0(k)\psi\bigr)_{n} = A_0(k)\psi_{n-1} + A_0(k)^*\psi_{n+1} + V_0(k)\psi_{n}\;,\nn %\label{1.5}
\ee 
where we set $\psi=\psi(k)$, $n=n_1$ and 
\be
A_0(k) = -t\begin{pmatrix} 0 & 1 \\ 0 & 0 \end{pmatrix}\;,\qquad V_0(k) = -t\begin{pmatrix} 0 & 1 + e^{-ik} \\ 1 + e^{ik} & 0 \end{pmatrix}\;.\label{1.6}
\ee
Likewise for $\widetilde H_0(k)$:
\be
\bigl(\widetilde H_0(k)\psi\bigr)_{n} = \widetilde A_0(k)\psi_{n-1} + \widetilde A_0(k)^*\psi_{n+1} + \widetilde V_0(k)\psi_{n}\;,\label{1.6c}
\ee
where now
\be
\widetilde A_0(k) = -t\begin{pmatrix} 0 & 1 \\ e^{ik} & 0 \end{pmatrix}\;,\quad \widetilde V_0(k) = -t\begin{pmatrix} 0 & 1 \\ 1 & 0 \end{pmatrix}\;.
\nn%\label{1.7}
\ee 
It should be noted that only the second Hamiltonian satisfies the condition $A(k)\in\GL(N)$ in Eq.~(\ref{1.1}). The two equivalent bulk Hamiltonians no longer are once they are turned into edge Hamiltonians by means of the Dirichlet boundary condition. Indeed, they correspond to (a) zigzag and (b) armchair boundary conditions, respectively. See again Fig.~\ref{fig1}.

\begin{prop} (i) The Hamiltonian $H_0^{\sharp}$ has $E=0$ as an eigenvalue. Actually, $H^{\sharp}_0(k)$ has it for $k\in(2\pi/3,4\pi/3)$. (ii) The Hamiltonian $\widetilde H_0^{\sharp}$ does not have any eigenvalue, \ie its spectrum is purely continuous.
\end{prop}

The result is known \cite{FWNK, Nak}, but perhaps not the argument below for (ii).\\

\noindent
{\it Proof.} i) Related to $A_0(k)$ being singular, $H_0^{\sharp}(k)\psi=0$ reduces to the first order equations 
\[
\psi^{B}_{n-1}+(1+e^{-ik})\psi^{B}_{n}=0\;,\qquad
\psi^{A}_{n+1}+(1+e^{ik})\psi^{A}_{n}=0\;,\qquad(n\in\mathbb{N})\;.
\]
The boundary condition $\psi_0=0$ implies $\psi^{B}=0$, but is ineffective for $\psi^{A}$, for which there is a non-trivial solution in $\ell^2(\mathbb{N})$, as long as $|1+e^{ik}|<1$, \ie for $k$ in the stated range. 

ii) Every second dimer of the armchair boundary is exposed (see Fig.~\ref{fig1} right) and the axis containing them is $n_1=0$. Let us consider, on the whole lattice $\Lambda$, the reflection ${\vec n}\mapsto r{\vec n}$ of lattice sites about the axis $n_1=0$ in Fig.~\ref{fig1}, as well as parity, $(P\psi)_{\vec n}=\psi_{r\vec n}$. Clearly, $[P,H_0]=0$. Any odd eigenfunction $\psi$ of $H_0$ satisfies the Dirichlet boundary condition on the line $n_1=0$, whence its restriction $\widetilde\psi$ to $n_1>0$ defines one for $\widetilde H_0^{\sharp}$. What matters more is that the converse is true as well: The odd extension $\psi$ of $\widetilde\psi$ satisfies the Schr\"odinger equation on $n_1=0$, and hence at all $\vec n$. The conclusion is by noting that $H_0$ has no eigenvalues; actually it has purely absolutely continuous spectrum. This is known and by the way has a short proof: It suffices to establish the property for the fiber Hamiltonian $\widetilde H_0(k)$. Since the latter has real analytic Bloch eigenvalues $\l(q)$, absolute continuity holds unless $\l(q)$ is constant in $q$. That, finally, is ruled out by $\widetilde A_0(k)\in\GL(N)$ in Eq.~(\ref{1.6c}) and the remark below.\qed

\begin{rem} (\cite{ASV}, Prop. 7) Suppose the Hamiltonian on $\ell^2(\mathbb{Z}; \mathbb{C}^{N})$
\ben
( H\psi)_{n} = A\psi_{n-1} + A^{*}\psi_{n+1} + V\psi_{n}
\een 
(note: $V$ independent of $n$) has a Bloch eigenvalue $\l(q)$ independent of the quasi-momentum $q$. Then $A$ is singular.
\end{rem}

\noindent
{\it Proof.} Let $\l$ be that eigenvalue. Then $\det(Az^{-1}+A^{*}z+V-\l)=0$
for $z=e^{iq}$, ($q\in\mathbb{R}$) and, by analyticity, for $z\neq 0$. Letting $z\to 0$ gives $\det A=0$.\qed.

\subsection{The Kane-Mele model}
In the next example we discuss a model with odd time-reversal symmetry. Wave functions are now of the form $\psi=(\psi_{\vec n})_{\vec n\in\mathbb{Z}^2}$, where $\vec n$ labels dimers, see Fig.~\ref{fig0}, and $\psi_{\vec n}\in\mathbb{C}^2\otimes \mathbb{C}^2$. The spin of the particle is represented by the Pauli matrices $\s_1$, $\s_2$, $\s_3$ acting on the second factor $\mathbb{C}^2$. The time-reversal operator is $\Th = -i(1\otimes \s_{2})C$, where $C$ denotes complex conjugation, which is the usual expression for a spin-$\half$ particle.

Let $H_0$ be the Schr\"odinger operator on the honeycomb lattice defined in Eq.~(\ref{1.4d}). We consider the operator
\be
H_\mathrm{KM}=H_0\otimes 1+H_1\otimes\s_3\;,\nn% \label{1.14da}
\ee
where the last term implements a {\em spin-orbit coupling}. There $H_1$ describes hopping between next-to-nearest neighbors; it thus acts diagonally on the pseudospin components of Eq.~(\ref{1.4c}), and specifically as 
\be
H_1\psi=\begin{pmatrix}h_1\psi^A\\-h_1\psi^B\end{pmatrix}\label{h1}
\ee
with 
\be
(h_1\psi)_{\vec n}=-t'\Bigl(i(\psi_{n_1-1,n_2+1} +\psi_{n_1+1,n_2} +\psi_{n_1,n_2-1})-i(\psi_{n_1,n_2+1}+\psi_{n_1+1,n_2-1}+\psi_{n_1-1,n_2})\bigr)\;.\nn
\ee
The grouping of terms reflects that, starting from a given $A$-site, its next-to-nearest neighbors are reached by turning right or left at a $B$-site. The turns go along with phases $\pm i=e^{\pm i\pi/2}$ modeling a magnetic flux $\pi/2$ through any (positively oriented) triangle $ABA$; yet the total flux through the hexagonal cell vanishes, since no phases are associated with the bonds forming its boundary, see Eq.~(\ref {1.4d}). Likewise for $B$ and $A$ interchanged, resulting in the sign in (\ref{h1}). Indeed, for a given dimer, the next-to-nearest neighbors form again dimers, but the turns linking $A$, resp. $B$-sites are opposite. 

The model is the special case of the {\em Kane-Mele model}, \cite{KM}, where two further terms (a Rashba term and a staggered chemical potential) have been set to zero. Even so, it exhibits a non-trivial topological phase, \ie $\mathcal{I}=-1$, for small $t'/t> 0$ and $\mu=0$. In fact, it is the direct sum of (a special case of) the {\em Haldane model} \cite{H}, $H_\mathrm{H}=H_0+H_1$, and of its time-reversed copy $H_0-H_1$. In such a situation we have $\mathcal{I}=(-1)^\mathcal{N}$, where $\mathcal{N}$ is the integer associated with the quantum Hall effect of $H_\mathrm{H}$. In the stated regime, a band gap of order $O(|t'|)$ opens, and $\mathcal{N}=1$, \cite{H}.

We conclude the example by giving the fibers of $H_\mathrm{KM}$ w.r.t. the Bloch decomposition (\ref{1.4f}). It will suffice to do so for $H_\mathrm{H}$: Instead of Eq.~(\ref{1.6}) we have
\be
A(k) = \begin{pmatrix} -i t'(e^{i k} - 1) & -t \\ 0 & i  t'(e^{i k} - 1) \end{pmatrix}\;,\qquad
V(k) =\begin{pmatrix} i  t'(e^{i k} - e^{-i k}) & -t(1 + e^{-ik}) \\ -t(1 + e^{ik}) & -i  t'(e^{i k} - e^{-i k})
\end{pmatrix}\;.\nn
\ee
\comment{
There are four energy bands, separated in pairs by a band gap $O(t')$, uniformly in $k\in S^{1}$.} 
\section{Time-reversal invariant bundles}\label{trib}
The purpose of this section is to define the index of bundles of the kind mentioned in Sect.~\ref{mr}, see Def.~\ref{def:bulk} below, as well as some auxiliary indices; and to formulate some of their properties. We refer the reader to Sect.~\ref{secproofs} for the proofs of the lemmas stated here. 

\subsection{The index of endpoint degenerate families}\label{edf}

We shall define an index for certain families of points on the unit circle. We will give the most general definition, in order to make evident the stability of the index under homotopy. At the end of this section we give a procedure to compute it in a more restrictive setting, which is suitable for our application.

Following (\cite{Ka}, Sect.~II.5.2) we consider unordered $N$-tuples $Z=(z_{1}\,,\ldots\,,z_{N})$ of complex numbers $z_{i}\in\mathbb{C}$. A distance is defined by
\be
d(Z',Z) = \min\max_{n}|z_{n} - z'_{n}|\;,\nn %\label{2.1.1}
\ee
where the minimum is taken over all possible relabellings of $Z$ or $Z'$. Let us recall (\cite{Ka}, Thm.~II.5.2): Given a family $Z(x)$, $(x\in[a,b])$, which is continuous w.r.t. $d$, there exists a (non-unique) continuous labeling $Z(x) = (z_{1}(x)\,,\ldots\,,z_{N}(x))$.

In the following we will consider $N$-tuples with $z_{i}\in S^{1}$. Any continuous labeling $z_{i}(x)$ induces continuous arguments $\th_{i}(x)$, \ie $z_{i}(x) = e^{i\th_{i}(x)}$. Two simple observations are in order: (i) $w(Z):= (2\pi)^{-1}\sum_{i=1}^{N}\th_{i}(x)\big|_{a}^{b}$ is independent of the choice of arguments, as well as of labeling. In fact $\prod_{i=1}^{N}z_{i}(x)$ is independent of the latter. (ii) Any family $Z(x)$ with $Z(a)=Z(b)$ has a {\em winding number} given as
\be
 \mathcal{N}(Z) := w(Z) \in \mathbb{Z}\;.\label{2.1.2}
\ee 

We next consider {\it endpoint degenerate families} $Z(\cdot)$: for $x=a,b$ each $z\in Z(x)$ occurs with even multiplicity. We may concatenate such a family with one, $\tilde Z(x)$, $(x\in [b,\,c])$, such that $\tilde Z(b) = Z(b)$ and $\tilde Z(c) = Z(a)$, while keeping $\tilde z_{i}(x)$ even degenerate. Clearly, $\tilde Z$ is not unique, but by (\ref{2.1.2})
\be
w(\tilde Z_{1}) - w(\tilde Z_{2}) \in 2\cdot  \mathbb{Z}\;.\nn %\label{2.1.3}
\ee
By the same reason, $\mathcal{N}(Z\#\tilde Z)$ is an integer; by $\mathcal{N}(Z\# \tilde Z) = w(Z) + w(\tilde Z)$ it is determined $\mod 2$ by $Z$. 
\begin{defi}{\em [Index]} We set
\be
\mathcal{I}(Z) = (-1)^{\mathcal{N}(Z\# \tilde Z)}\;,\label{2.1.4}
\ee
as the index of endpoint degenerate families.
\end{defi}

Consider now the special situation where there is $z\in S^{1}$ such that $z\in Z(x)$ occurs only at finitely many $x$, which moreover are simple crossings: $z = z_{j}(x)$ for a single $j$, and $z'_{j}(x)\neq 0$. Then
\be
\mathcal{I}(Z) = (-1)^{n}\;,\label{2.0.5}
\ee
where $n$ is the number of crossings of $z$. Indeed, one can choose $\tilde Z$ without crossings of $z$. Then $\mathcal{N}(Z\# \tilde Z)$ is the number of {\it signed} crossings of $z$ by $Z(\cdot)$ but that qualification is irrelevant for parity.

\subsection{The index of Kramers families of matrices} 

Let $\e$ be the matrix of even order $N$ given by the block diagonal matrix with blocks
\be
\begin{pmatrix} 0 & -1 \\ 1 & 0 \end{pmatrix}\;,\nn %\label{1.17}
\ee 
$C:\mathbb{C}^{N}\to \mathbb{C}^{N}$ the complex conjugation, and $\Th_{0} = \e C$ the standard time-reversal on $\mathbb{C}^{N}$. Suppose $T\in \GL(N)$ satisfies 
\be\label{2.6.1c}
\Th_0 T =  T^{-1}\Th_0\;.
\ee
Then the eigenvalues of $T$ come in pairs $\l$, $\bar \l^{-1}$ with equal algebraic multiplicity, which is moreover even if $\l=\bar \l^{-1}$. In particular their phases $z=\l/|\l|$ are even degenerate regardless. Indeed, 
$\Th_0(T-\l)^n=T^{-n}(1-\bar\l T)^n\Th_0$, as seen inductively for $n=0, 1, \ldots$; for $\l=\bar \l^{-1}$ the corresponding pairs of eigenvectors $v, \Th_0v$ remain linearly independent by $ \Th_0^2=-1$.

\begin{defi}{\em [Kramers property]}
We call Eq.~(\ref{2.6.1c}) the {\em Kramers property}. We say a family $T(\varphi) \in \GL(N)$, which is continuous in $0\leq\varphi\leq \pi$, has that property if the endpoints $T(0)$ and $T(\pi)$ have it.
\end{defi}
The repeated eigenvalues $\l_i(\varphi)$ of $T(\varphi)$ form a continuous family in the sense of the previous section (\cite{Ka}, Thm.~II.5.1), and so do the $z_i=\l_i/|\l_i|$. Moreover, $Z(\varphi)=(z_1(\varphi),\ldots z_N(\varphi))$ is an endpoint degenerate family. 

\begin{defi}{\em[Index]} We set
\be
\mathcal{I}(T) =\mathcal{I}(Z)\label{2.6.1b}
\ee
as the index of a Kramers family $T$. See Eq.~(\ref{2.1.4}).
\end{defi}
All it in fact takes for the definition is the endpoint degeneracy of $Z$ and not the stronger Kramers property of $T$. However we shall not need such an extension.
\begin{lemma}\label{lemmolt} Suppose 
\be
T_2(\varphi) =M_-(\varphi)T_1(\varphi)M_+(\varphi)^{-1}\nn
\ee
with continuous $M_\pm(\varphi) \in \GL(N)$ ($0\leq\varphi\leq \pi$), as well as
\be\label{2.6.1d}
M_-(0)\Th_0=\Th_0 M_+(0)\;,\qquad M_-(\pi)\Th_0=\Th_0 M_+(\pi)\;.
\ee
Then $M(\varphi) := M_{-}(\varphi)M_+(\varphi)^{-1}$ has the Kramers property, and $T_2(\varphi)$ has it iff $T_1(\varphi)$ does. If so,
\be
\mathcal{I}(T_2) = \mathcal{I}(T_1)\mathcal{I}(M)\;.\label{molt1}
\ee
\end{lemma}
The claims are of immediate verification, except for Eq.~(\ref{molt1}). However in the special case that $M_+(\varphi) = M_-(\varphi)$ the equality
\be
\mathcal{I}(T_2) = \mathcal{I}(T_1)\label{2.6.1e}
\ee
is also immediate, because $T_1(\varphi)$, $T_2(\varphi)$ then have the same eigenvalues. That case suffices for the basic result on bulk-edge duality.

\subsection{The index of time-reversal invariant bundles}\label{secchar}

Let $S^{1} = \mathbb{R}/ 2\pi\mathbb{Z}$ be the circle and $\mathbb{T} = S^{1}\times S^{1} \ni \varphi = (\varphi_1,\varphi_2)$ the torus with involution $\tau:\varphi\mapsto -\varphi$. It has four fixed points: $\varphi_0 = (0,0),\, (0,\pi),\, (\pi,0),\,(\pi,\pi)$. Let $E$ be a complex vector bundle with base $\mathbb{T}\ni \varphi$ and fibers $E_{\varphi}$ of dimension $N$. We say that $E$ is {\em time-reversal invariant} if there is a map $\Th: E\to E$ with $\Th^{2} = -1$ and $\Th: E_{\varphi}\mapsto E_{\tau\varphi}$ antilinear. 

We also consider the associated frame bundle $F(E)$ with the following operations induced on frames $v = (v_{1},\ldots, v_{N})\in F(E)_{\varphi}$:
\begin{itemize}
\item right multiplication by $M\in \GL(N)$:
\be
M: F(E)_{\varphi}\to F(E)_{\varphi},\qquad v\mapsto vM\;,\label{1.16}
\ee
with $(v M)_{j} = \sum_{i=1}^{N}v_{i}M_{ij}$. Any two frames are so related by a unique $M$.
\item $\Th: F(E)_{\varphi}\to F(E)_{\tau\varphi}$, $v\mapsto \Th v$, with $(\Th v)_{i} = \Th v_{i}$. Note that
\be
\Th(vM) = (\Th v)\overline{M}\;.\label{1.16b}
\ee
\end{itemize}
In order to classify time-reversal invariant bundles we consider the torus cut along the circle $\{\varphi_{1}\mid \varphi_{1} = \pi \cong -\pi\}\times S^{1}$, or more precisely $\dot{\mathbb{T}} = [-\pi,\pi]\times S^{1}$. Note that only $\varphi = (0,0),\, (0,\pi)$ remain among the fixed points of $\t$, and that any bundle on $\mathbb{T}$ naturally defines one on $\dot{\mathbb{T}}$.
\begin{lemma}{\em[Existence of time-reversal invariant sections]}\label{lemsec}
On the cut torus $\dot{\mathbb{T}}$, there are (smooth) sections $v:\dot{\mathbb{T}}\to F(E)$ of the frame bundle (whence $v(\varphi)\in F(E)_{\varphi}$) satisfying
\be
v(\t\varphi) = \Th v(\varphi)\e\;.\label{1.18}
\ee
As a result, they are compatible with right multiplication by $M(\varphi)\in \GL(N)$ iff
\be
 \Th_{0}M(\t\varphi) =M(\varphi)\Th_{0}\;.\label{1.19}
\ee
\end{lemma}
We remark that the condition (\ref{1.18}) was shown \cite{FK} to be obstructed on the (uncut) torus by the $\mathbb{Z}_{2}$-invariant.

Let $v_{\pm}(\varphi_2) := v(\pm \pi,\varphi_2)$ be the boundary values of $v(\varphi)$ along the two sides of the cut, and $T(\varphi_2)\in\GL(N)$ the transition matrix,
\be
v_{+}(\varphi_2) = v_{-}(\varphi_2)T(\varphi_2)\;,\qquad (\varphi_2\in S^{1})\;.\label{1.20}
\ee
\begin{lemma}\label{lemtrans}{\em [Time-reversal symmetry of the transition functions]}\label{lemtrsU}
\be
\Th_0^{-1}T(-\varphi_2)\Th_{0}T(\varphi_2) = 1\;.\nn %\label{1.21}
\ee
\end{lemma}
%
%We refer the reader to Section~\ref{prooftrans} for a proof. 
In particular, $T(\varphi_2)$ has the Kramers property on the interval $0\le \varphi_2\le\pi$, which hints at a the possibility of assigning an index to the bundle.
\begin{lemma}{\em [Independence of the index from the section]}\label{propind}
Let $v^{(i)}$, ($i=1,2$) be time-reversal invariant sections on the cut torus $\dot{\mathbb{T}}$, and let $T_i(\varphi_2)$ the corresponding transitions matrices across the cut. Then
\be
\mathcal{I}(T_1) = \mathcal{I}(T_2)\;.\nn %\label{2.6.7}
\ee
\end{lemma}
We may thus proceed to the following definition. 
\begin{defi}{\em[Index]} \label{def:bulk} 
We set
\be
\mathcal{I}(E) =\mathcal{I}(T) \label{2.6.7a}
\ee
as the index of a time-reversal invariant vector bundle $E$ over $\mathbb{T}$. See Eq.~(\ref{2.6.1b}).
\end{defi}
Such  bundles can hence be distinguished in two topologically distinct classes, according to the value of the index. 

We also retain the following remark, which appears as a byproduct of Lemma~\ref{lemsec}.
\begin{rem} \label{CN}
Let $\Th: \mathbb{C}^{N}\to \mathbb{C}^{N}$ satisfy the time-reversal symmetry condition (i) of Def.~\ref{trs}. Then there is a basis $v = (v_{1},\ldots, v_{N})$ of $\mathbb{C}^{N}$ such that $v = \Th v\e$. If condition (ii) applies too, the basis can be chosen orthonormal. Letting $U$ map $v$ to the standard basis of $\mathbb{C}^{N}$, a restatement is $U\Th=-\Th_0 U$ with $U$ unitary.
\end{rem}

\section{The bulk-edge correspondence}\label{sec2.1} 

It pays to look at first at wave-functions $\psi=(\psi_n)_{n\in\mathbb{Z}}$ as just sequences, \ie $\psi\in\mathcal{C}:=\mathbb{Z}\times\mathbb{C}^N$. It can be shown that for any $z\in \r(H(k))$ in the resolvent set, the Schr\"odinger equation $H(k)\psi = z\psi$ has $N$ linearly independent solutions which are square-integrable at $n\to +\io$. Let $E_{z,k}\subset\mathcal{C}$ the linear space they form. As explained in Sect.~\ref{mr} we obtain a vector bundle
\be
E = \{ ((z,k),\psi)\in \mathbb{T}\times \mathcal{C}\mid\psi\in E_{z,k} \}\label{2.1}
\ee
over the torus $\mathbb{T}=\gamma\times S^{1}$, which enjoys the property of being time-reversal invariant. The following definition is thus natural.
\begin{defi} We set
\be
\mathcal{I}=\mathcal{I}(E)
\label{bi}\ee
as the {\em bulk index}. See Eq.~(\ref{2.6.7a}).
\label{defbi}
\end{defi}
The edge index has been loosely introduced in Sect.~\ref{mr}. In preparation for precise definition let us take a closed interval $I$ containing the Fermi energy $\m$ in its interior and such that $\s(H(k))\cap I=\varnothing$ for $k\in S^{1}$. The spectrum of $H ^{\sharp}(k)$ is discrete and finite in $I$ (uniformly in $k$), and the eigenvalues are locally given by branches $\e_i(k)$ which are $C^{1}$ in $k$ (\cite{Ka}, Thm.~II.6.8). By possibly adjusting $\mu$ we can arrange that if $\e_i(k_*)=\mu$ for some $k_*$ then $\e_i'(k_*)\neq 0$. We note that the number of crossings at $k_*=0$ or $k_*=\pi$ is even by (\ref{1.14b}). 
\begin{defi} Let $n$ be the number of crossings $k_*\in[0,\pi]$, with those at endpoints counted half. Set
\be
\mathcal{I}^{\sharp}=(-1)^n\nn
\ee
as the {\em edge index}.
\label{defei}
\end{defi}
The next remark about the edge index is inessential for the following main result.
\begin{rem} The index $\mathcal{I}^{\sharp}$ could be also defined in terms of a Kramers family. To this end let $f(\e)$ be a continuous real function with $f(\e)=0$, ($\e<I$) and $f(\e)=1$, ($\e>I$), where $I$ is the aforementioned interval. Then, in view of $\Th g(H)\Th^{-1}=\bar g(\Th H\Th^{-1})$, the operator $T(k)=\exp(2\pi if(H ^{\sharp}(k)))$ has the Kramers property, albeit w.r.t. $\Th$, which is however irrelevant. The definition $\mathcal{I}^{\sharp}=\mathcal{I}(T)$ would be legitimate, since that of $w(Z)$ in Sect.~\ref{edf} extends to countable families $Z$ of points $z_i(x)\in S^{1}$, as long as only finitely many (uniformly in $x$) are $\neq 1$. It would agree with Def.~\ref{defei} by Eq.~(\ref{2.0.5}).
\end{rem}
\begin{thm}{\em [Bulk-edge correspondence for topological insulators]}\label{thm1}
Let the Bulk and the Edge Hamiltonian be as in Defs.~\ref{bh} and \ref{eh}. Assume the time-reversal symmetry conditions of Def.~\ref{trs} and the gap condition (\ref{1.13}). Then
\be
\mathcal{I}=\mathcal{I}^{\sharp}\;.\nn
\ee
\end{thm}
We can give the main steps of the proof right away. For given $(z,k)\in \r(H(k))\times S^{1}$ we consider, besides of $E_{z,k}$, also the linear space $E_{z,k}^{\sharp}$ of solutions $\psi^{\sharp}=(\psi_n^{\sharp})_{n\in \mathbb{Z}}$ decaying at $n\to\io$ of the Schr\"odinger equation $H^{\sharp}(k)\psi^{\sharp} = z\psi^{\sharp}$, without imposing boundary conditions at $n=0$. Eq.~(\ref{1.2a}) establishes a bijection 
\be\label{bij}
E_{z,k}\to E_{z,k}^{\sharp}\;, \qquad \psi\mapsto \psi^{\sharp}
\ee
determined by $\psi_n^{\sharp}=\psi_n$ for $n>n_0$. 

Frames $\Psi\in F(E)_{z,k}$ consists of $N$-tuples $\Psi=(\psi_1, \ldots, \psi_N)$ of linearly independent solutions $\psi_i\in  E_{z,k}$. Note that the index $i$ does not denote the lattice site $n\in \mathbb{Z}$. Since $\psi_{in}\in \mathbb{C}^N$, we may equivalently say: $\Psi=(\Psi_n)_{n\in \mathbb Z}$ with $\Psi_n\in M_{N}(\mathbb{C})$ belongs to $F(E)_{z,k}$ iff $\Psi$ is a solution of $H(k)\Psi=z\Psi$ decaying at $n\to+\io$ , which is moreover fundamental in the sense that for any $n$
\be
\Psi_{n}a=0\,,\; \Psi_{n+1}a=0\Rightarrow a=0\;, \qquad (a\in \mathbb{C}^{N}).\label{1.10}
\ee
The bijection (\ref{bij}) induces one between frame bundles, $F(E)_{z,k}\to F(E^{\sharp})_{z,k}$, $\Psi\mapsto \Psi^{\sharp}$. It is manifestly compatible with the right action (\ref{1.19}) of $\GL(N)$. The next lemma rests on the bijection. 
\begin{lemma}\label{lem1}
\begin{enumerate}
\item A point $(z_*,k_*)\in\mathbb{T}$ has $\det\Psi^{\sharp}_{0}=0$ for some (and hence all) $\Psi\in F(E)_{z_*,k_*}$ iff $z_*\in\s( H^{\sharp}(k_*))$. If so, then $z_*=\m$. For a dense set of Hamiltonians $H^{\sharp}$ near the given one, the points $k_*$ are isolated in $S^{1}$ and for each of them there is a simple eigenvalue branch $\e(k)$ with $\e(k_*)=\m$, $\e'(k_*)\neq 0$; moreover,
\be
\det \Psi^{\sharp}_{1}\neq 0\;.\label{2.7}
\ee
Density is meant with respect to the topology of the class of Hamiltonians specified at the beginning of Sect.~\ref{sec1.1}.
\item Let $\Psi(z,k)\in F(E)_{z,k}$ be a section defined in a neighborhood in $\mathbb{C}\times S^{1}\supset \mathbb{T}$ of any of the crossing points $(z_{*}=\m,k_*)$. The family of matrices 
\be
L(z,k) = -{\Psi_{1}^{\sharp}}^*(\bar z,k)A(k)\Psi^{\sharp}_{0}(z,k)\nn%\label{2.7a}
\ee
has the reflection property $L(z,k) = L(\bar z,k)^*$. Its eigenvalues are thus real for real $z$. There generically is a single eigenvalue branch $l(z,k)$ of $L(z,k)$ vanishing to first order at $(\m,k_*)$: There the derivatives $\partial l/\partial z$ and $\partial l/\partial k$ are real and non-zero. 
\comment{Its winding number there is
\be
w_{k_*} = -\sgn\Bigl( \frac{\partial l}{\partial z}\frac{\partial l}{\partial k} \Bigr)\Big|_{(z=\m, k= k_*)}\;.\nn %\label{2.9}
\ee}
\item At any of the points $(\m,k_*)$ we have
\be
\frac{\partial l}{\partial z}<0\;.\label{2.10}
\ee
\item As $k$ increases past $k_*$ the eigenvalue $\e(k)$ crosses $\m$ as an increasing function if
\be
\frac{\partial l}{\partial k}\Big|_{(z=\m, k=k_*)}>0\;,\nn %\label{2.11}
\ee
and as a decreasing one in the opposite case.
\end{enumerate}
\end{lemma}
The lemma, which is proven in Sect.~\ref{secproofs}, allows to complete the proof of the main result. As a matter of fact, only items (i, ii) matter to that end. For later use we mention that no use of time-reversal symmetry has been made in the lemma, in the sense that the statement about density of the Hamiltonians in (i) holds in either class, with and without that specification.\\

\noindent
{\it Proof of Thm.~\ref{thm1}.} As a preliminary we recall Remark~\ref{CN}: At the price of conjugating the Hamiltonians by $U$ in the internal space $\mathbb{C}^N$, which preserves the assumptions, we may assume
\be\label{th}
\Th=-\Th_0\;.
\ee
By density it will suffice to prove the theorem for Hamiltonians as specified in part (i) of the lemma. By the first sentence there, we can define a section $(z,k)\mapsto F(E)_{z,k}$ away from crossing points by requiring
\be
\Psi^{\sharp}_{0}(z,k) = 1\;.\label{2.3}
\ee
Clearly, the requirement can not be imposed at such points. In a small reflection symmetric disk $D=\bar D\subset\mathbb{T}$ containing the generic crossing point $(\m,k_*)$, Eq.~(\ref{2.7}) allows us to make an alternate choice $\widehat\Psi(z,k)$ by requiring 
\be
\widehat{\Psi}^{\sharp}_{1}(z,k) = 1\;.\label{2.3a}
\ee
\begin{figure}[hbtp]
\centering
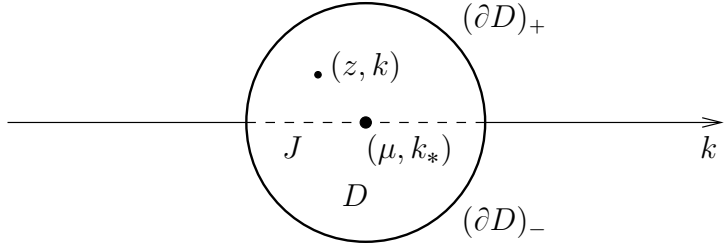
\caption{The neighborhood in $\mathbb{T}$ (see Fig.~\ref{fig2}) of a crossing point $(\m,k_*)$. Two different local sections are defined inside ($\widehat\Psi$) and outside ($\Psi$) of a small disk $D$ containing $(\m,k_*)$. They are glued across the two half-circles $(\partial D)_\pm$ by means of matrices $M_\pm(k)$.}\label{fig6}
\end{figure}

Based on the latter section we shall extend the former to the cut torus $\dot{\mathbb{T}}$, where the cut is the Fermi line $\{z=\m\}\times S^{1}$. On the boundary $\partial D$ both sections are defined, and we have 
\be
\widehat\Psi(z,k) = \Psi(z,k)M_\pm(k)\;,\label{2.4}
\ee
for $(z,k)\in (\partial D)_\pm=\partial D\cap \{\pm \Im z>0\}$ and some matrices $M_\pm(k)\in\GL(N)$, see Eq.~(\ref{1.16}), which are parametrized by $k$ in the interval $J$ resulting from the intersection of the Fermi line with $D$. The same matrices relate $\widehat\Psi^{\sharp}$ and $\Psi^{\sharp}$. We may then obtain the announced extension by (re)defining $\Psi(z,k)$ through Eq.~(\ref{2.4}) for $(z,k)\in D_\pm=D\cap \{\pm \Im z>0\}$. We observe that the section on $\dot{\mathbb{T}}$ so constructed satisfies Eq.~(\ref{1.18}); in fact its ingredients (\ref{2.3}, \ref{2.3a}) do by (\ref{th}) and $(-\Th_01)\e=1$. 

The transition matrix $T(k)$ across the cut differs from the identity only within the intervals $J$. Along such an interval the boundary values of the section are $\Psi_\pm(k)=\widehat\Psi(\m,k)M_\pm(k)^{-1}$, resulting in the transition matrix
\be
T(k)=M_-(k)M_+(k)^{-1}\;,\qquad(k\in I)\;.\label{2.4b}
\ee
We can compute the bulk index $\mathcal{I}(E)=\mathcal{I}(T)$ using Eq.~(\ref{2.0.5}). The theorem then reduces to the claim that, for each crossing point, the eigenvalues of $T(k)$ change with $k\in J$ from 1 to 1 without winding, except for a single one with winding number $\pm 1$.

It is with the proof of this claim that our choice of sections comes to fruition. In fact, by Eq.~(\ref{2.3}) and the sentence after Eq.~(\ref{2.4}) we have 
\be
\widehat\Psi^{\sharp}_{0}(z,k) = M_\pm(k)\;,\qquad((z,k)\in (\partial D) _\pm)\;.\nn %\label{2.6}
\ee
We apply part (ii) of the lemma to the section $\widehat\Psi^{\sharp}$, which is as required there. Then $L(z,k) = -A(k)\widehat\Psi^{\sharp}_{0}(z,k)$ by Eq.~(\ref{2.3a}) and hence 
\ben
T(k)=A(k)^{-1}L_-(k)L_+(k)^{-1}A(k)\;,\qquad(k\in I)
\een
with $L_\pm(k)=L(z,k)$ for $(z,k)\in (\partial D)_\pm$. Clearly, $A(k)$ can be dropped, as it does not affect the eigenvalues under investigation. We have 
\ben
L(z,k) = l(z,k)\Pi(z,k)\oplus \widetilde L(z,k)\;,
\een
where $\Pi(z,k)$ is the rank $1$ projection on the eigenspace corresponding to the eigenvalue $l(z,k)$, and $\widetilde L(z,k)$ is regular as a map on the range of the complementary projection. Clearly, $l(z,k)=O(\Delta)$ as $\Delta:=(z-\mu, k-k_*)\to 0$, but by (ii) above we also have $l(z,k)^{-1}=O(\Delta^{-1})$ on $\mathbb{T}$. We use the notation $f_0$ and $f_1$ for the value of a function $f$ and its gradient $(\partial_zf, \partial_kf)$ at the critical point. Then
\begin{gather}
L(z,k) = l(z,k)\Pi_0+\widetilde L_0+\widetilde L_1\cdot\Delta+O(\Delta^2)\;,\nonumber\\
L(z,k)^{-1}= l(z,k)^{-1}\Pi(z,k)\oplus \widetilde L(z,k)^{-1}=
l(z,k)^{-1}(\Pi_0 +\Pi_1\cdot\Delta)+\widetilde L_0^{-1}+O(\Delta)\;.\nonumber
\end{gather}
In view of $L_-(k)=L(\bar z,k)$ for $(z,k)\in (\partial D)_+$ we compute  
\begin{gather}
L(\bar z,k)L(z,k)^{-1}=\frac{l(\bar z,k)}{l(z,k)}\Pi_0+ R+(1-\Pi_0)+O(\Delta)\;,\label{ll}\\
R=l(z,k)^{-1}(\widetilde L_0\Pi_1\cdot\Delta+\widetilde L_1\cdot\bar\Delta\Pi_0)=l(z,k)^{-1}\widetilde L_0\Pi_1\cdot(\Delta-\bar\Delta)=O(1)\;.\nn
\end{gather}
In the second line we used $\widetilde L_0\Pi_1+\widetilde L_1\Pi_0=0$, as seen from expanding $\widetilde L\Pi=0$. Using $(1-\Pi_0)\Pi_1=\Pi_1\Pi_0$ from $\Pi=\Pi^2$, we find that $R=(1-\Pi_0)R\Pi_0$ is strictly block triangular. For small $D$ and $(z,k)\in (\partial D)_+$ the eigenvalues of (\ref{ll}) wind as if the error $O(\Delta)$ is omitted, provided those of the explicit part do not vanish. Those in turn  remain the same when $R$ is omitted, and are in fact equal to $1$ with multiplicity $N-1$, and to $l_-(k)/l_+(k)$ with multiplicity $1$, where $l_\pm(k)=l(z,k)$, ($(z,k)\in (\partial D)_\pm$). The winding number of $l_-(k)/l_+(k)$ along $J$ equals that of $l(z,k)$ along $\partial D$. Since the vanishing at the crossing point is of first order, $l(z,k)=l_1\cdot \Delta+O(\Delta^2)$, that number is
\be
\sgn((\partial_z l)(\partial_k l))=\pm 1\;,\nn
\ee
as claimed. Though the sign is irrelevant (so far), we observe that by (iv, v) it is $+1$ if the eigenvalue crossing is decreasing in $k$.\qed

\subsection{The bulk index as an index of Bloch bundles}\label{secbloch}

A feature of definition of the Bulk index is that it applies to Hamiltonians which are periodic just along the edge, see Eqs.~(\ref{1.1}, \ref{bi}). If the Hamiltonian is periodic in both directions and hence the Brillouin zone two-dimensional, the Bulk index allows for an alternate formulation in terms of the bundle of Bloch solutions, as we are about to explain. 

We temporarily suppress the longitudinal quasi-momentum $k$ in Eq.~(\ref{1.1}) and consider Hamiltonians of $\ell^{2}(\mathbb{Z},\mathbb{C}^{N})$ of the form
\be
(H\psi)_{n} = A\psi_{n-1} + A^{*}\psi_{n+1} + V_{n}\psi_{n}\;,\qquad (n\in\mathbb{Z},\psi_{n}\in\mathbb{C}^{N})\label{b1}
\ee
where $A\in\GL(N)$ and $V_{n}$ is periodic in $n$, \ie $V_{n+M} = V_{n}$. By once again considering sites $n$ within a period as labels of internal degrees of freedom, we may assume that the period is $1$. This amounts to the replacement of $\psi_{n}$, $A$, $V_{n}$ by
\be \label{b2}
\Psi = \begin{pmatrix}  \psi_{0} \\ \vdots \\ \psi_{M-1}   \end{pmatrix}\in \mathbb{C}^{MN}\;,\qquad
\mathcal{A} = \begin{pmatrix} 0 & \cdots & 0 & A \\ 0 & \cdots & & 0 \\ \vdots &&& \vdots\\ 0 & \cdots & & 0\end{pmatrix}\;,\qquad
\mathcal{V} = \begin{pmatrix} V_0 & A^{*} & 0 & \cdots & 0\\ A & \ddots &\ddots & \ddots & \vdots \\ 0 &\ddots&&& 0\\ \vdots &\ddots&\ddots&\ddots& A^{*} \\ 0 & \cdots & 0 & A & V_{M-1}\end{pmatrix}\;.
\ee
In particular $\mathcal{A}$ is singular, unlike $A$. A {\em Bloch solution} $(\psi_{n})_{n\in\mathbb{Z}}$ of quasi-periodicity $\xi\neq 0$,
\be \label{b2a}
\psi_{n+pM}=\xi^p\psi_n\;,
\ee
and of energy $z$ is then represented as a solution $\Psi$ of
\be
\mathcal{H}(\xi)\Psi \equiv (\mathcal{A}\xi^{-1} + \mathcal{A}^{*}\xi + \mathcal{V})\Psi = z\Psi\;,\qquad (\Psi\in\mathbb{C}^{MN})\;.\label{b3}
\ee
For $\kappa\in S^{1} = \mathbb{R}/2\pi\mathbb{Z}$ and $\xi = e^{i\kappa}$, the matrix $\mathcal{H}(\xi)$ is hermitian, since now $\bar\xi = \xi^{-1}$. It thus has eigenvalues $z=\l_{l}(\kappa)$ (real and increasingly ordered) and eigenvectors $\Psi_{l}(\kappa)$, $(l = 1,\ldots, MN)$. The ranges of the energy curves $\l_{j}(\kappa)$ are known as {\it energy bands}. Let
\be
\D_{l}(k) = \inf_{\kappa} \l_{l + 1}(\kappa) - \sup_{\kappa} \l_{l}(\kappa)\label{b4}
\ee
be the gap between successive bands of $H(k)$, where we temporarily reinstated the dependence on $k\in S^{1}$, implicit in (\ref{b1}). The gap is open if $\D_l(k)>0$. Let then $\D_{l} = \inf_{k}\D_{l}(k)$. In topological insulators the bands are degenerate for $k=0,\,\pi$, in fact at $\kappa=0,\,\pi$. A band thus can not be separated from the rest of the spectrum, but a pair of them can. We will assume so for the pairs $(2j-1,2j)$, $(j=1,\ldots, N_0/2)$, \ie
\be
\D_2,\,\D_4,\,\ldots \D_{N_0}>0\;,\label{b5}
\ee
where $N_0$ (even) is the uppermost band below the Fermi energy $\m$. For that band we actually retain the stronger assumption (\ref{1.13}) of a spectral gap. (It amounts to a positive gap in Eq.~(\ref{b4}) for $l = N_0$ even when extremizing jointly in $\kappa$, $k$.) 
\begin{defi}\label{blbun}% {\em [Bloch bundle]}
The {\em Bloch bundle} $E_j$ of the $j$-th pair of bands has the Brillouin zone $\mathbb{B}=S^{1}\times S^{1}\ni (\kappa,k)$ as base and the span $[\Psi_{2j-1}(\kappa,k),\Psi_{2j}(\kappa,k)]\subset \mathbb{C}^{MN}$ as fibers. 
\end{defi}
It should be noted that while the eigenvectors are not smooth in $\kappa$, $k$, their span is (\cite{Ka}, Sect.~II.1.4), since degeneracies occur within the pair. 

The result of this section is that the Bulk index can be expressed by means of the indices of the Bloch bundles of the filled pairs of bands.
\begin{thm}\label{thmbloch} Under the above assumption (\ref{b5}) and the gap condition (\ref{1.13}) we have
\be
\mathcal{I} = \prod_{j=1}^{N_0/2}\mathcal{I}(E_j)\;,\label{b6}
\ee
where $\mathcal{I} = \mathcal{I}(E)$ is the bulk index (\ref{bi}) and $\mathcal{I}(E_j)$ is defined in Eq.~(\ref{2.6.7a}) for the torus $\mathbb{B}$.
\end{thm}
Note the bundles seen on the two sides of Eq.~(\ref{b6}) have different base spaces and fibers; in fact the dimensions of the latter are $N$ and $2$, respectively. 

The proof we will give makes the simplifying assumption that the energy curves $\l_{l}(\kappa)$ do not have more critical points than required by the time-reversal symmetry of the Hamiltonian.

At first, we reduce the theorem to a lemma. The base space of the bundle $E$ in Eq.~(\ref{2.1}) can be extended from $\mathbb{T}$ to all of $\{ (z,k)\mid z\in\r(H(k)) \}$. In this notation, $\mathcal{I} = \mathcal{I}(E\upharpoonright\mathbb{T})$. The torus $\mathbb{T} = \gamma\times S^{1}$ may then be deformed and split into $N_0/2$ tori $\mathbb{T}_{j}$, $(j=1,\ldots, N_0/2)$ each surrounding a pair of bands $(2j-1,2j)$. Unlike $\mathbb{T}$, the $\mathbb{T}_j$ do not need to be a Cartesian product form, since (\ref{b5}) does not imply a spectral gap uniformly in $k$. We then consider the bundles $E^{(j)} := E\upharpoonright \mathbb{T}_{j}$. By homotopy and by the multiplicative property under splitting (Lemma~\ref{lemsplit}), we have:
\be
\mathcal{I}(E\upharpoonright\mathbb{T}) = \mathcal{I}\bigl(\bigcup_{j=1}^{N_0/2}E^{(j)}\bigr) = \prod_{j=1}^{N_0/2}\mathcal{I}(E^{(j)})\;,\nn%\label{b7}
\ee
and Eq.~(\ref{b6}) reduces to the following.

\begin{lemma}\label{lembloch}
\be
\mathcal{I}(E^{(j)}) = \mathcal{I}(E_j)\;.\label{b8}
\ee
\end{lemma}

The proof of this main lemma is deferred to Sect.~\ref{BlBu}.

\section{Quantum Hall systems}\label{QHI}

Much of what has been said in the previous sections has a counterpart for Hall systems. In that case the results are not completely new; nevertheless they generalize the bulk-edge correspondence of \cite{Hat0}.

The setting is the same as given by the Hamiltonians (\ref{1.1}, \ref{1.2}) with gap condition (\ref{1.13}), but without postulating a time-reversal symmetry. For the sake of brevity, definitions and statements given in the sequel rely on notations and contexts used for topological insulators.

\begin{defi}{\em[Index]} Let $T(\varphi)\in \GL(N)$, $(\varphi\in S^{1})$ be a continuous family. We set
\be
\mathcal{N}(T) = \mathcal{N}(Z)\;,\label{qhi1}
\ee
where $T(\varphi)$ determines $Z(\varphi)$ as in Eq.~(\ref{2.6.1b}) and $\mathcal{N}(Z)$ is defined in Eq.~(\ref{2.1.2}). Alternatively, $\mathcal{N}(T)$ is the winding number of $\det T(\varphi)$.
\end{defi}

Let $E$ be the vector bundle with base $\mathbb{T} = S^{1}\times S^{1}\ni (\varphi_{1},\varphi_{2}) = \varphi$ and fibers of dimension $N$. On the frame bundle $F(E)$, the right multiplication by $\GL(N)$ is defined as in (\ref{1.16}).

The classification of such bundles may again proceed by considering the cut torus $\dot{\mathbb{T}} = [-\pi,\pi]\times S^{1}$. In fact on $\dot{\mathbb{T}}$ there are smooth sections $v:\dot{\mathbb{T}}\rightarrow F(E)$; this is in analogy to Lemma~\ref{lemsec}, but with simpler proof, as observed in Remark~\ref{exQHI}.

\begin{defi}{\em[Index]} We set
\be
\mathcal{N}(E) = \mathcal{N}(T)\;,\label{qhi2}
\ee
where $T(\varphi_2)$ is the transition matrix introduced in Eq.~(\ref{1.20}). See Eq.~(\ref{qhi1}). The definition is again independent of the choice of the section $v$, as seen from the simplification of Lemma~\ref{propind}. It should be noted that the sign of the index would flip upon interchanging $\pm$ in Eq.~(\ref{1.20}).
\end{defi} 

The index $\mathcal{N}(E)$ is just the Chern number of $E$, but that will not be needed.
\comment{This fact, which will not be needed in the sequel, is seen as follows. [To be completed]}

The definitions of bulk and edge indices parallel Defs.~\ref{defbi} and \ref{defei}.

\begin{defi}
The {\em bulk index} is
\be
\mathcal{N} = \mathcal{N}(E)\;,\label{qhi3}
\ee
where $E$ is the bundle (\ref{2.1}). See Eq.~(\ref{qhi2}). The edge index $\mathcal{N}^{\sharp}$ is the number of signed crossings $k_{*}\in S^{1}$ of the Fermi energy $\m$ by eigenvalues of $H^{\sharp}(k)$. They are counted positively for decreasing eigenvalue branches.
\end{defi}
\begin{thm}\label{qhithm1}
Let bulk and edge index be defined as above. Then
\be
\mathcal{N} = \mathcal{N}^{\sharp}\;.\nn
\ee
\end{thm}

\noindent
{\it Proof.} The proof is contained in that of Thm.~\ref{thm1}. Now its last sentence matters.\qed

\subsection{The bulk index as an index of Bloch bundles}

We consider the case of doubly periodic Hamiltonians in close analogy to Sect.~\ref{secbloch}. As in Thm.~\ref{thmbloch} we will express the bulk index in terms of the Bloch bundles of the filled bands.

In the context of quantum Hall systems it is legitimate to assume that, for fixed longitudinal momentum $k$, the bands do not overlap:
\be
\D_1,\,\D_2,\,\ldots\D_{N_0} >0\;,\label{qhi4}
\ee
where $N_0$ is again the uppermost band below the Fermi surface, \cf~(\ref{b6}).

\begin{defi}
\label{bb_hall}
The Bloch bundle $E_{\ell}$ of the $\ell$-th band has the Brillouin zone $\mathbb{B} = S^{1}\times S^{1}\ni (\kappa,k)$ as base and the lines $[\Psi_{\ell}(\kappa,k)]\subset \mathbb{C}^{MN}$ as fibers.
\end{defi}

\begin{thm}\label{qhithm2}
Under the above assumption (\ref{qhi4}) and the gap condition (\ref{1.13}) we have
\be
\mathcal{N} = \sum_{\ell = 1}^{N_0}\mathcal{N}(E_{\ell})\;,\nn %\label{qhi5}
\ee
where $\mathcal{N} = \mathcal{N}(E)$ is the bulk index (\ref{qhi3}) and $\mathcal{N}(E_{\ell})$ is defined in (\ref{qhi2}) for the torus $\mathbb{B}$.
\end{thm}
\noindent
{\it Proof.} Consider the simplifying assumption that the energy curves do not have more critical points than necessary in absence of time-reversal symmetry, as explained after Eq.~(\ref{qhi6}) below. The proof then parallels that of Thm.~\ref{thmbloch}, but is much simpler. In fact the complex loop is of the type seen in the first case of Fig.~\ref{FigBl1}, but run through just once.\qed
\subsection{Bulk-edge correspondence through scattering theory}\label{scatt}
We propose a further perspective on the bulk-edge correspondence in the doubly periodic case. It does not rely on decaying bulk solutions, as Thm.~\ref{qhithm1} did. In contrast to that result, where edge states are intercepted at Fermi energy, here they are right at inception, \ie as they are born at band edges. That will be done by means of a result from scattering theory known as Levinson's theorem. In its usual form (\cite{RS}, Thm. XI.59) it computes the phase of the scattering matrix at thresholds. The version below computes the phase difference when a parameter is changed.

We focus on a single band $\ell$ which is separated from its neighbors,
\be
\D_{\ell-1},\,\D_{\ell}>0\;,\label{qhi6}
\ee
and on its Bloch bundle $E_{\ell}$, \cf~Def.~\ref{bb_hall}.
We also assume that, for fixed $k\in S^{1}$, the energy curve $\l = \l_{\ell}(\kappa,k)$ has as a function of $\kappa$ just two critical points, both non-degenerate, namely a maximum $\kappa_{+}(k)$ and a minimum $\kappa_{-}(k)$ (That assumption would not be consistent with topological insulators, \cf~Fig.~\ref{FigBl1}.) The curves $\kappa_{\pm}(k)$ cut the Brillouin zone $\mathbb{B}$ into two open domains $\mathbb{B}_+$ (resp. $\mathbb{B}_-$) where $\l(\kappa,k)$ is increasing (resp. decreasing) in $\kappa$ w.r.t. the orientation of $S^{1}$. 
\begin{figure}[hbtp]
\centering
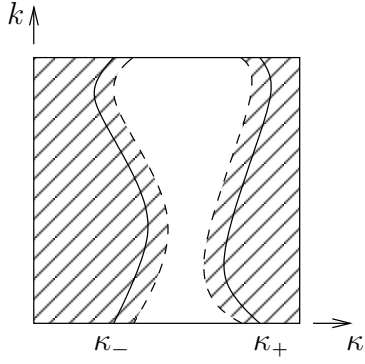
\caption{The Brillouin zone $\mathbb{B}$ with the domain $\widetilde{\mathbb{B}}_{-}$ (shaded) and the curves $\kappa_\pm(k)$.}
\label{FigHall}
\end{figure}
\begin{lemma}\label{lemrefl}
\begin{enumerate}
\item
There is a map $r: \mathbb{B}\to \mathbb{B}$ defined by $\l(\kappa,k) = \l(r(\kappa,k),k)$ and the property that is has the two extrema $\kappa = \kappa_{\pm}(k)$ as its only fixed points. It interchanges $\mathbb{B}_{\pm}$ and is real analytic in $\kappa$. \comment{Moreover, 
\be
\frac{\partial r}{\partial k}(\kappa_{\pm}(k),k)=-1\;.\nn
\ee}
\item
There is a domain $\widetilde{\mathbb{B}}_{-}\supset\overline{\mathbb{B}_{-}}$ and on there a section $\Psi^{-}(\kappa,k)\neq 0$ of the Bloch bundle $E_{\ell}$ which is smooth in $k$ and analytic in $\kappa$.
\end{enumerate}
\end{lemma}

We will occasionally omit $k$ from the notation in the sequel. In order to ensure that edge states are indeed ``born at band edges'', we make the assumption that there are none embedded in the band,
\be
\s_\mathrm{pp}(H^{\sharp})\cap [\l(\kappa_{-}), \l(\kappa_{+})] = \emptyset\;.\label{qhi6b}
\ee
The assumption is generically satisfied, but counterexamples can be constructed by taking the direct sum of two Hamiltonians, such that the pure point edge spectrum of one overlaps the band of the other.
\begin{lemma}\label{lemedge}
At energies $\l = \l(\kappa)$, $(\kappa\in \mathbb{B}_-)$ the edge Hamiltonian $H^{\sharp}$ has a bounded eigensolution $\psi^{\sharp} = (\psi_{n}^{\sharp}(\kappa))_{n\in\mathbb{N}}$ which is unique up to multiples. It satisfies
\be
\psi^{\sharp}_{n}(\kappa) = \psi^{-}_{n}(\kappa) + \psi^{+}_{n}(r(\kappa)) + o(1)\;,\quad (n\to +\io)\label{qhi7}
\ee
where the Bloch solution $\psi^{+}(r(\kappa))\neq 0$ is uniquely determined by $\psi^{-}(\kappa)$. (We recall the relation (\ref{b2}) between $\Psi$ and $\psi$.) Hence $\Psi^{+}$ is a section of $E_\ell$ on $\mathbb{B}_+=r(\mathbb{B}_-)$.
\end{lemma}
The bounded solution $\psi^{\sharp}$ ought to be interpreted as a scattering solution for the reflection at the boundary $n=0$. In fact, since $\l'(\kappa)<0$ on $\mathbb{B}_-$, $\psi^{-}$ represents an incoming wave, and $\psi^{+}$ an outgoing one.

Since $\mathbb{B}_+$ and $\widetilde{\mathbb{B}}_-$ overlap near $\kappa_{\pm}$ we may introduce {\em scattering amplitudes} $S_{\pm}(\kappa,k)$ for $\kappa$ near $\kappa_{\pm}(k)$ and $\kappa>\kappa_{+}(k)$, resp. $\kappa<\kappa_{-}(k)$:
\be
\Psi^{+}(r(\kappa)) = S_{\pm}(\kappa)\Psi^{-}(r(\kappa))\;.\label{qhi7af}
\ee
The scattering amplitudes $S_{\pm}(\kappa_{\pm}\pm\d)$, which by the lemma do not vanish, play the role of transition matrices (\ref{1.20}) for the line bundle $E_{\ell}$. Therefore, by (\ref{qhi2}),
\be
\mathcal{N}(E_{\ell}) = \mathcal{N}(S_{+}) - \mathcal{N}(S_{-})\;,\nn %\label{qhi9}
\ee
where $\mathcal{N}(S_{\pm})$ is defined in (\ref{qhi1}).

\begin{defi}
We say that $H^{\sharp}$ has a semi-bound state at the upper band edge $\l(\kappa_+)$ if it admits a bounded solution $\psi^{\sharp} = (\psi^{\sharp}_n)_{n\in\mathbb{N}}$ of $H^{\sharp}\psi^{\sharp} = \l(\kappa_+)\psi^{\sharp}$.
\end{defi}

\begin{lemma}\label{lemsemi}
Suppose that a branch $\e(k)$ of discrete eigenvalues of $H^{\sharp}(k)$ touches the $\ell$-th band from above at $k_*$, \ie
\be
\e(k) - \l(\kappa_+(k),k)\to 0\;,\qquad (k\to k_*)\;.\label{qhi10}
\ee
Then $H^{\sharp}(k_*)$ has a semi-bound state.
\end{lemma}

\begin{thm}{\em[Relative Levinson Theorem]}\label{thmlev}
Let $k = k_{i}\in S^{1}$, $(i=1,2)$ not correspond to semi-bound states of $H^{\sharp}(k)$. Then 
\be
\lim_{\d\to 0}\arg S_{+}(\kappa_+(k) + \d,k)\mid^{k_2}_{k_1} = 2\pi N_+\;,\label{qhi11}
\ee
where $\arg$ denotes a continuous argument and $N_+$ is the signed number of discrete eigenvalue branches of $H^{\sharp}(k)$ emerging $(-)$ or disappearing $(+)$ at the upper band edge, as $k$ runs from $k_1$ to $k_2$ in the orientation of $S^1$. Likewise for the lower band edge and $S_-$, except for a reversed count of signs in $N_-$.
\end{thm}

In particular the theorem may be applied to $k_1 = k_2$, \ie to a full circle $S^1$. Then it states $\mathcal{N}(S_{\pm}) = N_{\pm}$, $(\d>0)$. If $\ell = N_0$ is the uppermost band below the Fermi energy, then $N_{+} = \mathcal{N}^{\sharp}$. If the same assumptions hold true for all bands below it, then the bulk-edge correspondence
\be
\sum_{\ell=1}^{N_0}\mathcal{N}(E_{\ell}) = \mathcal{N}^{\sharp}\nn
\ee
is recovered in view of $N^{(\ell)}_{-} = N^{(\ell - 1)}_{+}$.

The proofs of the results of this section are found in Sect.~\ref{bec}.
\subsection{Comparison between two approaches}\label{comp}

In this section we compare the independent approaches to bulk-edge correspondence underlying Thms.~\ref{qhithm1} and \ref{thmlev}.

\begin{figure}[h]
\centering
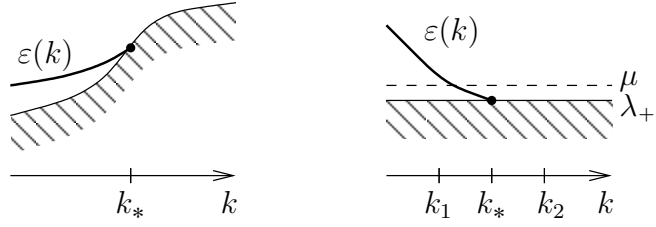
\caption{
Left: A discrete edge eigenvalue $\e(k)$ disappearing into the uppermost band below the Fermi energy $\m$. Right: Same, after energy shift.}
\label{FigComp}
\end{figure}

Let $H^{\sharp}(k)$ have a semi-bound state at the isolated point $k_*$. Suppose that a branch $\e(k)$ of discrete eigenvalues disappears there into the $N_0$-th band from above, \cf~Eq. (\ref{qhi10}) with $k\uparrow k_*$, whereas none emerges; see Fig.~\ref{FigComp}. By a suitable, $k$-dependent energy shift we may assume that the upper band edge $\l(\kappa_+(k)) \equiv \l_+$ is constant, and pick the Fermi energy $\m>\l_+$ arbitrarily close to it. The branch $\e(k)$ is then decreasing and crosses the (dashed) Fermi line. For $k_1$, $k_2$ near $k_*$ with $k_1 < k_* < k_2$ we have
\begin{align*}
\lim_{\d\to 0} \arg S_+(\kappa_+(k) + \d,k)|_{k_1}^{k_2} &= 2\pi\;,\\
\arg\det T(k)|_{k_1}^{k_2} &= 2\pi
\end{align*}
for $\m$ close enough to $\l_+$, where $T$ is the transition matrix (\ref{2.4b}) across the Fermi line, as used in the proof of Thms.~\ref{thm1} and \ref{qhithm1}. In fact, the first equation is the specialization of Eq. (\ref{qhi11}) and the second one follows from the claim made after Eq. (\ref{2.4b}). In case of an emerging branch the r.h.s. of both equations would change sign.

Both expressions on the l.h.s. arose as winding numbers. The point we wish to make here is that their equality can also be seen by homotopy, rather than by separate evaluation. We thus emphasize:
\begin{prop}\label{propcomp}
Under the above assumptions we have
\be
\lim_{\d\to 0}\arg S_{+}(\kappa_+(k)+\d,k)|_{k_1}^{k_2} = \arg\det T(k)|_{k_1}^{k_2}\;.\label{comp1}
\ee
\end{prop}
The proof is given in Sect.~\ref{bec}.

We close this section with a remark. The duality of Sect.~\ref{scatt} relied on the assumption that, for fixed $k$, the energy curve $\l_l(\kappa,k)$ had a single local maximum and minimum in $\kappa$. That naturally assigns a unique outgoing wave to a given incoming one, with transversal momenta $\kappa\mapsto r(\kappa)$. Should the assumption fail, multiple reflected waves could arise, preventing that essential assignment; at least without further ado, like dealing with scattering matrices instead of amplitudes. That failure is in fact unavoidable for time-reversal symmetric Hamiltonians (see Fig.~\ref{FigBl1} left). However, Prop.~\ref{propcomp} still indicates that the index of the Bloch bundle $E_j$ (associated to the pair of bands $l=2j-1, 2j$) can be determined by the glitches of $\arg S_+(\kappa,k)$ along the lines $\kappa=\kappa_\pm(k)$ of global maxima (or minima). In fact points where edge state emerge or disappear generically occur only on those lines, see Eq.~(\ref{qhi6b}), whence at nearby energies scattering remains described by a single amplitude $S_+$.

\section{Basic bulk-edge correspondence: Proofs}\label{secproofs}
The present section contains the proofs of our basic results for topological insulators (Thm.~\ref{thm1}) and for integer quantum Hall systems (Thm.~\ref{qhithm1}). We begin by supplying those relating to Sect.~\ref{trib} (except for Lemma~\ref{lemmolt}) and concerning the index of an abstract time-reversal invariant bundle. \\

\noindent
{\it Proof of Lemma~\ref{lemsec}.}\label{proofsec}
The main claim (\ref{1.18}) states that there are linearly independent vectors $v_{1}(\varphi),\,\ldots, v_{N}(\varphi)\in E_{\varphi}$ such that
\ben
\bigl(v_{i-1}(\t\varphi), v_{i}(\t\varphi)\bigr) = \bigl( \Th v_{i-1}(\varphi), \Th v_{i}(\varphi) \bigr)\begin{pmatrix} 0 & -1 \\ 1 & 0 \end{pmatrix}\;,\nn %\label{2.0.1}
\een
$(i = 2,4,\ldots, N)$, \ie
\be
v_{i-1}(\t\varphi) = \Th v_{i}(\varphi)\;,\qquad v_{i}(\t\varphi) = -\Th v_{i-1}(\varphi)\;.\label{2.0.2}
\ee
We first consider either fixed point $\varphi_0 = \t\varphi_0\in\dot{\mathbb{T}}$. We will prove by induction in $n = 0,2,\ldots, N$ that there are linearly independent vectors $v_{k} = v_{k}(\varphi_0)\in E_{\varphi_0}$, $(k = 1,2,\ldots, n-1,n)$ such that (\ref{2.0.2}) holds true for $i=2,\ldots, n$. Indeed, pick $v_{n+1}$ such that $v_{1},\ldots, v_{n}, v_{n+1}$ are linearly independent and set $v_{n+2} = -\Th v_{n+1}$. Then Eqs.~(\ref{2.0.2}) also hold for $i=n+2$ by $\Th^{2} = -1$. Moreover $v_{1}, \ldots, v_{n+2}$ are still linearly independent: $\sum_{k=1}^{n+2}\l_{k}v_{k}=0$ implies, by $v = \Th v \e$, the same with $\tilde\l_k = \sum_{\ell = 1}^{n+2}\e_{k\ell}\bar\l_{\ell}$; or for short for $\tilde\l = \Th_0\l$ instead of $\l = (\l_1,\ldots, \l_{n+2})$. Since the rank of those vectors is at least $n+1$ we have
\ben
a\l + \tilde a\tilde \l =0\;,\nn %\label{2.0.3}
\een
with $(a,\tilde a)\in \mathbb{C}^{2}$, $(a,\tilde a)\neq 0$. Applying $\Th_0$ yields $-\bar{\tilde a}\l + \bar a\tilde \l=0$ and together $\l=\tilde\l =0$ by $|a|^2 + |\tilde a|^2 \neq 0$.

The frames $v(\varphi_0)$ at the two fixed points may be interpolated over the line $\{\varphi_1 = 0\}\times [0,\pi]$. In fact, there the bundle $E$ is trivial (\cite{Na}, Cor.~9.5), since $[0,\pi]$ is contractible to a point; and $\GL(N)$ is connected. Eq.~(\ref{1.18}) is then used as a definition to extend $v(\varphi)$ to a section over the circle $\{\varphi_1 =0 \}\times S^{1}$. 

The section $v(\varphi)$ may then be extended further to the half torus $\dot{\mathbb{T}}_{1/2} := [0,\pi]\times S^{1}$ by homotopy. (We still denote by $E$ the bundle restricted to it.) More precisely, two maps $\dot{\mathbb{T}}_{1/2}\to \dot{\mathbb{T}}_{1/2}$ are clearly homotopic, namely the identity map and $f:(\varphi_1,\varphi_2)\mapsto (0,\varphi_2)$. Hence $F(E)$ and $f^{*}F(E)$ are equivalent bundles (\cite{Na}, Thm.~9.4). Since the latter has fibers $(f^{*}F(E))_{\varphi} = F(E)_{f(\varphi)}$, the above section over the circle extends trivially to $\dot{\mathbb{T}}_{1/2}$; by equivalence the same holds true for $F(E)$. Finally, Eq.~(\ref{1.18}) extends the section $v$ to all of $\dot{\mathbb{T}}$. An inspection of the procedure shows that the necessary smoothness of the section can be ensured. Eq.~(\ref{1.19}) then follows in the equivalent form $\e M(\t\varphi) =\overline{M(\varphi)}\e$ using (\ref{1.16b}).\qed

\begin{rem} In absence of time-reversal symmetry, see Sect.~\ref{QHI}, the existence of a global section on $\dot{\mathbb{T}}$ is quite obvious. In view of the above we may just note that it exists on the circle $\{\varphi_1 =0 \}\times S^{1}\equiv S^{1}$. In fact we may pick any frame $v(0)=v(2\pi)$ and interpolate along $[0,2\pi]$.\label{exQHI}
\end{rem}

\noindent
{\it Proof of Lemma~\ref{lemtrans}.}\label{prooftrans}
Note that Eq.~(\ref{1.18}) states $v_{\pm}(\varphi_2) = \Th v_{\mp}(-\varphi_2)\e$ for the boundary values. We thus have
\bea
v_{+}(\varphi_2) &=& v_{-}(\varphi_2)T(\varphi_2) = (\Th v_{+}(-\varphi_2))\e T(\varphi_2)\nn\\
&=& (\Th v_{-}(-\varphi_2))\overline{T(-\varphi_2)}\e T(\varphi_2) = (\Th^2 v_+(\varphi_2))\e\overline{T(-\varphi_2)}\e T(\varphi_2)\;.\nn %\label{2.0.4}
\eea
Since $\Th^{2} = -1$ and the right action is transitive, the claim follows.\qed\\

\noindent
{\it Proof of Lemma~\ref{propind}.}\label{proofind}
The two sections obey the relation $v^{(2)}(\varphi) = v^{(1)}(\varphi)M(\varphi)$ for some matrix $M(\varphi)\in \GL(N)$ continuous on $\dot{\mathbb{T}}$. Let $M_{\pm}(\varphi_2) := M(\pm\pi ,\varphi_2)$ be its boundary values. The two transition matrices are then related by
\be
T_1(\varphi_2) = M_-(\varphi_2)T_2(\varphi_2)M_{+}(\varphi_2)^{-1}\;.\label{2.6.6a}
\ee
More generally, we consider 
\be
T(\varphi_1,\varphi_2) := M(-\varphi_1,\varphi_2)T_2(\varphi_2)M(\varphi_1,\varphi_2)^{-1}\nn %\label{2.6.6c}
\ee
as a homotopy in $0\le\varphi_1\le\pi$ of families in $\varphi_2$. Eq.~(\ref{1.19}) implies that, at fixed $\varphi_1$, the matrices $M(\pm\varphi_1,\cdot)$ satisfy (\ref{2.6.1d}). Hence the family $T(\varphi_1,\cdot)$ has the Kramers property. Clearly the index of $T(\varphi_1,\cdot)$ is constant in $\varphi_1$ by continuity. Note that one end of the homotopy is 
\be
T(0,\varphi_2) = M(0,\varphi_2)T_2(\varphi_2)M(0,\varphi_2)^{-1}\;,\nn %\label{2.6.6d}
\ee
which by (\ref{2.6.1e}) has the same index as $T_2$, while the other is (\ref{2.6.6a}). We conclude $\mathcal{I}(T_2)=\mathcal{I}(T_1)$.\qed\\

We next come to the the proof of the main technical lemma in relation with the bulk-edge correspondence.\\

\noindent
{\it Proof of Lemma~\ref{lem1} i).} The proof of the lemma follows quite closely that of (\cite{BGO}, Lemma~3). 

Consider the finite difference equation $(H^{\sharp}-z)\psi^{\sharp} = 0$ with $z\in\mathbb{C}$ and without imposing the boundary condition $\psi_0^{\sharp}=0$. Solutions $\psi^{\sharp}$ are square-summable at $n\to+\infty$ iff $\psi^{\sharp}=\Psi^{\sharp}a$ for some $a\in\mathbb{C}^N$. Hence $0$ is eigenvalue of $\Psi^{\sharp}_{0}$ iff $z$ is an eigenvalue of the operator $H^{\sharp}$, which now includes the boundary condition. In particular $z$ is then real, because $H^{\sharp}$ is self-adjoint. For $\g$ as in Fig.~\ref{fig2} we have $\g\cap\s(H^{\sharp})\subset\{\m\}$.

We may arrange for the absence of ``flat'' crossings, $\e(k_*)=\m$, $\e'(k_*)= 0$, by adding to $V_n(k)$ and $V^{\sharp}_n(k)$ in Eqs.~(\ref{1.1}, \ref{1.2}) an arbitrarily small constant. In particular, the points $k_*$ are isolated, as claimed. Moreover, they are generically simple. To show this, we perturb $V$, $V^{\sharp}$ by $tW_n(k)$ and determine the splitting $\m+t\tilde \m+o(t^2)$, $(t\to 0)$ of a degenerate eigenvalue $\m$ of $H(k_*)$. On general grounds $\tilde \m$ is found by diagonalizing $W$ after orthogonally projecting it onto the unperturbed eigenspace of $\m$. Here, that eigenspace is the image of $\ker\Psi^{\sharp}_{0}\subset\mathbb{C}^N$ under $\Psi^{\sharp}$. Hence the eigenvalue problem reads
\begin{equation}\label{eq:evlp}
P_0\Bigl(\sum_{n=0}^\infty (\Psi^\sharp_n)^*W_n(k_*)\Psi^\sharp_n\Bigr)P_0 a=
\tilde \mu P_0\Bigl(\sum_{n=0}^\infty(\Psi^\sharp_n)^*\Psi^\sharp_n\Bigr)P_0 a\; ,
\qquad (a \in \mathbb{C}^N)\,,
\end{equation}
where $P_0$ is the orthogonal projection onto
$\ker\Psi^{\sharp}_{0}$. By Eq.~(\ref{1.10}) the matrix in brackets on the r.h.s. is positive definite on $\mathbb{C}^N$, while that on the l.h.s. may take arbitrary Hermitian values, along with $W_n(k_*)$. As a result, the eigenvalues $\tilde \m$ are generically distinct and, since $\e'(k_*)\neq 0$, the points $k_*$ split into non-degenerate ones. Similarly, points $k_*$ with $\det\Psi^{\sharp}_{1}=0$ correspond to $\m$ being a Dirichlet eigenvalue for $n=1$. Its perturbative splitting is determined by (\ref{eq:evlp}) with the replacement of $0$ by $1$. By (\ref{1.10}), $\ker \Psi^{\sharp}_{0}\cap \ker \Psi^{\sharp}_{1}=\{0\}$, so that the operator on the l.h.s. can be chosen with independent projections under $P_0$ and $P_1$. Any coincidence between eigenvalues of the two Dirichlet problems is thus generically lifted.

The above argument did not pay attention to the time-reversal symmetry of the Hamiltonian and hence of $W(k)$. We do so now: A crossing at $k_*=0,\pi$ may be displaced by perturbing by an arbitrarily small constant; one at $k_*\neq 0,\pi$ by letting $W(k)$ satisfying Eq.~(\ref{1.14b}). It still remains arbitrary at $k_*$.\qed\\

In preparation for the proof of part (ii) let us introduce the Casoratian, which is to finite difference equations what the Wronskian is to ordinary differential equations (both linear and of second order). Let $H$ be as in Eq.~(\ref{b1}). Given $\psi = (\psi_n)_{n\in\mathbb{Z}}$, $\varphi = (\varphi_n)_{n\in\mathbb{Z}}$ with $\psi_n,\varphi_{n}\in\mathbb{C}^{N}$ viewed as column, resp. row vectors, let
\be
C_n(\varphi,\psi) = \varphi_n A^* \psi_{n+1} - \varphi_{n+1} A\psi_{n}\;.
\label{cas}
\ee
Suppose that $\psi,\varphi$ satisfy the Schr\"odinger equation in the form
\be
(H-z)\psi=0\;,\qquad \varphi(H-z)=0\;,\label{qhi7a}
\ee
where $(\varphi H)_{n} = \varphi_{n-1} A^* + \varphi_{n+1} A + \varphi_n V_n$. Then
\begin{enumerate}
\item[C1)] $C_n(\varphi,\psi)$ is independent of $n\in\mathbb{Z}$ and denoted $C(\varphi,\psi)$.
\item[C2)] Let $\psi(z)$ solve the first Eq.~(\ref{qhi7a}). Then 
$\psi(\bar z)^*$ solves the second. In particular $C(\psi(\bar z)^*, \psi(z))$ is well-defined.
\item[C3)] Items (C1--2) apply if $H,\psi,\mathbb{Z}$ are replaced by $H^{\sharp}, \psi^{\sharp}, \mathbb{N}$. 
\end{enumerate}
Property (C1) follows from the identity
\be
C_n(\varphi,\psi) -C_{n-1}(\varphi,\psi)=
\varphi_n(A\psi_{n-1}+A^*\psi_{n+1})-(\varphi_{n-1}A^*+\varphi_{n+1}A)\psi_n
\label{cas1}
\ee
and Eq.~(\ref{qhi7a}); the others are straightforward. The Casoratian may be extended literally to matrix solutions $\Psi, \Phi$ of Eqs.~(\ref{qhi7a}), in which case $C_n(\Phi, \Psi)$ is itself a matrix. Its entries are the Casoratians of the rows and columns of $\Phi$, resp. $\Psi$. Properties (C1--3) hold correspondingly.\\  

\noindent
{\it Proof of Lemma~\ref{lem1}, continued.} ii) We drop $k$ and notice that
\be
C_0(\Psi^{\sharp}(\bar z)^*,\Psi^{\sharp}(z))
=\Psi^{\sharp}_0(\bar z)^*A^*\Psi^{\sharp}_1(z) - \Psi^{\sharp}_1(\bar z)^*A\Psi^{\sharp}_0(z)=L(z)-L(\bar z)^*\;.\nn
\ee
By (C2) the l.h.s. equals $\lim_{n\to\infty}C_n(\Psi^{\sharp}(\bar
z)^*,\Psi^{\sharp}(z))=0$; whence the reflection property. The
statement about the eigenvalue branch follows from part (i) and the
definition of $L$. 

iii) We drop $k=k_*$. Let $u\in \mathbb{C}^{N}$ be the normalized eigenvector of $L(\m)$ with eigenvalue $l(\m)=0$, whence $\Psi^{\sharp}_0(\m)u=0$ by (\ref{2.7}). Thus,
\be
\frac{\partial l}{\partial z}\Big|_{\m} = \bigl( u,\frac{\partial L}{\partial z}\Big|_{\m}u \bigr)
= -\bigl( u, \Psi^{\sharp}_1(\m)^*A(\partial_{z} \Psi^{\sharp}_0(\m))u \bigr)
=\bigl( u, C_0( \Psi^{\sharp}(\m)^*,\partial_{z}\Psi^{\sharp}(\m))u \bigr)\;.\nn
\ee
Next we observe that
\be
C_{n}( \Psi^{\sharp}(\m)^*,\partial_{z}\Psi^{\sharp}(\m)) - C_{n-1}( \Psi^{\sharp}(\m)^*,\partial_{z}\Psi^{\sharp}(\m)) = \Psi^{\sharp}_{n}(\m)^*\Psi^{\sharp}_{n}(\m)\geq 0\;,\nn
\ee
because of (\ref{cas1}) and of $(H^{\sharp}-z)\partial_{z}\Psi^{\sharp}=\Psi^{\sharp}$. Since $C_{n}( \Psi^{\sharp}(\m)^*,\partial_{z}\Psi^{\sharp}(\m))\to 0$, $(n\to\infty)$, we conclude $(\partial_{z}l)(\m)\le 0$. Actually, equality is excluded, because $\sum_{n=1}^\infty \Psi^{\sharp}_{n}(\m)^*\Psi^{\sharp}_{n}(\m)$ is positive definite, as remarked earlier.

iv) Let $\e(k)$ the eigenvalue branch crossing $\m$ at $k_*$, whence $l(\e(k),k)=0$ near $k_*$. The claim follows from 
\be
\frac{d}{dk}l(\e(k),k)\Big|_{k=k_*} = 
\frac{\partial l}{\partial z}\Big|_{\m,k_*}\e'(k) +\frac{\partial l}{\partial k}\Big|_{\m,k_*} \nn
\ee
together with (\ref{2.10}). \qed\\

This concludes the proof of Thms.~\ref{thm1} and \ref{qhithm1}.

\subsection{Supplementary results}\label{sr}
This section contains a few details related to Sect.~\ref{trib}, some of independent interest and some needed in connection with Thm.~\ref{thmbloch}, but none with the proof of the basic results, which is by now complete. Also addressed is the relation with other indices found in the literature.\\

{\it Polar decomposition.} Given a matrix $T\in \GL(N)$, let $T=PU$ be its (unique, left) polar decomposition, \ie $P= P^{*}>0$, $U^* U = 1$.

\begin{lemma}\label{lemdef}
Let $T\in\GL(N)$ satisfy Eq. (\ref{2.6.1c}). Then so does $U$ and the following deformations of $T$ retaining that property are possible: (i) $T$ to $U$, while keeping the polar part fixed; (ii) $U$ to  $1$, while preserving unitarity; (iii) and hence $T$ to $1$.
\end{lemma}

\noindent {\it Proof.} The right polar decomposition is $T=U\tilde P$ with $\tilde P=U^{-1}PU$. Eq.~(\ref{2.6.1c}) states $T=\Th_0^{-1}T^{-1}\Th_0=(\Th_0^{-1}U^{-1}\Th_0)(\Th_0^{-1}P^{-1}\Th_0)$ and is hence equivalent to
\be
U=\Th_0^{-1}U^{-1}\Th_0\;,\qquad U^{-1}PU=\Th_0^{-1}P^{-1}\Th_0\;.\label{eq:pol}
\ee
In particular the preliminary claim holds true. 

i) Denoting by $P=\sum_i\l_i \Pi_i$, ($\l_i>0$) the spectral decomposition of $P$, the second Eq.~(\ref{eq:pol}) is equivalent to the existence of an involution $i\mapsto\bar\imath$ such that $\l_{\bar\imath}=\l_i^{-1}$ and $U^{-1}\Pi_{\bar\imath} U=\Th_0^{-1}\Pi_i\Th_0$. By interpolating the eigenvalues but not the eigenprojections, it becomes clear that $T$ can be deformed as stated. 

ii) Denoting now by $U = \sum_{i}z_i \Pi_i$, ($|z_i|=1$) the spectral decomposition of $U$, the first Eq.~(\ref{eq:pol}) is equivalent to $\Pi_i = \Th_0^{-1}\Pi_i \Th_0$. By again interpolating eigenvalues only we arrange for $U$ going to $1$.\qed\\

Let us also mention the following consequence:
\begin{rem}\label{polar} Let $T(\varphi)=P(\varphi)U(\varphi)$ be the polar decomposition of $T(\varphi)\in\GL(N)$. If the family $T$ has the Kramers property, then so does $U$. Even though the $z_i(\varphi)$ are generally not the eigenvalues of $U(\varphi)$, it holds true that $\mathcal{I}(T) =\mathcal{I}(U)$.
\end{rem}
\noindent
{\it Proof.} By (i) of the previous lemma, $T(\varphi)$ can be continuously extended to the right on an interval $\varphi\in(\pi, b]$ in such a way that there Eq.~(\ref{2.6.1c}) holds throughout, $U(\varphi)\equiv U(\pi)$, and $P(b)=1$. Likewise on an interval $[a,0)$. By continuity neither index, $\mathcal{I}(U)$ or $\mathcal{I}(T)$, changes in the process. At this point they are manifestly equal. In fact $\{P\mid P>0\}$ is a convex set; hence any continuous map $P:[a,b]\ni x\mapsto P(x)>0$ with $P(a)=P(b)=1$ is homotopic to $P\equiv 1$. \qed\\

\noindent
{\it Proof of Lemma~\ref{lemmolt}.} As mentioned, only Eq. (\ref{molt1}) requires proof. Consider the families
\be
T_2'(\varphi) = M_-(0)T_1(\varphi)M_+(0)^{-1}\;,\quad T''_2(\varphi) = M_-(\varphi)T_1(\pi)M_+(\varphi)^{-1}\;.\nn
\ee
They enjoy the Kramers property along with $T_1$; they can be concatenated, since $T'_2(\pi) = T_2''(0)$; and $T_2$ is homotopic to $T'_2 \# T''_2$, as seen by postponing the change of $M_{\pm}$ till after that of $T_1$. Hence
\be
\mathcal{I}(T_2) = \mathcal{I}(T'_2\# T''_2) = \mathcal{I}(T'_2)\mathcal{I}(T''_2)\;,\nn
\ee
where $\mathcal{I}(T'_2) = \mathcal{I}(T_1)$ by deforming $M_\pm(0)$ to $1$ while preserving (\ref{2.6.1d}); and $\mathcal{I}(T''_2) = \mathcal{I}(M)$ by Lemma~\ref{lemdef} (iii).\qed\\

{\it Alternate definition.} There is an alternative way to Def.~\ref{def:bulk} of computing the index $\mathcal{I}(E)$ of a time-reversal invariant bundle $E$. Let $u(\varphi):\ddot{\mathbb{T}}\to F(E)$ be a section satisfying (\ref{1.18}) on the twice cut torus $\ddot{\mathbb{T}}=( [-\pi,0]\sqcup[0,\pi])\times S^{1}$, where $\tau$ exchanges the left and right halves of $\ddot{\mathbb{T}}$. The section gives rise to four boundary values and two transition matrices parametrized by $\varphi_2\in S^1$:
\be
u(0+,\varphi_2)=u(0-,\varphi_2)T_0(\varphi_2)\;,\qquad
u(+\pi,\varphi_2)=u(-\pi,\varphi_2)T_\pi(\varphi_2)\;.\label{jump}
\ee
\begin{lemma} In this situation, the families $T_0, T_\pi$ enjoy the Kramers property and
\be\label{i2cuts}
\mathcal{I}(E)=\mathcal{I}(T_0)\mathcal{I}(T_\pi)\;.
\ee
\end{lemma}
Def.~\ref{def:bulk} corresponds to the special case of a single cut at $\varphi_1=\pm\pi$, whence $T_0\equiv 1$; a single cut at $\varphi_1=0$ could have been used there instead. \\

\noindent
{\it Proof.} The two families are Kramers because Lemma~\ref{lemtrsU} still applies. Let $v:\dot{\mathbb{T}}\to F(E)$ be a time-reversal invariant section as in Lemma~\ref{lemsec} and let $M(\varphi)$, ($\varphi\in\ddot{\mathbb{T}}$) be the change of frame $v(\varphi)=u(\varphi)M(\varphi)$. It satisfies (\ref{1.19}). Since $v(\varphi_1,\varphi_2)$ is continuous along $\varphi_1=0\pm$ and satisfies (\ref{1.20}) along $\varphi_1=\pm\pi$ we have
\be
1=M_{-0}(\varphi_2)T_0(\varphi_2)M_{+0}(\varphi_2)^{-1}\;,\qquad
T(\varphi_2)=M_{-\pi}(\varphi_2)T_\pi(\varphi_2)M_{+\pi}(\varphi_2)^{-1}\;,\nn
\ee
where $M_{\pm0}(\varphi_2)=M(0\pm,\varphi_2)$, $M_{\pm\pi}(\varphi_2)=M(\pm\pi,\varphi_2)$. By Eq.~(\ref{molt1}), 
\be
1=\mathcal{I}(T_0)\mathcal{I}(M_{-0}M_{+0}^{-1})\;,\qquad
\mathcal{I}(T)=\mathcal{I}(T_\pi)\mathcal{I}(M_{-\pi}M_{+\pi}^{-1})\;.\nn
\ee
As remarked in the proof of Lemma~\ref{propind}, the family $M(-\varphi_1,\cdot)M(\varphi_1,\cdot)^{-1}$ is Kramers for fixed $\varphi_1\in[0,\pi]$. It is continuous in $\varphi_1$ in that interval, and its index constant. Hence the claim.\qed\\
  
{\it Splitting the torus.} We consider a time-reversal invariant bundle $E$ on the torus $\mathbb{T}$, as described in Sect.~\ref{secchar}. We suppose moreover that $\varphi_{1} = \pm\pi/2$ is a distinguished pair of lines in the sense that
\be
E_{(-\frac{\pi}{2},\varphi_2)} = E_{(\frac{\pi}{2},\varphi_2)}\;,\qquad (\varphi_2\in S^{1})\label{sp1}
\ee 
(In applications, that identification of fibers may occur because the bundle $E$ is effectively the pull-back of another one under a map $f$ with $f(-\pi/2,\varphi_2) = f(\pi/2,\varphi_2)$.) Let us define the torus $\mathbb{T}_1 = \{ (\varphi_1,\varphi_2)\mid |\varphi_1|\leq \pi/2,\, \varphi_2\in S^{1} \}$ with identified edges $\varphi_1 = \pm \pi/2$, whence it still has four time-reversal invariant points. By (\ref{sp1}) $E_1 = E\upharpoonright \mathbb{T}_{1}$ is a well-defined time-reversal invariant bundle. Similarly for $\mathbb{T}_2 = \{ (\varphi_1,\varphi_2)\mid |\varphi_1 - \pi|\leq \pi/2,\,\varphi_2\in S^{1} \}$ and $E_2 = E\upharpoonright \mathbb{T}_2$. 

\begin{lemma}{\em[Splitting lemma]}\label{lemsplit}
In the situation set by Eq.~(\ref{sp1}) we have
\be
\mathcal{I}(E) = \mathcal{I}(E_1)\mathcal{I}(E_2)\;.\label{sp2}
\ee
\end{lemma}

\noindent
{\it Proof.} By Lemma~\ref{lemsec} there exists a section $v_{1}(\varphi)$ of $F(E_1)$ satisfying (\ref{1.18}) on $\dot{\mathbb{T}}_1$, the torus $\mathbb{T}_1$ cut along $\varphi_1 = \pm\pi/2$; likewise, $v_2(\varphi)$ on $\dot{\mathbb{T}}_2$ with cut $\varphi_1 = \pm \pi$. Let $T_1(\varphi)$, $T_2(\varphi)$ be the corresponding transition matrices (\ref{1.20}). By (\ref{sp1}) we may also introduce transition matrices $M_{\pm}(\varphi_2)$, ($\varphi_2\in S^{1}$) by
\be
v_{2}(\pm\frac{\pi}{2},\varphi_2) = v_{1}(\pm \frac{\pi}{2},\varphi_2)M_{\pm}(\varphi_2)\;.\label{sp2b}
\ee
Multiplying by $\Th$, $\e$ from the left, resp. right we obtain
\be
v_{2}(\mp \frac{\pi}{2},-\varphi_2) = v_{1}(\mp \frac{\pi}{2},-\varphi_2)\Th_0^{-1}M_{\pm}(\varphi_2)\Th_0\nn
\ee
and thus
\be
\Th_0 M_{\pm}(-\varphi_2) = M_{\mp}(\varphi_2)\Th_0\;.\label{sp3}
\ee
Moreover, the l.h.s. of (\ref{sp2b}) is independent of $\pm$, whence
\be
T_{1}(\varphi_2) = M_{-}(\varphi_2)M_{+}(\varphi_2)^{-1}\;.\label{sp4}
\ee
On $\dot{\mathbb{T}}$ we define the section
\be
v(\varphi) =  \begin{cases} v_{1}(\varphi)\;, & (|\varphi_1|\leq \pi/2) \\ v_{2}(\varphi)M_{\pm}(\varphi_2)^{-1}\;, & (\frac{\pi}{2}\leq \pm \varphi_1 \leq \pi)\;.
 \end{cases}\nn
\ee
It is continuous by (\ref{sp2b}) and satisfies (\ref{1.18}): by hypothesis for $|\varphi_1|\leq \pi/2$, but also for $\pi/2\leq \pm \varphi_1\leq \pi$ by (\ref{sp3}). Its transition matrix is read off as
\be
T(\varphi_2) = M_{-}(\varphi_2)T_2(\varphi_2)M_+(\varphi_2)^{-1}\;.\nn
\ee
Now Eq.~(\ref{sp2}) follows in the form $\mathcal{I}(T) = \mathcal{I}(T_1)\mathcal{I}(T_2)$ from (\ref{molt1}, \ref{sp4}).\qed\\

{\it Indices in the literature.} The $\mathbb{Z}_2$-index has been introduced in various other forms before \cite{KM, FK}, but to our knowledge not in the form (\ref{2.6.7a}). We shall first recall a formulation \cite{FK} (there connected to \cite{KM}), which rests on an additional, metric structure, and later establish equivalence with ours when there is overlap. Like our index, it first deals with continuous families of matrices $W(\varphi)$, ($0\leq \varphi\leq \pi$) which, in lieu of the Kramers property, enjoy antisymmetry $W^{T} = - W$ at endpoints $W = W(0),\,W(\pi)$. There,
\be
\det W = (\pf W)^{2}\;,\nn
\ee
where $\pf W$ is the Pfaffian of $W$; in between, consider continuous branches $\pm\sqrt{\det W(\varphi)}$. One of them will connect $\pf W(0)$ to $\widehat{\mathcal{I}}(W)\pf W(\pi)$, where $\widehat{\mathcal I}(W) = \pm 1$ defines the index of the family, \cf~(\cite{FK}, Eq.~(3.24)). In the special case that $\pf W(0) = \pf W(\pi)$, and hence $\det W(0) = \det W(\pi)$, the index reduces to
\be
\widehat{\mathcal{I}}(W) = (-1)^{n}\;,\label{pf1}
\ee
where $n$ is the winding number of $\det W(\varphi)$, ($0\leq\varphi\leq \pi$).

Let us move on to time-reversal invariant bundles $E$. They are assumed equipped with a compatible hermitian metric, \ie with an inner product $\langle \cdot, \cdot \rangle$ on any fiber $E_{\varphi}$ such that $\Th^{*}\Th = 1$. Use is made of the fact that $E$ is trivial if $\Th$ is disregarded. There thus is a section $v(\varphi)$ of $F(E)$ on the torus $\mathbb{T}\ni\varphi=(\varphi_1,\varphi_2)$. We stress: not just on the cut torus $\dot{\mathbb{T}}$, but at the price of forgoing time-reversal symmetry. It can be taken to consist of orthonormal frames $v = (v_1,\ldots v_N)$. Let
\be
W_{ij}(\varphi) = \langle v_{i}(\varphi),\Th v_j(\t\varphi) \rangle\;;\nn
%\label{pf2}
\ee
see \cite{HK} and (\cite{FK}, Eq.~(3.16) with $\varphi, \t\varphi$ switched). Then $W(\varphi)^*W(\varphi) = 1$ and $W(\varphi)^T = -W(\t\varphi)$. By this last property, one may define the index as
\be
\widehat{\mathcal{I}}(E) = \widehat{\mathcal{I}}(W_0)\widehat{\mathcal{I}}(W_\pi)\;,\nn
\ee
where $W_0(\varphi_2) = W(0,\varphi_2)$, $W_\pi(\varphi_2)= W(\pi,\varphi_2)$, $(0\leq \varphi_2\leq \pi)$; see (\cite{HK}, Eq.~(10)). The relation to the indices of the present work is as follows.

\begin{prop}{\em [Relation between indices]}
\begin{itemize}
\item[i)] Suppose $T(\varphi)$, $(0\leq \varphi \leq \pi)$ has the Kramers property and $W(\varphi) := T(\varphi)\e$ is antisymmetric at endpoints. Then
\be
\mathcal{I}(T) = \widehat{\mathcal{I}}(W)\;.\label{pf3}
\ee
\item[ii)] For a time-reversal invariant vector bundle $E$ as above we have
\be
\mathcal{I}(E) = \widehat{\mathcal{I}}(E)\;.\nn
\ee
\end{itemize}
\end{prop}

\noindent
{\it Proof.} We begin with a preliminary remark. Consider the three properties of a matrix $T\in\GL(N)$:
\be
\overline{T}\e T\e = -1\;,\qquad T^* T = 1\;,\qquad T^{T} = -\e T \e\;,\nn
\ee
called Kramers, unitarity and antisymmetry, the first one being indeed a restatement of Eq.~(\ref{2.6.1c}). In terms of $W = T\e$ they respectively read
\be
\overline{W}W = -1\;,\qquad W^* W = 1\;,\qquad W^T = -W\;.\nn
\ee
Clearly any two of them imply the third.

i) The assumptions imply that $T(0),\,T(\pi)$ are unitary. By Lemma~\ref{lemdef} (ii) we may deform the family at its endpoints in such a way that $T(0) = T(\pi)$, while retaining the Kramers property and unitarity, and hence antisymmetry. In the process both indices (\ref{pf3}) remain defined and constant. At this point, see Eqs.~(\ref{2.1.4}, \ref{2.6.1b}), $\mathcal{I}(T) = (-1)^{n}$, where $n$ is the winding number of $\det T(\varphi)$. The claim follows by (\ref{pf1}) and $\det T(\varphi) = \det W(\varphi)$.

ii) We define a time-reversal invariant section $u(\varphi)$, see Eq.~(\ref{1.18}), on the twice cut torus $\ddot{\mathbb{T}}$ as
\be
u(\varphi)=\begin{cases}v(\varphi)\;,&(-\pi\leq \varphi_1\leq 0)\\
\Th v(\t\varphi) \e\;,& (0 \le \varphi_1 \leq \pi).\end{cases}\nn
\ee
It has $\varphi_1=0\pm$ and $\varphi_1=\pm \pi$ as lines of discontinuity, where Eq.~(\ref{jump}) becomes
\be
\Th v(0,-\varphi_2) \e=v(0,\varphi_2)T_0(\varphi_2)\;,\qquad
\Th v(\pi,-\varphi_2) \e=v(\pi,\varphi_2)T_\pi(\varphi_2)\;,\nn
\ee
by the continuity of $v$. Taking the inner product with the vectors of the orthonormal frames $v(0,\varphi_2)$ resp. $v(\pi,\varphi_2)$ yields $W_0(\varphi_2)\e=T_0(\varphi_2)$ and $W_\pi(\varphi_2)\e=T_\pi(\varphi_2)$. The claim follows by (\ref{i2cuts}, \ref{pf3}).\qed\\
\comment{
We shall construct by means of $u(\varphi)$ a section $v(\varphi)$ on the cut torus which satisfies the symmetry (\ref{1.18}). There is no loss in assuming that $u(\varphi)$ already does on the circle $\{ \varphi_1 = 0 \}\times S^1$. We make the ansatz
\be
v(\varphi) = u(\varphi)M(\varphi)\;,\qquad (\varphi = (\varphi_1,\varphi_2)\in \dot{\mathbb{T}})\nn
\ee
where $M(\varphi)\in\GL(N)$ satisfies $M(\varphi) = 1$ for $-\pi\leq \varphi_1\leq 0$. The requirement thus is
\be
u(\t\varphi) = \Th u(\varphi) M(\varphi) \e\;,\qquad (0 < \varphi_1 \leq \pi)\nn
\ee
and its solution $M(\varphi)$ is continuous on $\dot{\mathbb{T}}$. Taking the inner product with the vectors of the orthonormal frame $u(\t\varphi)$ yields by (\ref{pf2})
\be
W(\varphi)M(\varphi)\e = 1\;.\nn
\ee
By $u_{+}(\varphi_2) = u_-(\varphi_2)$ the transition matrix (\ref{1.20}) for $v(\varphi)$ is $T(\varphi_2) = M(\pi,\varphi_2)$. In particular, $W_+(\varphi_2)T(\varphi_2)\e = 1$. We conclude by (i)
\be
\mathcal{I}(E) = \mathcal{I}(T) = \widehat{\mathcal{I}}(W_+^{-1}) = \widehat{\mathcal{I}}(W_+) = \widehat{\mathcal{I}}(E)\;.\nn
\ee
\qed
}
\section{The bulk index as an index of Bloch bundles: Proofs}\label{BlBu}
As a preliminary to the proof of Lemma~\ref{lembloch}, we consider Eq.~(\ref{b3}) not just for $|\xi| = 1$, but for $\xi\neq 0$, together with its characteristic polynomial
\be
P(\xi,z)  = \det(\mathcal{H}(\xi)-z)=\det(\mathcal{A}\xi^{-1} + \mathcal{A}^*\xi + \mathcal{V} - z)\;.\label{b9}
\ee
Its basic properties, to be proven later, are as follows.
\begin{lemma}\label{lembl2}
$P$ is a polynomial of degree $MN$ in $z$ and a Laurent polynomial in $\xi$ of degrees $N$ and $-N$. Moreover
\be
\overline{P(\xi,z)} = P(\bar\xi^{-1},\bar z)\label{b10}
\ee
and, in the time-reversal invariant case $\Th \mathcal{H}(\xi) \Th^{-1} = \mathcal{H}(\bar\xi)$, also
\be
\overline{P(\xi,z)} = P(\bar\xi,\bar z)\;.\label{b11}
\ee
(As a function of $k$ that case holds true for $k=0,\,\pi$.)\end{lemma}

We then consider the Riemann surface ({\em Bloch variety}) defined by
\be
B =\{ (\xi,z)\in\mathbb{C}^*\times\mathbb{C}\mid P(\xi,z)=0 \}\;,\label{b12}
\ee
where $\mathbb{C}^* = \mathbb{C}\setminus\{0\}$. We make some
assumptions, which are typically true: $B$ is non-singular as a Riemann surface, meaning that $(\partial P/\partial \xi, \partial P/\partial z)\neq 0$ at all points $(\xi_0,z_0)\in B$; and, if either partial derivative vanishes, the corresponding second derivative does not. As a result, near a point $(\xi_0,z_0)$ where $\partial P/\partial z\neq 0$ we can solve $P(\xi,z)=0$ as $z=z(\xi)$ with $z$ analytic; and if $\partial P/\partial \xi= 0$ at a {\em critical point} $\xi=\xi_0$, then $z'(\xi_0)=0$, but $z''(\xi_0)\neq 0$. Similarly for interchanged roles of variables and at {\em branch points}. We conclude that eigenvalue branches $z=z(\xi)$ have only non-degenerate critical points, and only branch points of order 2.

For $k=0,\,\pi$ we will have to exceptionally allow singular points $(\xi_0,z_0)\in B$ where both derivatives of $P$ vanish. However they shall be ordinary double points, \ie the Hessian of $P$ is non-degenerate.

Of importance is also the real Bloch variety
\be
B_0 = \{ (\xi,z)\in B\mid |\xi| = 1 \}\;. \label{b13}
\ee
In fact, for any $z\in\mathbb{C}$ let $m\geq 0$ be the number of $\xi$ with $|\xi|=1$ and $P(\xi,z)=0$, \ie of points $(\xi,z)\in B_0$. Then $m>0$ implies $z\in\mathbb{R}$, since now $\bar\xi = \xi^{-1}$ and $P(\xi,\cdot)$ becomes the characteristic polynomial of a hermitian matrix. Moreover $m$ is the multiplicity of $z$ as a point of the spectrum $\s(H)$. By (\ref{b10}) the remaining $2N-m=2(N-m/2)$ points $(\xi,z)$ with $|\xi|\neq 1$ come in pairs $(\xi,z)$, $(\bar\xi^{-1},z)$. In particular $m$ is even. 

For each $(\xi, z)\in B$ we consider the (geometric) eigenspace $\widetilde E_{\xi,z}$ of the eigenvalue $z$ of the matrix seen in Eq.~(\ref{b9}). 
\begin{lemma}\label{lembl2bis} For $k\neq 0,\pi$
\be
\widetilde E:=\{(\xi,z,\Psi)\mid \text{Eq.~(\ref{b3}) holds}\}\nn
\ee
is a line bundle with base $B$ and fibers $\widetilde E_{\xi,z}$.
\end{lemma}
In order to properly state the relation between this bundle and $E$ we make the following definition.
\begin{defi} \label{gBl}
A {\em generalized Bloch solution} $(\widetilde{\psi}_{n})_{n\in\mathbb{Z}}$ of energy $z$ and quasi-periodicity $\xi\neq 0$ satisfies $(H-z)\widetilde{\psi}=0$ and 
\be \label{gBl1}
\widetilde{\psi}_{n+pM}=\xi^p(\widetilde{\psi}_n+p\xi^{-1}\psi_n)\;,
\ee
where $(\psi_{n})_{n\in\mathbb{Z}}$ is a Bloch solution for the same $\xi, z$.
\end{defi}

Eq.~(\ref{gBl1}) characterizes a generalized eigenvector (of order 2) of the translation operator. In terms of Eq.~(\ref{b2}) a generalized Bloch solution $\widetilde{\psi}$ corresponds to $\widetilde{\Psi}\in\mathbb{C}^{MN}$ with 
\be 
\mathcal{H}(\xi)\widetilde{\Psi}+(\mathcal{A}^*\xi - \mathcal{A}\xi^{-1})\xi^{-1}\Psi=z\widetilde{\Psi}\;.\nn
\ee
\begin{lemma}\label{lembl2ter}
Let the eigenvalue branch $z=z(\xi)$ have a critical point at $\xi=\xi_0$ and let $\Psi(\xi)$ be a local section of $\widetilde{E}$ at $(\xi_0, z_0=z(\xi_0))$. Then $\Psi'(\xi_0)$ corresponds to a generalized Bloch solution for $\xi_0, z_0$.
\end{lemma}
\noindent
{\it Proof.} Follows by differentiating in $\xi$ Eqs.~(\ref{b2a}, \ref{b3}). In fact,
\be \label{gBl2}
(\mathcal{H}(\xi)-z)\Psi'+\frac{d\mathcal{H}}{d\xi}\Psi=\frac{dz}{d\xi}\Psi
\ee 
with $z'(\xi_0)=0$ and $\xi d\mathcal{H} / d\xi = \mathcal{A}^*\xi - \mathcal{A}\xi^{-1}$.\qed\\
 
Finally, we consider
\be
B_- = \{ (\xi, z)\in B \mid |\xi|<1 \}\nn%\label{b14}
\ee
$(\partial B_- = B_0)$ equipped with the projection
\be
\pi: B_- \to \mathbb{C},\, (\xi,z)\mapsto z\;.\label{b14bis}
\ee
The preimage $\pi^{-1}(z)$ of a point $z\in\mathbb{C}$ consists of $N$ points iff $z\in\r(H)$. In that case we have
\be
\pi_*(\widetilde E_{\pi^{-1}(z)}) = E_z\;,\nn
\ee
where the l.h.s. is to be understood as a direct image: The fiber at $z\in\r(H)$ is $\oplus \widetilde E_{\xi,z}$ with sum over $(\xi,z)\in\pi^{-1}(z)$; the equation itself on rests on the relation (\ref{b2}) between $\Psi\in \widetilde E_{\xi,z}$ and $\psi\in E_z$. Some care applies at points $(\xi,z)$ where $\partial P/\partial \xi= 0$, \ie to $\xi$ a multiple (in fact, double) preimage of $z$, because this seemingly results in a missing solution on the l.h.s.. It should however be read so as to include the there existing generalized Bloch solution. Equivalently, for $z'$ near $z$ the fiber $\pi_*(\widetilde E_{\pi^{-1}(z')})$ is of dimension $N$ and has a limit (of the same dimension) as $z'\to z$. With this reading both sides are continuous in $z\in\r(H)$. 

Likewise, the preimage $\pi^{-1}(\g)$ of a loop $\gamma\subset \r(H)$ is a cycle in $B_-$, possibly consisting of several loops, and we have
\be
\pi_*(\widetilde E \upharpoonright \pi^{-1}(\g)) = E\upharpoonright \g\;.\label{b15}
\ee
Besides of $\pi$ we consider the projection
\be
\sigma: B_- \to \{\xi\in\mathbb{C}\mid |\xi|<1\},\, (\xi,z)\mapsto \xi\nonumber
\ee
and the bundle $\sigma_*(\widetilde E_{\sigma^{-1}(\cdot)})$. Its fibers are of dimension $MN$, even at branch points by continuous interpretation.\\

\begin{figure}[hbt]
\begin{minipage}[b]{0.45\linewidth}
\centering
\includegraphics[width=0.5\textwidth]{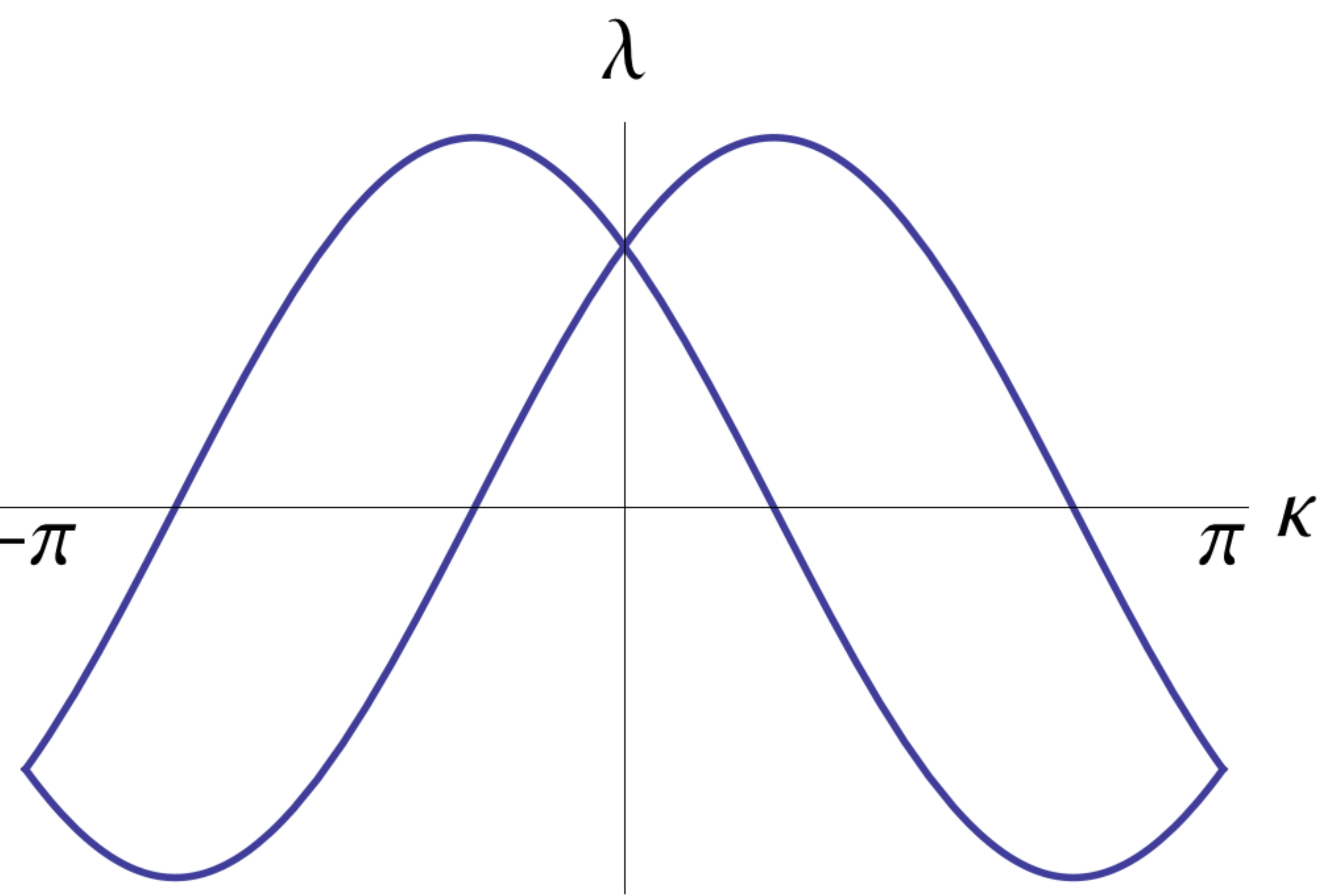}
\vskip 1.6cm
\includegraphics[width=0.5\textwidth]{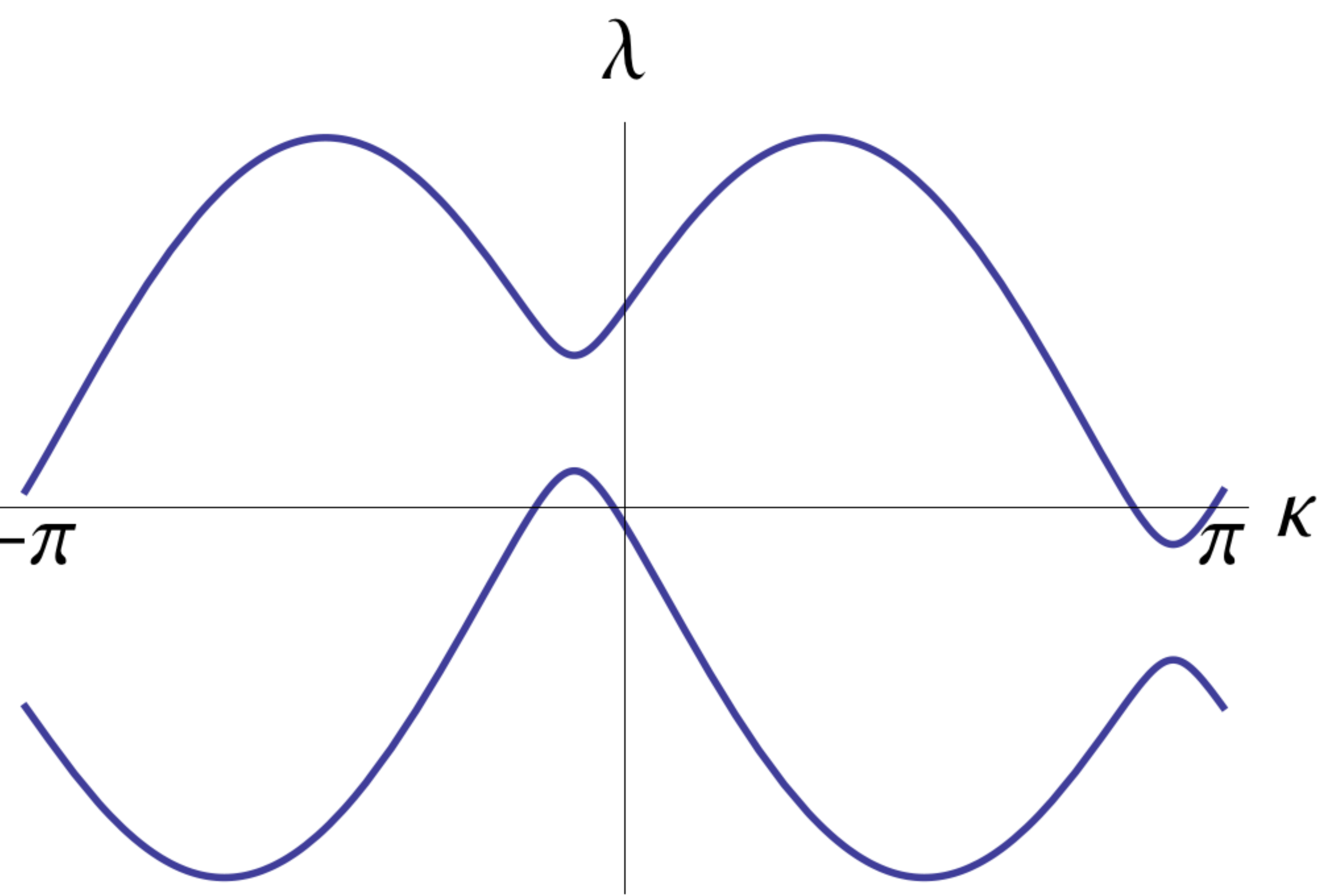}
\vskip 1.6cm
\includegraphics[width=0.5\textwidth]{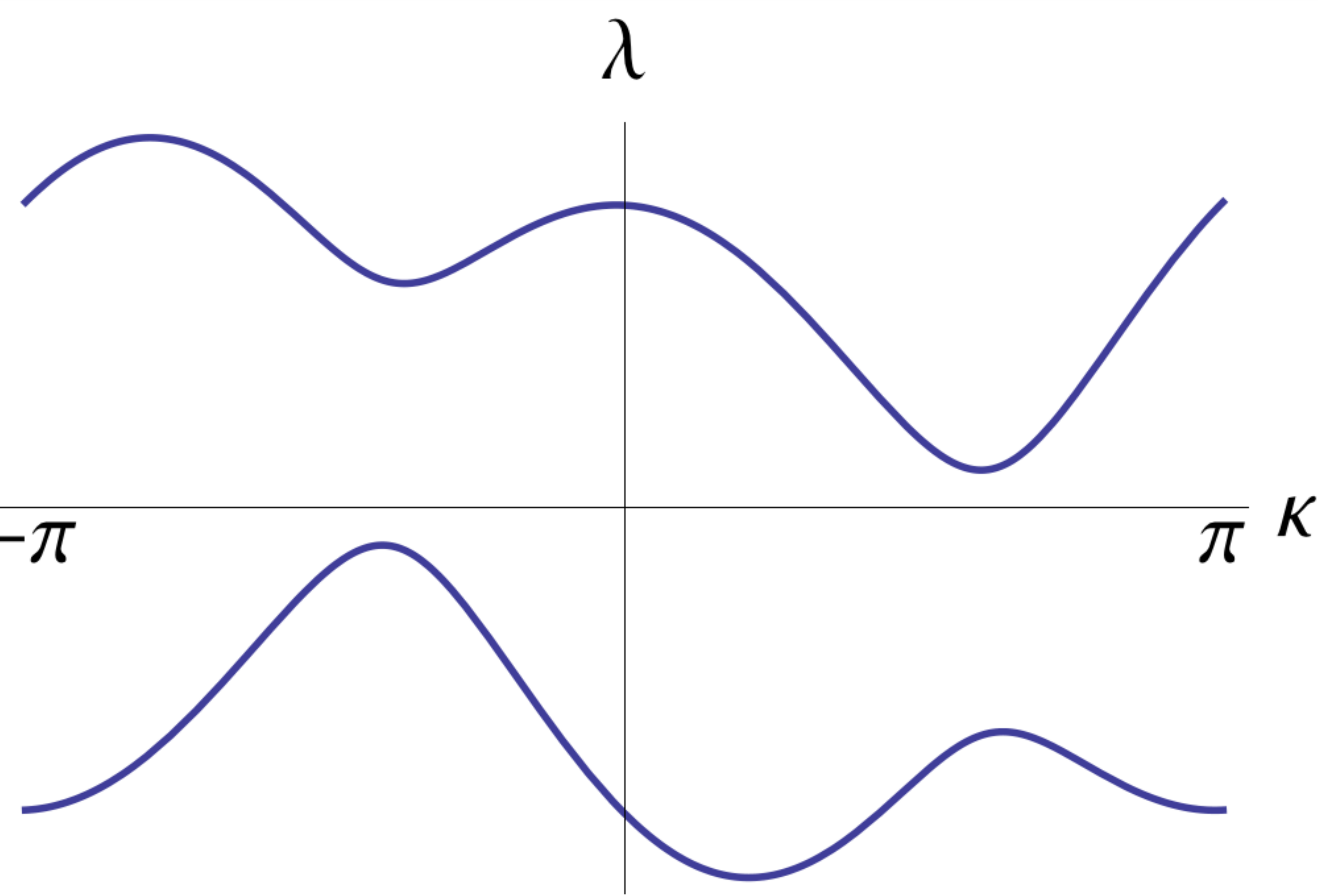}
\end{minipage}
\hspace{0.5cm}
\begin{minipage}[b]{0.5\linewidth}
\centering
\makebox[0.5cm][l]{}\hskip 0.5cm
\includegraphics[width=0.7\textwidth]{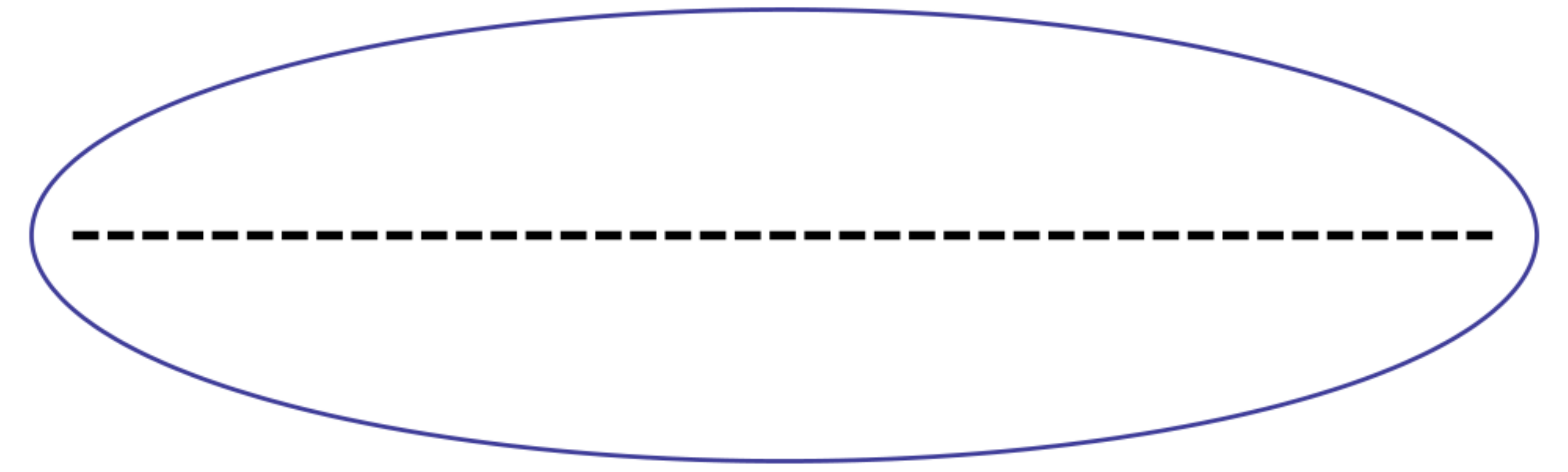}
\vskip 0.6cm
\makebox[0.5cm][l]{\raisebox{0.7cm}{(a)}}\hskip 0.5cm
\includegraphics[width=0.7\textwidth]{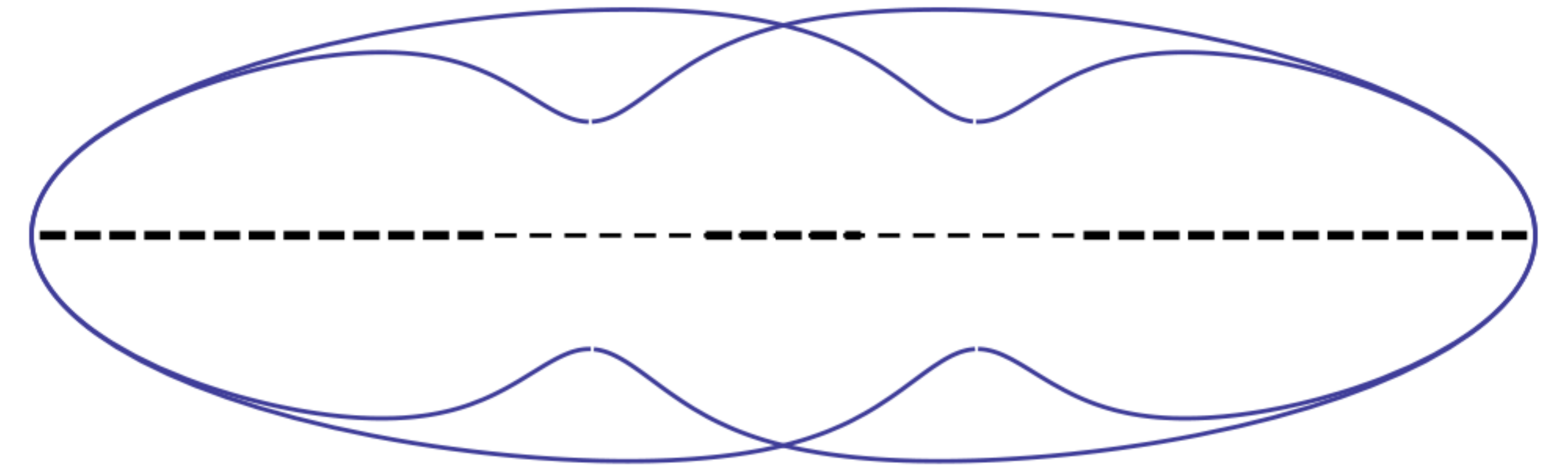}
\vskip 0.3cm
\makebox[0.5cm][l]{\raisebox{0.7cm}{(b)}}\hskip 0.5cm
\includegraphics[width=0.7\textwidth]{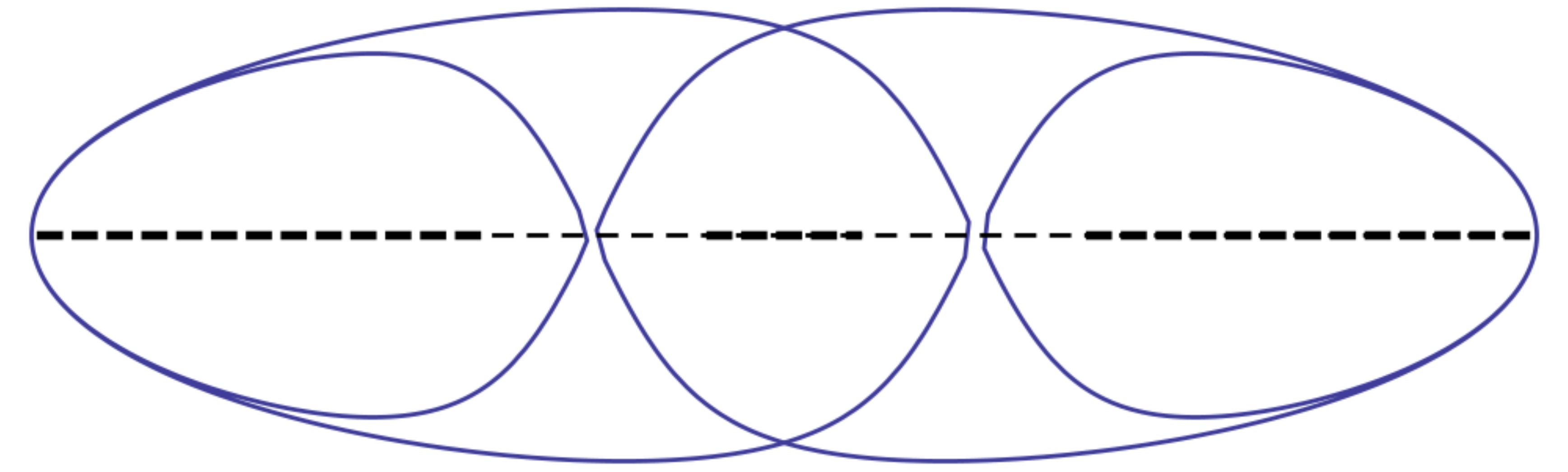}
\vskip 0.3cm
\makebox[0.5cm][l]{\raisebox{0.7cm}{(c)}}\hskip 0.5cm
\includegraphics[width=0.7\textwidth]{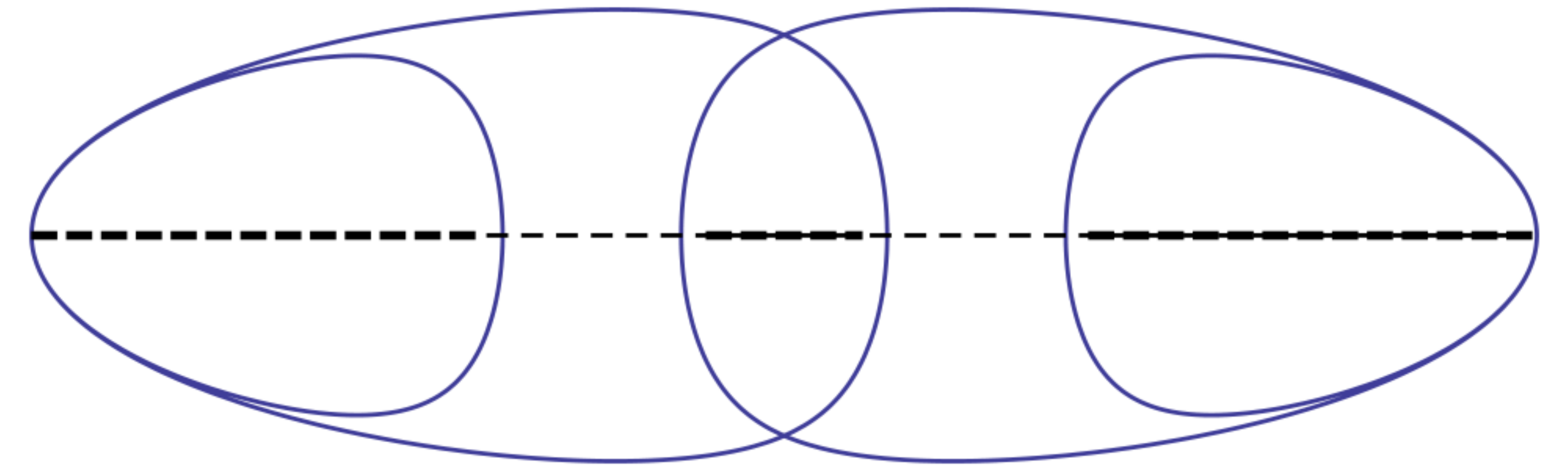}
\vskip 0.6cm
\makebox[0.5cm][l]{}\hskip 0.5cm
\includegraphics[width=0.7\textwidth]{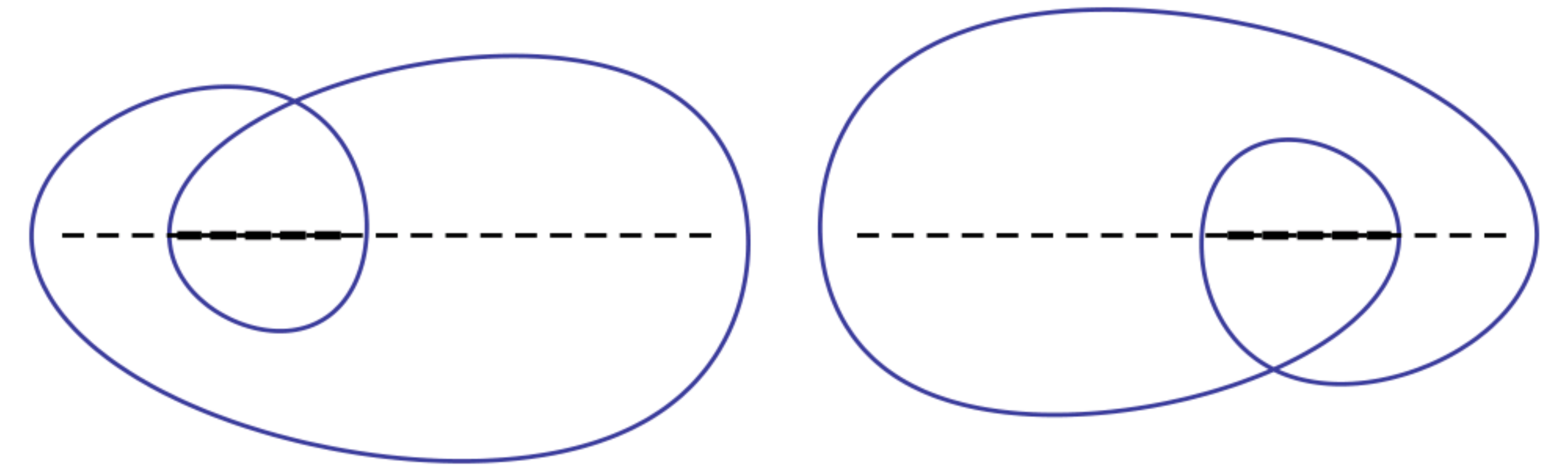}
\end{minipage}
\caption{
Left: Real energy curves $\l_{l}(e^{i\kappa})$, $(\kappa\in S^{1}, l = 2j-1,2j)$ for various values of $k$. The first case corresponds to time-reversal invariant points $k=0,\pi$. Right: Loops $\l_{l}(\xi)$, $(|\xi| = q<1)$ in the complex plane. In the first case the loop is run through twice. The second case is illustrated thrice (a--c) for different values of $q$ (or $k$). The dashed lines represent energy bands of multiplicity $2$, resp. $4$ if thick.}\label{FigBl1}
\end{figure}

\noindent
{\it Proof of Lemma~\ref{lembloch}.} We discuss a typical case only. Let us first recall the base spaces of the two bundles, $E^{(j)} = E\upharpoonright\mathbb{T}_j$ and $E_j$, seen in Eq.~(\ref{b8}): On the l.h.s. it is $\mathbb{T}_j\ni(z,k)$ and on the r.h.s. $\mathbb{B}\ni(\kappa,k)$. In a nutshell, we shall show that each side of the equation is associated with a cycle in $B_-\ni(\xi=e^{i\kappa},k)$ or actually with a family of cycles parametrized by $k\in S^{1}$; and, more precisely, with the bundle $\widetilde E$ on $B_-$ of Lemma~\ref{lembl2bis} restricted to that family. The proof then proceeds by an interpolation deforming one family into the other, up to contractible cycles. Most of the discussion occurs at fixed $k$, which is hence once more omitted. 

\begin{figure}[hbtp]
\centering
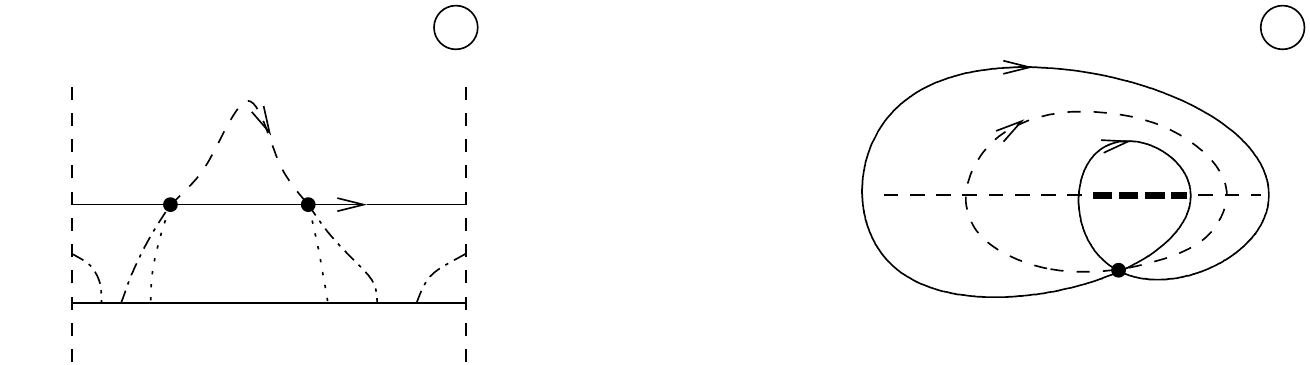
\caption{
{\em Basic elements.} Left: The sheet $l=2j$ of $B_-$ near $B_0$ and the path $I_{(2j)}=I_1\cup I_2:|\xi|=q$ parametrized as $\xi = e^{i(\kappa+ i\e)}$, ($\kappa\in S^1$) for $q=e^{-\e}<1$. Right: Its image under $\xi\mapsto \l_{2j}(\xi)$ is a cycle $\g_{1}\cup \g_{2}$ with loops $\g_{i} = \l_{2j}(I_{i})$. {\em Further elements.} Annulus $A= A_1\cup A_2 = \{ \kappa \mid 0\leq |\Im\kappa|<\e \}\cong \{ \xi \mid q<|\xi|\leq 1 \}$ and its subsets $A_i$, $(i=1,2)$, where $\l_{2j}$ is $1$ to $1$, resp. $2$ to $1$. The two sets are shown separated by a line (dotted). The set $A_2$ is mapped onto the inside of $\g_2$; $A_1$ onto the complementary subset of the inside of $\g_1$. Interpolating curve $\g_\l$ and one of its preimages, $\G_\l$ (both dashed); another preimage (dashed-dotted).}
\label{FigBl2}
\end{figure}

For $E^{(j)} = E\upharpoonright\mathbb{T}_j$ on the l.h.s. of Eq.~(\ref{b8}) the association is by (\ref{b15}) with $\gamma= \{ z\mid (z,k)\in\mathbb{T}_j \}$. For $E_j$ on the r.h.s. a longer discussion is needed. The Bloch bundle $E_j$ (see Def.~\ref{blbun}) can be analytically continued from real $\kappa$ to the path $I=\{\kappa\in\mathbb{C}\mid \Im\kappa =\e\}$, as long as $\e$ is small enough. By $\xi=e^{i\kappa}$ there correspond two paths $I_{(l)}=\{(\xi,\l_l(\xi))\mid |\xi|=e^{-\e}\}$, in the sense that $I_{(2j-1)}\cup I_{(2j)}\subset\sigma^{-1}(I)$ is a subcycle lying in the sheets $l=2j-1,2j$ of $B_-$. Then
\be
E_j=\sigma_*(\widetilde E\upharpoonright(I_{(2j-1)}\cup I_{(2j)})\;.\label{b16a}
\ee
For the sake of illustration Fig.~\ref{FigBl1} shows the real energy curves $\l_{l}(\kappa)\equiv \l_{l}(e^{i\kappa})$, $(\kappa\in S^{1}, l = 2j-1,2j)$ for various values of $k$ (left) next to the complex loops $\l_{l}(\xi)$, $(|\xi| = q)$ with $q<1$ close to $1$ (right). The loops are understood on the observation that a portion of the energy curve $\l_{l}(\kappa)$ increasing in $\kappa$ is flanked by a nearby branch running in the upper half-plane. The loops can be mutually and self-intersecting, and intersect the spectrum. However the outer boundary $\g_1$ of both loops does not. In the last case of Fig.~\ref{FigBl1} $\g_1$ consists of two disconnected loops, each surrounding one of the bands $l = 2j-1,\,2j$, the second one being shown in Fig.~\ref{FigBl2}, right.

The contour $\g\subset\rho(H)$ underlying $E^{(j)}$ may be picked as $\g_1$, at least in the cases where the latter is connected.

We claim:

\begin{lemma}\label{lembl3}
The bundle $E\upharpoonright \g$ contains a subbundle $\widehat E$ which is homotopic to the Bloch bundle $E_j \upharpoonright I \cong E_j$. Moreover
\be
\widehat E = \pi_*(\widetilde E\upharpoonright \widehat \G)\;, \label{b16}
\ee
where $\widehat \G\subset \pi^{-1}(\g)$ is a subcycle of the cycle $\pi^{-1}(\g)\subset B_-$. It covers $\g$ twice. 
\end{lemma}
As a result, $\widecheck\G := \pi^{-1}(\g)\setminus \widehat\G$ is a cycle too, and
\be
\widecheck E := \pi_*(\widetilde E\upharpoonright \widecheck \G)\nn %\label{b17}
\ee
provides a complementary subbundle, 
\be
E\upharpoonright \g = \widehat E\oplus \widecheck E\;.\nn %\label{b18}
\ee
\begin{lemma}\label{lembl4}
$\widecheck\G$ is contractible within $B_-$.
\end{lemma}

As a result $\widetilde E\upharpoonright \widecheck\G$, and hence $\widecheck E$, are trivial.

This is the core of the  proof of Lemma~\ref{lembloch}. In summary,
\be
E^{(j)} \cong E\upharpoonright \g \cong \widehat E \cong E_j\nn %\label{b19}
\ee
where $\cong$ is by homotopy or by dropping trivial subbundles. Some further comments in relation with Lemma~\ref{lembl3} are appropriate.

First, one ought to worry about whether these equivalences persist once the dependence on $k\in S^{1}$ is included into the picture. This will be understood by inspection of the proofs of the two lemmas. Indeed, the construction of $\widehat \G$, and hence that of $\widehat E$, $\widecheck E$, will not involve any choices, except among homotopically equivalent constructions. In particular, as $k$ runs from $0$ to $2\pi$, the cycles $\widehat \G$ match at the endpoints, whence the family glues up to tori; likewise for deformed cycles used in the interpolation. 

Second, no deformation is needed at $k=0,\pi$, see Fig.~\ref{FigBl1}
(first case), where
\be
\widehat \G=I_{(2j-1)}\cup I_{(2j)}\;.\label{b19a} 
\ee
Third, a cycle $\mathcal{C}\subset B_-$ (parametrized by $k$) may give rise to a (i) torus (still called $\mathcal{C}$) and to a (ii) bundle $\widetilde E\upharpoonright \mathcal{C}$, which are both time-reversal invariant. For (i) it suffices that the map $\tau: (\xi,z,k)\mapsto  (\bar\xi,\bar z,-k)$ on $B_-\times S^1$ leaves $\mathcal{C}$ invariant. For (ii) the bundle, being of dimension $1$, does not qualify as such. An additional structure is needed, namely an involution $\alpha$ on $\mathcal{C}$ preserving $k$, which moreover commutes with $\tau$. Then $\tau$ descends to the quotient $\mathcal{C}/\alpha$, which consists of pairs of points related by the involution  as well as by fixed points. Now $\widetilde E\upharpoonright \mathcal{C}$ shall actually stand for $\widetilde E\upharpoonright (\mathcal{C}/\alpha)$, whose fibers $\widetilde E_{\xi,z}\oplus\widetilde E_{\alpha(\xi,z)}$ are, by continuous interpretation, of dimension $2$ even at fixed points. Both bundles $\widetilde E\upharpoonright \widehat \G$ and $\widetilde E\upharpoonright(I_{(2j-1)}\cup I_{(2j)})$ induce such involutions, $\alpha_1$ on $\widehat \G$, resp. $\alpha_2$ on $I_{(2j-1)}\cup I_{(2j)}$: Points $(\xi,z)$ and $(\xi',z')$ are related by $\alpha_1$ if $z=z'$; and by $\alpha_2$ if $\xi=\xi'$. We remark that $\alpha_1, \alpha_2$ are consistent with Eqs.~(\ref{b16}, \ref{b16a}), respectively. At $k=0,\pi$, where the cycles agree by (\ref{b19a}), the involutions $\alpha_1, \alpha_2$ nevertheless differ. We will prove the homotopy in Lemma~\ref{lembl3} by constructing cycles $\mathcal{C}_\l$, $(1\le\l\le 2)$ interpolating between $\mathcal{C}_1=\widehat \G$ and $\mathcal{C}_2=I_{(2j-1)}\cup I_{(2j)}$. The bundle $\widetilde E\upharpoonright \mathcal{C}_\l$ goes with the ride because of the following fact.
\begin{lemma}\label{lembl5} Let $\alpha_1, \alpha_2$ be as above. Then there is an interpolating involution $\alpha_\l$ on $\mathcal{C}_\l$, $(1\le\l\le 2)$ commuting with $\t\upharpoonright\mathcal{C}_{\l}$.
\end{lemma}
\qed
\subsection{Further details}
\noindent
{\it Proof of Lemma~\ref{lembl2}.} We expand the determinant (\ref{b9}) along the first row of blocks. Then, by the Laplace rule,
\be
P(\xi,z) = (-1)^{M(N+1)}(\det A)\xi^{-N}\cdot \left| \begin{array}{ccccc} A & V_1-z & A^* & &  \\ 0 & \ddots & \ddots & \ddots &  \\ &\ddots&\ddots&\ddots&A^*\\ & & \ddots & \ddots &  V_{M-2}-z \\ A^*\xi & & & 0 & A \end{array} \right| + P_+(\xi,z)\nn %\label{b20}
\ee
where $P_+$ contains only powers of $\xi$ of degree $-(N-1)$ or higher. When looking for the leading term in $\xi^{-1}$ we may replace $A^*\xi$ by $0$, which leaves an upper block triangular matrix. That term thus is $(-1)^{M(N+1)}(\det A)^{M}$ $\xi^{-N}$. Similarly, the leading term in $\xi$ is $(-1)^{M(N+1)}(\det A^*)^{M}\xi^{N}$. Eq.~(\ref{b10}) follows from $\overline{\det \mathcal{M}}=\det\mathcal{M}^*$. In the time-reversal invariant case we use $\overline{\det \mathcal{M}}=\det(\Theta\mathcal{M}\Theta^{-1})$, where $\Theta=\diag(\Theta,\ldots, \Theta)$ in the notation of Eq.~(\ref{b2}). Hence the last claim. \qed\\

\noindent
{\it Proof of Lemma~\ref{lembl2bis}.} To be shown is that, for each $(\xi, z)\in B$, the eigenvalue $z$ of $\mathcal{H}(\xi)$, \cf~Eq.~(\ref{b3}), is geometrically simple.

If $\partial P/\partial z\neq 0$ the eigenvalue $z=z(\xi)$ is even algebraically simple. In that case the eigenspace depends analytically on $\xi$. If $\partial P/\partial z=0$ at $(\xi_0,z_0)\in B$, meaning a branch point, then we have by our general assumptions,
\be
\frac{\partial^2 P}{\partial z^2}\neq 0\;,\qquad \frac{\partial P}{\partial \xi}\neq 0\;;\nn
\ee
in particular $z_0$ is an eigenvalue of $\mathcal{H}(\xi_0)$ of algebraic multiplicity $2$ and $\xi'(z_0)=0$, $\xi''(z_0)\neq 0$. It follows that nearby eigenvalues of $\mathcal{H}(\xi)$ are given as
\be
z - z_0 = \pm\sqrt{\frac{2}{\xi''(z_0)}(\xi-\xi_0)} + O(\xi - \xi_0)\;, \qquad(\xi\to\xi_0)\;.\label{ba21}
\ee
In order to show that the geometric multiplicity of $z_0$ is nevertheless $1$, we denote by $\Pi(\xi)$ the projection onto the group (\ref{ba21}) of eigenvalues $z$ (it is analytic in $\xi$; \cite{Ka}, Sect.~II.1.4), as well as by $\Pi_0 = \Pi(\xi_0)$ and by
\be
\mathcal{N}_0 = (\mathcal{H}(\xi_0) - z_0)\Pi_0\nn
\ee
the eigenprojection and eigennilpotent of $z_0$. The claim amounts to $\mathcal{N}_0\neq 0$, and we prove it by contradiction: $\mathcal{N}_0 = 0$ implies that
\be
\frac{\mathcal{H}(\xi) - z_0}{\xi - \xi_0}\Pi(\xi)\nn
\ee
has a removable singularity at $\xi=\xi_0$. Hence $z-z_0 = O(\xi - \xi_0)$, in violation of (\ref{ba21}).\qed\\

The proofs of Lemmas~\ref{lembl3} and~\ref{lembl4} will mostly deal with a single $k$ at a time and will be given them in two parts. The first part is, so to speak, a test run which avoids the complications due to the overlapping bands. It is thus restricted to the last case of Fig.~\ref{FigBl1} and pretends that $\g=\g_1$ even there. The limitations will be overcome in a second part.\\

\noindent
{\it Proof of Lemma~\ref{lembl3}, first part.} We consider a homotopy of curves $\g_{\l}$, $(2\geq \l\geq 1)$ which lie between $\g_2$ and $\g_1$. In particular, we keep the endpoints of $\g_{\l}$ fixed, as indicated by the dashed line in Fig.~\ref{FigBl2} (right). Among the many curves in $\pi^{-1}(\g_{\l})$ let us track the curve, $\G_\l\subset \overline{B_-}$, arising from $I_2$ at $\l=2$ by continuity, \ie $\G_2 = I_2$. The curve $\G_\l$ has to stay outside of $A$. Indeed, it can cross neither $I_2$, as $\pi(\G_\l)$ would end up inside $\g_2$, nor $I_1$, as $\pi(\G_\l)$ would approach $\g_1$ from the outside. As $\G_{\l}$ does not penetrate into $A$, it does not approach the real $\kappa$ axis (\ie $|\xi| = 1$) on the branch of $\l_{2j}$. Moreover it does not either on any other branch of $B_0$, since those map to other, disjoint bands. That ensures that $\mathcal{C}_\l=I_{1}\cup \G_{\l}$ is a cycle in $B_-$ and that the bundle $\widetilde E\upharpoonright (I_1\cup\G_{\l})$ is defined and continuous in $\l$. Finally, the homotopy of bundles $\pi_*(\widetilde E\upharpoonright (I_1\cup \G_{\l}))$, $(2\geq\l\geq 1)$ does the job: For $\l=2$ we have
\be
\pi_*(\widetilde E\upharpoonright (I_1\cup I_2)) = E_j\upharpoonright I\nn %\label{b21}
\ee
by construction; for $\l=1$, $\pi(I_1\cup\G_{1}) = \g_1\cup\g_1 = \g_1$ and
\be
\pi_*(\widetilde E\upharpoonright(I_1\cup \G_1)) \subset \pi_*(\widetilde E\upharpoonright \pi^{-1}(\g_1)) = E\upharpoonright \g_1\nn
\ee
by (\ref{b15}). Thus Lemma~\ref{lembl3} holds with $\widehat\G = I_1\cup \G_1$.\qed\\

Let us recall the count of preimages $\pi^{-1}(z)$ done in relation with Eq.~(\ref{b14bis}). We find it convenient to pretend that all points $z\in\mathbb{C}$, \ie including $z\in\s(H)$, have precisely $N$ preimages under $\pi$. This is enforced by including the $m$ preimages on $B_0$, but by counting them with weight $1/2$, as explained in the sequel of Eq.~(\ref{b13}). The preimages of a loop intersecting the spectrum then is a {\em pseudo-cycle}, where it is tolerated that curves break up at pairs of points in $B_0$ with same $\l_l(\kappa)$ (see the joined dashed and dashed-dotted curves in Fig.~\ref{FigBl2}, left).\\

\noindent
{\it Proof of Lemma~\ref{lembl4}.} We first observe from the previous proof that any point $z$ inside $\g_1$ has $N-2$ preimages $(\xi, z)$ outside of and away from $I_1\cup \G_1$. We next extend $\g_{\l}$ from $1\leq \l\leq 2$ to $2\leq \l\leq 3$ so that it contracts $\g_2$ to a point $\g_3 = \{ z_0 \}$ inside of it. We then consider the deformation $\widehat\G_{\l}$, $(1\leq \l\leq 3)$ of pseudo-cycles covering $\g_{\l}$ twice and starting with $\widehat\G_1 =  I_1\cup \G_1$. By continuity in $\l$, $\widehat\G_{\l}$ moves inside of $I_1\cup \G_1$, while $N-2$ preimages of $\g_\l$ remain outside. As a result $\widecheck \G_{\l} := \pi^{-1}(\g_\l)\setminus \widehat\G_{\l}$ stays away from $B_0$ and is a cycle in $B_-$.\qed\\

\comment{We shall construct a deformation $\widehat\G_{\l}$, $(1\leq \l\leq 2)$ of cycles covering $\g_{\l}$ twice and starting with $\widehat\G_1 = \G_1$. One curve (with fixed endpoints in $B_-$) contained in $\widehat\G_{\l}$ is just $\G_\l$: it backtracks the deformation used in the previous proof and provides one cover of $\g_\l$.

Exactly one curve (with same but opposite endpoints as $\G_\l$) lies in $A_1$, as follows from the definition of $A_1$ mentioned in the caption of Fig. 3. 

It though has 2 further endpoints on $B_0$, as mentioned in the previous proof. However that second cover of $\g_\l$ exhausts $m=2$ preimages of the spectral points swept in the process. As a result, $\widecheck \G_{\l} := \pi^{-1}(\g_\l)\setminus \widehat\G_{\l}$ stays away from $B_0$ and is a cycle in $B_-$.

Next we consider a deformation $\g_{\l}$, $(2\leq \l\leq 3)$ which contracts $\g_2$ to a point $\g_3 = \{ z_0 \}$ inside it. Moreover, we continuously extend $\widehat\G_{\l}$ in $2\leq \l\leq 3$ as twice covering $\g_{\l}$. Now the two coverings are provided by two curves contained in $\widehat \G_{\l}$, both lying in $A_2$ by the definition of that set. Again, no further curves in $\pi^{-1}(\g_\l)$ penetrate $A$, nor touch $B_0$. We conclude that $\widecheck \G_{\l}$ contracts to $\widecheck \G_{3}$, which consists of points $\pi^{-1}(\{z_0\})\cap B_-$.\qed\\}

\noindent
{\it Proof of Lemma~\ref{lembl3}, second part.} We extend the proof to general values of $k$. They are illustrated by the three cases seen in Fig.~\ref{FigBl1} with the middle one having three variants (a--c). The argument will rely on two sheets ($l=2j-1, 2j$) of $B_-$, as opposed to just one before (see Fig.~\ref{FigBl2} left). Each comes with an annulus in $\kappa$ bounded above by $I_{(2j-1)}, I_{(2j)}$. The sheets share two branch points: In the first case ($k=0$ or $\pi$) they rather are, to be precise, two double points at $\kappa=0$ and $\pi$. In the variant (a) each of them gives way to two branch points, out of which one is in $B_-$, and actually in the annuli; in (b) the branch points have reached the paths $I_{(2j-1)}, I_{(2j)}$ (this might occur at different $k$ for the two points), and in (c) they have left the annuli. This is seen in Fig.~\ref{FigBl1} (right) as a rearrangement of loops through an intersection point at (b). It should be noted that the Bloch bundle with fibers $[\Psi_{2j-1}(\kappa,k),\Psi_{2j}(\kappa,k)]$ remains continuous as the branch point reaches the paths $I_{(2j-1)}, I_{(2j)}$, at least if properly interpreted: That point is then common to the two paths, resulting in a just 1-dimensional fiber, \cf~Lemma~\ref{lembl2bis}; however the 2-dimensional fiber has a limit at the branch point, as remarked after Eq.~(\ref{ba21}), and unlike the two spanning vectors individually. That limit provides the appropriate fiber at the branch point.

In the first case of Fig.~\ref{FigBl1}, where the curves do not intersect the spectrum, the conclusion is reached as before, but without need for deformations. We next turn to the three variants (a--c) of the middle case. Let us specify the deformations of the curves in $\mathbb{C}$ and describe their lifts in $B$.

(a-b) We deform the inner parts $\g_2$ of the curve till they reach the outermost parts $\g_1$, while keeping the intersection points fixed. By the same arguments as before the preimage selected by continuity remains outside the annuli of both sheets. 

(c) Let us expand the inner loops. Quite soon the situation will look as in the undeformed variant (b) with the inner loops touching essentially the same curves as they did there. By continuity from (b) the touching also takes places between their respective lifts in $B$. From there on the deformation proceeds as in variant (b). As a whole it includes and reverts the rearrangement of loops which occurred between (a) and (c). 

In the last case we first perform the deformation described in the first part of this proof. The two components of $\g_1$, which at this point are run through twice, still require a deformation to a countour $\g$ encircling both bands at once. When the two components of $\g_1$ first touch, so do the two lifts of each,  as inherited from (c) by continuity.\qed\\

Likewise the proof of Lemma~\ref{lembl4} extends to the other cases, too. In the first case ($k=0,\pi$) the loops $\g_\l$ ($2\le \l\le 3$) are to be chosen so that $\overline{\g_\l}=\g_\l$. \\

\noindent
{\it Proof of Lemma~\ref{lembl5}.} We begin with some preliminary observations about circle homeomorphisms $\beta: S^1\to S^1$, $p\mapsto \beta(p)$. Their degree is $d(\beta)=\pm 1$ and any two of them are homotopic iff they have the same degree ($\beta_1\sim\beta_2\Leftrightarrow d(\beta_1)=d(\beta_2)$); if so, the homotopies $\beta_\l$, $(1\le\l\le 2)$ between them fall into connected components which, relatively to one another, are labeled by $n\in\mathbb{Z}$. Indeed, for $\beta_1=\beta_2$ and fixed $p$ the map $\l\mapsto\beta_\l(p)$ has a winding number $n$ (independent of $p$).

Continuous maps $\beta: S^1\to S^1$ are called involutions if $\beta\circ\beta=1$. They are homeomorphisms. Involutions have either all or no points $p\in S^1$ as fixed points, or else just two \cite{Pf}. The two cases correspond to $d(\beta)=\pm 1$, respectively. The above statements about homotopy remain true within the class of involutions. 

We remark that the properties to be proven,
\be
\alpha\circ\alpha=1\;,\qquad\alpha\circ\tau=\tau\circ\alpha\;,
\label{aprop}
\ee
are relations within or between the fibers at $k$ and $-k$, respectively. 

We first consider the cases $k=0,\pi$. There the cycle $\mathcal{C}_\l=S^1_+\sqcup S^1_-$ is independent of $\l$ and consists of two loops $S^1_\pm$ with the involution $\tau: S^1_\pm\to S^1_\mp$, $(\xi,z)\mapsto (\bar\xi,\bar z)$ (see (\ref{b11}, \ref{b12})) interchanging them. Maps $\alpha: S^1_\pm\to S^1_\mp$ are expressible in terms of $\beta=\alpha\circ\tau: S^1_\pm\to S^1_\pm$ and properties (\ref{aprop}) are (jointly) equivalent to $\beta\circ\beta=1$, $\beta\circ\tau=\tau\circ\beta$; actually the latter property just determines $\beta\upharpoonright S^1_-$ in terms of $\beta\upharpoonright S^1_+$. It is clear from the definitions of $\alpha_1, \alpha_2:S^1_\pm\to S^1_\mp$ that they satisfy (\ref{aprop}). Moreover it is impossible that $\alpha_1(\xi,z)=\alpha_2(\xi,z)\equiv(\xi', z')$ for some $(\xi,z)\in S^1_\pm$, since that would imply $\xi=\xi'$, $z=z'$ in contradiction with $(\xi', z')\in S^1_\mp$. Hence $\alpha_1\circ\alpha_2$ has no fixed points. But that map equals $\beta_1\circ\beta_2$, whence $d(\beta_1)d(\beta_2)=d(\beta_1\circ\beta_2)=1$ and $d(\beta_1)=d(\beta_2)$. Therefore there is an interpolation $\beta_\l$, $(1\le\l\le 2)$ with $\alpha_\l$ satisfying (\ref{aprop}). In addition, there are more of them, differing from it by any given winding $n\in\mathbb{Z}$.

We now turn to $0<k<\pi$. The second property (\ref{aprop}) will simply determine $\alpha_\l$ at $-k$ in terms of its value at $k$. Hence only the first one matters in the following construction of $\alpha_\l$, $(1<\l< 2)$. To begin, for (i) $k$ near $0$ the cycle $\mathcal{C}_\l$ can still be identified with $S^1_+\sqcup S^1_-$; further on it may (ii) collapse into a ''figure 8'' (for $\lambda=2$ this occurs when a single branch point intersects $\mathcal{C}_2$; but does not for $\l=1$), and thereafter (iii) become a single loop. Eventually, the cycle reverts to two separate loops before $k$ reaches $\pi$ . In range (i) $\alpha_\l$ ought to interchange the two loops and (\ref{aprop}) just determines $\alpha_\l\upharpoonright S^1_- $ in terms of $\alpha_\l\upharpoonright S^1_+$. That half of $\alpha_\l$ can simply be chosen by continuity from $k=0$. At (ii) continuity requires that the intersection point be a fixed point of $\alpha_\l$, which can be arranged for on the side of (i). In the range (iii) $\alpha_\l$ is an involution on a single loop with just two fixed points. At the end of that range, the two fixed points are again to coalesce into an intersection point, but that does not obstruct the construction of $\alpha_\l$ (see the observations at the beginning of the proof). 

The whole construction could have begun from $k=\pi$ instead. We have to ensure that the two interpolations $\alpha_\l$ constructed from the two ends match at some $0<k<\pi$. This can be arranged, because their relative winding $n$ can be chosen to vanish.\qed
 
\section{Bulk-edge correspondence through scattering theory: Proofs}
\label{bec}
\noindent
{\it Proof of Lemma~\ref{lemrefl}.}
i) The map $r(\kappa) \equiv r(\kappa,k)$ is well-defined by the required properties, and it remains to show that it is real analytic in $\kappa$. Away from critical points, $\kappa\neq \kappa_{\pm}$, let $F(\kappa,r) = \l(\kappa) - \l(r)$, so that $F(\kappa,r(\kappa))=0$. Since $\partial F/\partial r = -\l'(r)\neq 0$ for $r=r(\kappa)$, the claimed analyticity follows by the implicit function theorem. Near a critical point, say $\kappa_+$, the argument must be modified. Let there
\be
F(\kappa,r) = \begin{cases} \frac{\l(\kappa) - \l(r)}{\kappa - r}, & (\kappa\neq r) \\ \l'(\kappa), & (\kappa=r).  \end{cases}\nn
\ee
Note that $F(\kappa,r)=0$ still has the solution $r=r(\kappa)$, but no longer $r=\kappa$, except for $\kappa = \kappa_+$. Since
\be
\frac{\partial F}{\partial r} = \frac{\l(\kappa) - \l(r) - \l'(r)(\kappa - r)}{(\kappa - r)^2}\nn
\ee
equals $\l''(\kappa_{+})/2\neq 0$ at $(\kappa,r) = (\kappa_+,r(\kappa_+))$ the solution $r=r(\kappa)$ is analytic also near $\kappa=\kappa_+$.\\

ii) We recall the Def.~\ref{bb_hall} of the Bloch bundle $E_{\ell}$. As shown in Fig.~\ref{FigHall}, $\widetilde{\mathbb{B}}_-$ does not contain any non-contractible loop winding around $\kappa\in S^{1}$. Thus there is no obstruction for a smooth section $\Psi(\kappa,k)\neq 0$ of $E_{\ell}$ on $\widetilde{\mathbb{B}}_-$, or even on small complex neighborhood in $\kappa$ thereof. To be shown is that there is one, $\Psi^{-}(\kappa,k)$, which at fixed $k$ is analytic in $\kappa$ for $(\kappa,k)$ in that neighborhood. As a preliminary, let $P(\kappa,k)$ be the Riesz projection onto the fiber $(E_{\ell})_{\kappa,k}\in\mathbb{C}^{NM}$,
\be
P(\kappa) = -\frac{1}{2\pi i}\oint (\mathcal{H}(\kappa) - z)^{-1}dz\;,\nn
\ee 
where $\mathcal{H}(\kappa)\equiv \mathcal{H}(e^{i\kappa})$ is defined in Eq.~(\ref{b3}), the variable $k$ is suppressed, and the integration contour surrounds once the eigenvalue $\l_{\ell}(\kappa)$ of $\mathcal{H}(\kappa)$, and no further ones. We observe that $P$ is analytic in $\kappa$.

We construct $\Psi^{-}$ first by setting $\Psi^{-}(\kappa,k) = \Psi(\kappa,k)$ for, say, $\kappa = \kappa_+(k)$; then by extending it in $\kappa$ (at fixed $k$) through parallel transport:
\be
P\overline{\partial_{\kappa}}\Psi^{-}=0\;,\qquad \Psi^{-} = P\Psi^{-}\;.\label{qhi6a}
\ee
This is feasible. In fact, the ansatz
\be
\Psi^{-}(\kappa) = \Psi(k)\l(\kappa)\;,\qquad (\l(\kappa)\in\mathbb{C}^*)\nn
\ee
reduces both equations to
\be
P(\overline{\partial_{\kappa}}\Psi) + (P\Psi)\overline{\partial_{\kappa}}\log\l=0\;.\nn
\ee
Here $P(\overline{\partial_{\kappa}}\Psi) = f(P\Psi)$ for some function $f(\kappa)\in\mathbb{C}$, since the fibers are lines. The equation thus reads $\overline{\partial_{\kappa}}\log\l = -f$, which can be integrated starting from $\l(\kappa_+)=1$. It remains to verify the Cauchy-Riemann condition $\overline{\partial_{\kappa}}\Psi^{-}=0$. By the Eqs.~(\ref{qhi6a}) and $\overline{\partial_{\kappa}}P=0$ we indeed have $\overline{\partial_{\kappa}}\Psi^{-}= P(\overline{\partial_{\kappa}}\Psi^{-})=0$.\qed\\

In preparation for the proof of Lemma~\ref{lemedge}, or actually of an extension thereof, let us list two further properties of the Casoratian, Eq.~(\ref{cas}). This time, the energy $z$ in Eq.~(\ref{qhi7a}) is real.
\begin{enumerate}
\item[C4)] Let $\psi$ be a Bloch solution of quasi-periodicity $\xi = e^{i\kappa}$, $(\kappa\in S^{1})$ and energy $z=\l(\kappa)$. Then
\be
C(\psi^*,\psi) = -i\l'(\kappa)\sum_{n=0}^{M-1}\psi^*_{n}\psi_{n}\;.\label{qhi7'}
\ee
\item[C5)] Let $\psi = \psi(\kappa)$ be a Bloch solution as in (C4) for $\kappa$ near $\kappa_{\pm}$. Then $\psi'(\kappa_\pm)$ is a generalized Bloch solution by Lemma~\ref{lembl2ter} and
\be
C(\psi'^*,\psi) - C(\psi'^*,\psi)^* = -i\l''(\kappa_+)\sum_{n=0}^{M-1}\psi_{n}^*\psi_{n}\;.\label{qhi7aa}
\ee
\comment{
\item[C6)] If $\psi^{\sharp}_0 = 0$ then
\be
C_0(\psi^{\sharp *},\psi^{\sharp})=0\;.\nn
\ee
}
\end{enumerate}
\comment{
Only Eqs.~(\ref{qhi7'}, \ref{qhi7aa}) are not straightforward enough to deserve an explanation. }
The derivation is as follows. We recall that $\Psi$ in Eq.~(\ref{b2}) satisfies Eq.~(\ref{b3}) with $\mathcal{H}(\xi) = \mathcal{H}(\xi)^*$.
\comment{Differentiating
\be
(\Psi,(\mathcal{H}(e^{i\kappa}) - \l(\kappa))\Psi)=0\nn
\ee
w.r.t. $\kappa$ yields the Feynman--Hellmann equation}
Eq.~(\ref{gBl2}) thus yields the Feynman--Hellmann equation
\be
(\Psi, \xi \frac{d \mathcal{H}}{d\xi}\Psi) = -i\frac{d\l}{d\kappa}(\Psi,\Psi)\nn
\ee
by $\xi(d/d\xi)=-id/d\kappa$, the inner product being the standard one on $\mathbb{C}^{MN}$. The l.h.s. equals 
\comment{In view of $\xi d\mathcal{H} / d\xi = \mathcal{A}^*\xi - \mathcal{A}\xi^{-1}$}
\be
\xi\psi^*_{M-1}A^*\psi_{0} - \bar\xi \psi^{*}_0A\psi_{M-1} = \psi^{*}_{M-1}A^*\psi_{M} - \psi^*_{M}A\psi_{M-1} = C_{M-1}(\psi^*,\psi)\;,\nn
\ee
as claimed in (C4).
\comment{Clearly
\be
\psi'(\kappa_+) = \lim_{\kappa\to\kappa_+}\frac{\psi(\kappa) - \psi(r(\kappa))}{\kappa - r(\kappa)}\;,\nn %\label{qhi7ab}
\ee
is a solution for $z = \l(\kappa_+)$, because $\l(\kappa) = \l(r(\kappa))$. }
Differentiation of (\ref{qhi7'}) yields (\ref{qhi7aa}) because of $C(\psi^*,\varphi^*)= -C(\varphi,\psi)^*$ and $\l'(\kappa_{\pm})=0$. \\

Let us consider a complex neighborhood of the band $\l$ under consideration:
\be
N(\d_0) = \{ \l(\kappa)\mid |\Im\kappa|<\d_0 \}\;,\qquad (\d_0>0)\nn
\ee
The following lemma manifestly implies Lemma~\ref{lemedge}, which therefore will not require a separate proof. 

\begin{lemma}{\em [Solutions of at most small exponential growth]}\label{lemsol}
For small enough $\e_0>0$ there is $\d_0>0$ such that we have: For $z=\l(\kappa)\in N(\d_0)$ there is (up to multiples) a unique solution $\psi^{\sharp}\neq 0$ of
\be
(H^{\sharp} - z)\psi^{\sharp}=0\;,\qquad \psi^{\sharp}_0=0\nn%\label{qhi7ah}
\ee
with $\psi^{\sharp}_n = O(e^{\e_0 n})$, $(n\to +\io)$. It is of the form
\be
\psi^{\sharp}_n = \psi^{(12)}_n + O(e^{-\e_0 n})\;,\quad (n\to +\io)\label{qhi7ac}
\ee
where
\be
\psi^{(12)}_n  = \psi^{(1)}_n + \psi^{(2)}_n \neq 0\label{qhi7ad}
\ee
and
\begin{enumerate}
\item[i)] for $z\neq \l(\kappa_{\pm})$: $\psi^{(1)}_n$, $\psi^{(2)}_n$ are Bloch solutions for $H$ with $\xi = e^{i\kappa}$, resp. $\xi = e^{ir(\kappa)}$;
\item[ii)] for $z = \l(\kappa_{\pm})$: $\psi^{(1)}_n$, $\psi^{(2)}_n$ are Bloch, resp. generalized Bloch solutions with $\xi = e^{i\kappa} = e^{ir(\kappa)}$.
\end{enumerate}
\comment{:
\be
\psi^{(2)}_{n+M} = \xi(\psi^{(2)}_{n} + \psi^{(1)}_n)\;.\label{qhi7ae}
\ee
}
Moreover, if $z=\l(\kappa)$, $(\kappa\in S^1$), then $\psi_{n} = O(1)$ and $\psi^{(1)}_n,\psi^{(2)}_n\neq 0$ in case (i); and $\psi_{n} = O(n)$ in case (ii).
\end{lemma}

\noindent
{\it Proof.} At first, let $z$ belong to the band under consideration, \ie $z=\l(\kappa)$, $(\kappa\in S^1$). As explained in connection with the Bloch variety Eq.~(\ref{b12}), we have $P(\xi,z)=0$ for $m=2$ values of $\xi$ with $|\xi|=1$ (counting multiplicities) and for $N-1$ values with $|\xi|\le e^{-\e_0}$ and small enough $\e_0>0$. By possibly making it smaller the same remains true for $z\in N(\d_0)$ and small $\d_0>0$, provided $|\xi|=1$ is replaced by $e^{-\e_0}<|\xi|<e^{\e_0}$. Any eigensolution $\psi$ of $H$, which remains $O(e^{\e_0 n})$, $(n\to +\io)$ is thus of the form
\be
\psi_n = \psi^{(1)}_n + \psi^{(2)}_{n} + \sum_{j=3}^{N+1}\psi^{(j)}_{n}\;,\quad (n\in\mathbb{Z})\label{qhi7ag}
\ee
where $(\psi^{(j)}_{n})$ is a (generalized) Bloch solution corresponding to the above $\xi$'s, in the stated order. Disregarding the boundary condition, the same applies to eigensolutions $\psi^{\sharp}$ of $H^{\sharp}$ for $n>n_0$, by means of the map (\ref{bij}).

We claim that up to multiples there is precisely one such solution $\psi^{\sharp}$, once the boundary condition $\psi^{\sharp}_0 = 0$ is imposed. Without any conditions the space of solutions $\psi^{\sharp}$ has dimension $2N$. The subspaces determined by $\psi^{\sharp}_0=0$, resp. $O(e^{\e_0 n})$, have dimensions $N$, $N+1$. Their intersection $V$ thus has dimension at least $1$. We make two claims: $(a)$ $V\cap \{\psi^{\sharp}\mid \psi^{(12)}=0 \} = \{ 0 \}$, \cf~(\ref{qhi7ac}), which implies Eq.~(\ref{qhi7ad}) and $\dim V\leq 2$; and the stronger: $(b)$ $\dim V\leq 1$, and hence $\dim V = 1$, which is the statement of uniqueness of $\psi^{\sharp}$. Properties $(a,b)$ are stable, because the nullity $\dim(V_1\cap V_2)$ of a pair of subspaces is an upper semi-continuous function of them (\cite{Ka}, Thm.~IV.4.24); hence it suffices to prove them for $z= \l(\kappa)$ with $\kappa$ real.

We begin with $(a)$: The opposite would amount to an embedded eigenvalue, which is ruled out by (\ref{qhi6b}). To prove $(b)$, together with the remaining claims, we consider first the case $\kappa \neq \kappa_{\pm}$, where $\xi = e^{i\kappa}, e^{ir(\kappa)}$ are different. Suppose, indirectly, $\dim V=2$. Then, by taking a suitable linear combination of solutions we could arrange for $\psi^{(1)}\neq 0$, $\psi^{(2)}=0$ (or viceversa). This would imply the contradiction 
\be
0 = C_{0}(\psi^{\sharp*},\psi^{\sharp}) = \lim_{n\to \io} C_{n}(\psi^{\sharp*},\psi^{\sharp}) = C(\psi^{(1)*},\psi^{(1)}) \neq 0\nn
\ee
by $\psi^{\sharp}_0 = 0$ and the preliminary remark (C4). Finally we are left with $\kappa = \kappa_+$ (or $\kappa_-$). This time $\dim V = 2$ would imply that there exist two solutions $\psi^{\sharp}_\mathrm{a},\, \psi^{\sharp}_\mathrm{b}\in V$: one with $\psi^{(1)}_\mathrm{a} = \psi$, $\psi^{(2)}_\mathrm{a} = 0$ in Eq.~(\ref{qhi7ag}) and the other with $\psi^{(1)}_\mathrm{b}=0$, $\psi^{(2)}_\mathrm{b} = \psi'$. The resulting contradiction is
\be
0 = C_0(\psi^{\sharp*}_\mathrm{b},\psi^{\sharp}_\mathrm{a}) = C(\psi'^*,\psi)\neq 0\nn
\ee
by (C5) and $\l''(\kappa_+)\neq 0$. 
\comment{Differentiating $\psi_{n + Mp}(\kappa) = e^{i\kappa p}\psi_{n}(\kappa)$, $(n,p\in\mathbb{Z})$ yields
\be
-i\psi'_{n+Mp}(\kappa) = e^{i\kappa p}(-i\psi'_{n}(\kappa) + p\psi_{n}(\kappa))\nn
\ee
and hence (\ref{qhi7ae}) and the bound $O(n)$.}
Moreover, the bounds $O(1)$ and $O(n)$ follow from Eqs.~(\ref{b2a}, \ref{gBl1}) with $|\xi|=|e^{i\kappa}|=1$. \qed\\

\noindent
{\it Proof of Lemma~\ref{lemsemi}.} Let again $k$ be fixed till further notice. We consider $z$ in a complex neighborhood of the upper band edge $\l(\kappa_+)$. There the expanded part of the solution (\ref{qhi7ac}) may alternatively be written as a linear combination of
\be
\psi^{(1)}(\kappa) = \psi^{-}(\kappa) + \psi^{-}(r(\kappa))\;,\qquad
\psi^{(2)}(\kappa) =\begin{cases}  \frac{\psi^{-}(\kappa) - \psi^{-}(r(\kappa))}{\kappa - r(\kappa)}\,, & (\kappa\neq\kappa_{+}) \\ \frac{\partial\psi^{-}}{\partial\kappa}\,, & (\kappa = \kappa_+) \end{cases}\nn
\ee
\comment{
\bea
\psi^{(1)}(\kappa) &=& \psi^{-}(\kappa) + \psi^{-}(r(\kappa))\nn\\
\psi^{(2)}(\kappa) &=&\begin{cases}  \frac{\psi^{-}(\kappa) - \psi^{-}(r(\kappa))}{\kappa - r(\kappa)}\,, & (\kappa\neq\kappa_{+}) \\ \frac{\partial\psi^{-}}{\partial\kappa}\,, & (\kappa = \kappa_+) \end{cases}\nn
\eea
}
where $\psi^{-}(\kappa)$ is the section of Bloch solutions of Lemma~\ref{lemrefl} (ii), rather than as linear combination of Bloch solutions $\psi^{-}(\kappa)$, $\psi^{-}(r(\kappa))$. The advantage of the basis $\{ \psi^{(1)}(\kappa), \psi^{(2)}(\kappa) \}$ is that it does not degenerate as $\kappa\to\kappa_+$. The solution (\ref{qhi7ac}) is unique up to a multiple, which we shall fix by means of a prescription independent of $\kappa$. For instance, since $\psi^{\sharp}_1\neq 0$, there is a linear functional $\ell$ on $\mathbb{C}^{N}$ such that $\ell(\psi^{\sharp}_{1}(\kappa))\neq 0$ for $\kappa$ near $\kappa_+$. We impose the normalization $\ell(\psi^{\sharp}_1(\kappa))=1$. We then have
\be
\psi^{\sharp}_n(\kappa) = \a(\kappa)\psi^{(1)}_n(\kappa) + \b(\kappa)\psi^{(2)}_n(\kappa) + O(e^{-\e_0 n})\;,\quad (n\to+\io)\label{qhi11a}
\ee
with $\a,\b$ analytic near $\kappa_+$. Moreover, $\a,\b$ are even under $r$ since $\psi^{\sharp}$, $\psi^{(1)}$, $\psi^{(2)}$ are. A semi-bound state is tantamount to $\b(\kappa_+)=0$. For $\kappa\neq\kappa_+$ we may also write
\be
\psi^{\sharp}_n(\kappa) = f(\kappa)\psi^{-}_n(\kappa) + f(r(\kappa))\psi^{-}_n(r(\kappa)) + O(e^{-\e_0 n})\nn%\label{qhi11e}
\ee
where
\be
f(\kappa) = \a(\kappa) + \frac{\b(\kappa)}{\kappa - r(\kappa)}\label{qhi11b}
\ee
is analytic in a punctured neighborhood of $\kappa_+$. Comparison with Eqs.~(\ref{qhi7}, \ref{qhi7af}) yields for later use
\be
S_{+}(\kappa) = \frac{f(r(\kappa))}{f(\kappa)}\;,\qquad (\kappa > \kappa_+)\;.\label{qhi11c}
\ee
For $\Im \kappa < 0$ the Bloch solution $\psi^{-}_{n}(\kappa)$ is exponentially diverging for $n\to \io$ due to $|\xi| = e^{-\Im \kappa}$. Based on $z= \l(\kappa)$, edge state energies $\e>\l(\kappa_+)$ close to $\l(\kappa_+)$ occur iff $f(\kappa)=0$ for some $\kappa$ close to $\kappa_+$ with $\Im \kappa <0$. For that to happen $\b(\kappa)$ has to be correspondingly small: 
\be
\b(\kappa)=\a(\kappa)(r(\kappa) -\kappa)\;. 
\label{qhi11d}
\ee
Let us now reintroduce the parameter $k$ of the lemma and of its assumption Eq.~(\ref{qhi10}). Then $\b(k_*,\kappa_+(k_*))=0$ follows by continuity. This concludes the proof.\qed\\

\noindent
{\it Proof of Theorem~\ref{thmlev}.} Let us first deal with the simple case that the interval $[k_1,k_2]$ does not contain any $k_*$ where $H^{\sharp}(k_*)$ has a semi-bound state. Then $N_+$ vanishes by Lemma~\ref{lemsemi} and so does the l.h.s. of Eq.~(\ref{qhi11}). In fact, by (\ref{qhi11b}, \ref{qhi11c}) we have
\be
\lim_{\d\to 0}S(\kappa_+(k) + \d,k) = -1\nn
\ee
uniformly in $k\in[k_1,k_2]$, since $\b(\kappa_+(k),k)\neq 0$. We consider next the case where such points $k_*$ are present. Though they are generically isolated, we will treat the general case, where they form intervals $I\subset [k_1,k_2]$. Those are closed, countably many and possibly consisting of single points. By the first case and by compactness it suffices to consider such an interval $I$ and $k_1, k_2$ sufficiently close to its endpoints. 

Let first $k\in I$. Then $\a(\kappa_+(k),k)\neq 0$, $(k\in I)$ since $\b(\kappa_+(k),k)=0$; see Eq.~(\ref{qhi11a}). We claim there is a punctured disk of fixed radius centered at $\kappa_+(k)$, which remains free of zeros $\kappa$ of $f(\kappa,k)$. Indeed, if that radius is small, Eq.~(\ref{qhi11d}) can not hold true there, because its r.h.s. is linearly large in $\kappa-\kappa_+(k)\to 0$, while the l.h.s. is quadratically small, as $\beta$ is even. The claim extends by continuity to $k\in[k_1,k_2]$, provided the punctured disk is replaced by an annulus $A(k)$ of fixed radii.  

\begin{figure}[hbtp]
\centering
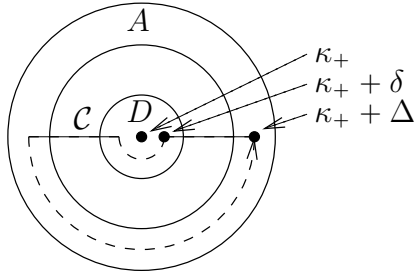
\caption{The disk $D$, the annulus $A$ and the contour $\mathcal{C}$ in the $\kappa$-plane.}
\label{FigHall2}
\end{figure}

We recall that by (\ref{qhi6b}) real $\kappa\neq\kappa_+(k)$ do not occur as zeros either. Let $N(k)$ be the number of zeros inside the annulus and having $\Im \kappa <0$. Thus
\be
-N_+ = N(k_2) - N(k_1)\;.\nn
\ee 
Eq.~(\ref{qhi11}) then follows from the claim that
\be
\lim_{\d\to 0} \arg S_{+}(\kappa_{+}(k_i) + \d, k_i) = -2\pi \bigl(N(k_i) + \frac{1}{2}\bigr)\;,\quad (i=1,2)\;.\label{qhi12}
\ee
In proving it we drop $k_i$ from the notation, \eg $A=A(k_i)$. We note that by assumption $\b(\kappa_+)\neq 0$, whence there is a disk $D$ centered at $\kappa_+$ which is free of zeros of $f$. In particular, $\kappa_+ + \d\in D$ for small $\d>0$, and $\kappa_+ + \D\in A$ for suitable $\D>\d$; likewise for the images under $r$, see Fig.~\ref{FigHall2}.

We consider the contour $\L \cup \G \cup r(\L) \cup \g$, where $\L = [\kappa_+ + \d, \kappa_+ + \D]$, $\G \subset A \cap \{ \Im \kappa <0 \}$ joins $\kappa_+ + \D$ to $r(\kappa_+ + \D)$, and $\gamma\subset D\cap \{ \Im\kappa <0 \}$ joins $r(\kappa_+ + \d)$ to $\kappa_+ + \d$. Let us denote by $\mathcal{C}$ that contour with opposite (positive) orientation. By the argument principle we have
\be
\int_{\mathcal{C}} f(\kappa)^{-1}\frac{\partial f}{\partial\kappa} d\kappa = 2\pi i N\;.\nn
\ee
The contour may also be split into two parts, and their contributions computed otherwise. First, by (\ref{qhi11c})
\be
\arg S(\kappa_+ + \d) =\arg f(r(\kappa_+ + \d)) - \arg f(\kappa_+ + \d)
= \Im \int_{\L \cup \G \cup r(\L)} f(\kappa)^{-1}\frac{\partial f}{\partial \kappa} d\kappa\;.\nn
\ee
(Note that the first equality is consistent with $\arg$ being a continuous argument in $k\in[k_1,k_2]$; so is the second, because the path remains free of zeros as $k$ changes, unlike $\g$). Second, by Eq.~(\ref{qhi11b})
\be
\lim_{\d\to 0}\int_{\g}f(\kappa)^{-1}\frac{\partial f}{\partial \kappa}d\kappa = i\pi\;.\nn
\ee
Together, this proves (\ref{qhi12}) and hence the theorem for the upper band edge. The case of the lower band edge is similar, except that incoming states are found at $\kappa < \kappa_-(k)$. This explains the reversed count of signs in $N_-$. \qed\\

\noindent
{\it Proof of Proposition~\ref{propcomp}.} We write $k$ only when necessary. Since $\kappa_+$ is a non-degenerate maximum, the energy curve $z=\l(\kappa)$ bijectively maps a neighborhood of $\kappa_+$ in the half-plane $\Im \kappa <0$ to one of the band edge $\l_+$, however slit by the band itself. It will be understood that $z$ and $\kappa$ are so related. We first consider the case of a disappearing branch and reformulate the statement using the notation from the proofs of Thms.~\ref{thmlev} and \ref{thm1}. Since no eigenvalue branch is present at $k_2 > k_*$, the two sides of Eq.~(\ref{comp1}) are, after dividing by $2\pi$, those of
\be
\frac{1}{2\pi i}\int_{\mathcal{C}} f(\kappa, k_1)^{-1} df = \frac{1}{2\pi i}\int_{\partial D}l(z,k)^{-1}d l\;,\label{prfcmp1}
\ee
where
\begin{itemize}
\item $\mathcal{C}$ is the contour in the $\kappa$-plane described in Fig.~\ref{FigHall2};
\item $\partial D \subset \mathbb{T}$ is the contour encircling the crossing point with the Fermi line and described in Fig.~\ref{fig6};
\item $l(z,k)$ is the eigenvalue of $L(z,k)$ described in Lemma~\ref{lem1} (ii). (We recall that the use made of $L(z,k)$ rested on Eq.~(\ref{2.7})). 
\end{itemize}
$L(z)$ has an eigenvalue $l(z)=0$ iff $f(\kappa)=0$ (with both zeros being of first order), since both conditions are equivalent to $z=\e$. Hence Eq.~(\ref{prfcmp1}) holds provided the contours are homotopic under $z=\l(\kappa)$.

To show this, let us visualize $\partial D \subset \mathbb{T}$ in the setting of Fig.~\ref{fig2}. That contour can be rotated without intersecting the discrete eigenvalue branch, and so as to lie in a $z$-plane at fixed $k=k_1$. The contour remains positively oriented w.r.t. the orientation of that plane and, after substitution $z=\l(\kappa)$, it is homotopic to $\mathcal{C}$ at $k_1<k_*$.

In the case of an emerging eigenvalue branch, the l.h.s. of (\ref{prfcmp1}) is evaluated at $k_2 > k_*$ and acquires a minus sign, \cf~(\ref{comp1}). However, when $\partial D$ is rotated as just prescribed it ends up negatively oriented in the $z$-plane at $k=k_2$. Hence the modified Eq.~(\ref{prfcmp1}) still holds.\qed\\

\noindent
{\bf Acknowledgments.} We thank Y. Avron, C. Cedzich, M. Fraas, J. Fr\"ohlich, H. Schulz-Baldes, and D. W\"ursch for discussions. The work of M.P. is supported by the Swiss National Science Foundation. 

\begin{spacing}{0.9}

\end{spacing}

\end{document}

%% file: torus2basic.pdf_t
\begin{picture}(0,0)%
\includegraphics{torus2basic.pdf}%
\end{picture}%
\setlength{\unitlength}{4144sp}%
\begingroup\makeatletter\ifx\SetFigFont\undefined%
\gdef\SetFigFont#1#2#3#4#5{%
  \reset@font\fontsize{#1}{#2pt}%
  \fontfamily{#3}\fontseries{#4}\fontshape{#5}%
  \selectfont}%
\fi\endgroup%
\begin{picture}(5069,2542)(-571,-4213)
\put(389,-3135){\makebox(0,0)[lb]{\smash{{\SetFigFont{12}{14.4}{\familydefault}{\mddefault}{\updefault}{\color[rgb]{1,0,0}$0$}%
}}}}
\put(1019,-2325){\makebox(0,0)[lb]{\smash{{\SetFigFont{12}{14.4}{\familydefault}{\mddefault}{\updefault}{\color[rgb]{1,0,0}$\pi$}%
}}}}
\put(-556,-3900){\makebox(0,0)[lb]{\smash{{\SetFigFont{12}{14.4}{\familydefault}{\mddefault}{\updefault}{\color[rgb]{1,0,0}$-\pi$}%
}}}}
\put(1829,-1830){\makebox(0,0)[lb]{\smash{{\SetFigFont{12}{14.4}{\familydefault}{\mddefault}{\updefault}{\color[rgb]{1,0,0}$k\in S^1$}%
}}}}
\put(416,-2008){\makebox(0,0)[lb]{\smash{{\SetFigFont{12}{14.4}{\familydefault}{\mddefault}{\updefault}{\color[rgb]{0,0,0}$\Im z$}%
}}}}
\put(901,-2761){\makebox(0,0)[lb]{\smash{{\SetFigFont{12}{14.4}{\familydefault}{\mddefault}{\updefault}{\color[rgb]{1,0,0}$\g$}%
}}}}
\put(3577,-3280){\makebox(0,0)[lb]{\smash{{\SetFigFont{12}{14.4}{\familydefault}{\mddefault}{\updefault}{\color[rgb]{0,0,0}$\Re z$}%
}}}}
\put(2677,-3216){\makebox(0,0)[lb]{\smash{{\SetFigFont{12}{14.4}{\familydefault}{\mddefault}{\updefault}{\color[rgb]{0,0,0}$\mu$}%
}}}}
\end{picture}%

%% file: honeycomb1.pdf_tex
%% Creator: Inkscape inkscape 0.48.1, www.inkscape.org
%% Accompanies image file 'honeycomb1.pdf' (pdf, eps, ps)

\begingroup
  \makeatletter
  \providecommand\color[2][]{%
    \errmessage{(Inkscape) Color is used for the text in Inkscape, but the package 'color.sty' is not loaded}
    \renewcommand\color[2][]{}%
  }
  \providecommand\transparent[1]{%
    \errmessage{(Inkscape) Transparency is used (non-zero) for the text in Inkscape, but the package 'transparent.sty' is not loaded}
    \renewcommand\transparent[1]{}%
  }
  \providecommand\rotatebox[2]{#2}
  \ifx\svgwidth\undefined
    \setlength{\unitlength}{1545.21586914pt}
  \else
    \setlength{\unitlength}{\svgwidth}
  \fi
  \global\let\svgwidth\undefined
  \makeatother
  \begin{picture}(1,0.66487123)%
    \put(0,0){\includegraphics[width=\unitlength]{honeycomb1.pdf}}%
    \put(0.82949872,0.55613434){\color[rgb]{0,0,0}\makebox(0,0)[lb]{\smash{$A$}}}%
    \put(0.72731523,0.55613434){\color[rgb]{0,0,0}\makebox(0,0)[lb]{\smash{$B$}}}%
    \put(0.61484235,0.10420386){\color[rgb]{0,0,0}\makebox(0,0)[lb]{\smash{$\vec{a}_{2}$}}}%
    \put(0.76445949,0.02460333){\color[rgb]{0,0,0}\makebox(0,0)[lb]{\smash{$\vec{a}_{1}$}}}%
    \put(0.09903324,0.4863275){\color[rgb]{0,0,0}\makebox(0,0)[lb]{\smash{$(n_{1},n_{2})$}}}%
    \put(0.20529583,0.63948839){\color[rgb]{0,0,0}\makebox(0,0)[lb]{\smash{$(n_{1}+1,n_{2})$}}}%
    \put(-0.00257852,0.21020641){\color[rgb]{0,0,0}\makebox(0,0)[lb]{\smash{$(n_{1},n_{2}-1)$}}}%
  \end{picture}%
\endgroup

%% file: honey_boundary.pdf_tex
%% Creator: Inkscape inkscape 0.48.2, www.inkscape.org
%% Accompanies image file 'honey_boundary.pdf' (pdf, eps, ps)
\begingroup%
  \makeatletter%
  \providecommand\color[2][]{%
    \errmessage{(Inkscape) Color is used for the text in Inkscape, but the package 'color.sty' is not loaded}%
    \renewcommand\color[2][]{}%
  }%
  \providecommand\transparent[1]{%
    \errmessage{(Inkscape) Transparency is used (non-zero) for the text in Inkscape, but the package 'transparent.sty' is not loaded}%
    \renewcommand\transparent[1]{}%
  }%
  \providecommand\rotatebox[2]{#2}%
  \ifx\svgwidth\undefined%
    \setlength{\unitlength}{2101.21979399bp}%
    \ifx\svgscale\undefined%
      \relax%
    \else%
      \setlength{\unitlength}{\unitlength * \real{\svgscale}}%
    \fi%
  \else%
    \setlength{\unitlength}{\svgwidth}%
  \fi%
  \global\let\svgwidth\undefined%
  \global\let\svgscale\undefined%
  \makeatother%
  \begin{picture}(1,0.41103051)%
    \put(0,0){\includegraphics[width=\unitlength]{honey_boundary.pdf}}%
    \put(0.01381971,0.33743231){\color[rgb]{0,0,0}\makebox(0,0)[lb]{\smash{$n_{2}$}}}%
    \put(0.73498142,0.35270111){\color[rgb]{0,0,0}\makebox(0,0)[lb]{\smash{$n_{1}$}}}%
    \put(0.35263639,0.3078149){\color[rgb]{0,0,0}\makebox(0,0)[lb]{\smash{$n_{1}$}}}%
    \put(0.24436458,0.38941005){\color[rgb]{0,0,0}\makebox(0,0)[lb]{\smash{$A$}}}%
    \put(0.18440073,0.38941006){\color[rgb]{0,0,0}\makebox(0,0)[lb]{\smash{$B$}}}%
    \put(0.13426202,0.07861691){\color[rgb]{0,0,0}\makebox(0,0)[lb]{\smash{$\vec{a}_{2}$}}}%
    \put(0.24094211,0.02531454){\color[rgb]{0,0,0}\makebox(0,0)[lb]{\smash{$\vec{a}_{1}$}}}%
    \put(0.79522535,0.07643793){\color[rgb]{0,0,0}\makebox(0,0)[lb]{\smash{$\vec{a}_{1}$}}}%
    \put(0.69416954,0.076352){\color[rgb]{0,0,0}\makebox(0,0)[lb]{\smash{$\vec{a}_{2}$}}}%
    \put(0.763398,0.23235413){\color[rgb]{0,0,0}\makebox(0,0)[lb]{\smash{$\vec{a}_{1} + \vec{a}_{2}$}}}%
    \put(-0.00157607,0.39660488){\color[rgb]{0,0,0}\makebox(0,0)[lb]{\smash{$a)$}}}%
    \put(0.45689091,0.39659949){\color[rgb]{0,0,0}\makebox(0,0)[lb]{\smash{$b)$}}}%
    \put(0.48108077,0.33743178){\color[rgb]{0,0,0}\makebox(0,0)[lb]{\smash{$n_{2}$}}}%
  \end{picture}%
\endgroup%

%% file: FigCr.pdf_t
\begin{picture}(0,0)%
\includegraphics{FigCr.pdf}%
\end{picture}%
\setlength{\unitlength}{3947sp}%
\begingroup\makeatletter\ifx\SetFigFont\undefined%
\gdef\SetFigFont#1#2#3#4#5{%
  \reset@font\fontsize{#1}{#2pt}%
  \fontfamily{#3}\fontseries{#4}\fontshape{#5}%
  \selectfont}%
\fi\endgroup%
\begin{picture}(4524,1547)(139,-1125)
\put(3001,239){\makebox(0,0)[lb]{\smash{{\SetFigFont{12}{14.4}{\familydefault}{\mddefault}{\updefault}{\color[rgb]{0,0,0}$(\partial D)_+$}%
}}}}
\put(3001,-1036){\makebox(0,0)[lb]{\smash{{\SetFigFont{12}{14.4}{\familydefault}{\mddefault}{\updefault}{\color[rgb]{0,0,0}$(\partial D)_-$}%
}}}}
\put(2401,-586){\makebox(0,0)[lb]{\smash{{\SetFigFont{12}{14.4}{\familydefault}{\mddefault}{\updefault}{\color[rgb]{0,0,0}$(\m,k_*)$}%
}}}}
\put(2176,-61){\makebox(0,0)[lb]{\smash{{\SetFigFont{12}{14.4}{\familydefault}{\mddefault}{\updefault}{\color[rgb]{0,0,0}$(z,k)$}%
}}}}
\put(2251,-886){\makebox(0,0)[lb]{\smash{{\SetFigFont{12}{14.4}{\familydefault}{\mddefault}{\updefault}{\color[rgb]{0,0,0}$D$}%
}}}}
\put(1876,-586){\makebox(0,0)[lb]{\smash{{\SetFigFont{12}{14.4}{\familydefault}{\mddefault}{\updefault}{\color[rgb]{0,0,0}$J$}%
}}}}
\put(4501,-586){\makebox(0,0)[lb]{\smash{{\SetFigFont{12}{14.4}{\familydefault}{\mddefault}{\updefault}{\color[rgb]{0,0,0}$k$}%
}}}}
\end{picture}%

%% file: FigHall.pdf_t
\begin{picture}(0,0)%
\includegraphics{FigHall.pdf}%
\end{picture}%
\setlength{\unitlength}{4144sp}%
\begingroup\makeatletter\ifx\SetFigFont\undefined%
\gdef\SetFigFont#1#2#3#4#5{%
  \reset@font\fontsize{#1}{#2pt}%
  \fontfamily{#3}\fontseries{#4}\fontshape{#5}%
  \selectfont}%
\fi\endgroup%
\begin{picture}(2096,2186)(-194,-942)
\put(1850,-864){\makebox(0,0)[lb]{\smash{{\SetFigFont{12}{14.4}{\familydefault}{\mddefault}{\updefault}{\color[rgb]{0,0,0}$\kappa$}%
}}}}
\put(338,-864){\makebox(0,0)[lb]{\smash{{\SetFigFont{12}{14.4}{\familydefault}{\mddefault}{\updefault}{\color[rgb]{0,0,0}$\kappa_-$}%
}}}}
\put(1293,-864){\makebox(0,0)[lb]{\smash{{\SetFigFont{12}{14.4}{\familydefault}{\mddefault}{\updefault}{\color[rgb]{0,0,0}$\kappa_+$}%
}}}}
\put(-179,1085){\makebox(0,0)[lb]{\smash{{\SetFigFont{12}{14.4}{\familydefault}{\mddefault}{\updefault}{\color[rgb]{0,0,0}$k$}%
}}}}
\end{picture}%

%% file: FigComp.pdf_t
\begin{picture}(0,0)%
\includegraphics{FigComp.pdf}%
\end{picture}%
\setlength{\unitlength}{4144sp}%
\begingroup\makeatletter\ifx\SetFigFont\undefined%
\gdef\SetFigFont#1#2#3#4#5{%
  \reset@font\fontsize{#1}{#2pt}%
  \fontfamily{#3}\fontseries{#4}\fontshape{#5}%
  \selectfont}%
\fi\endgroup%
\begin{picture}(3796,1350)(-135,-1264)
\put( 46,-286){\makebox(0,0)[lb]{\smash{{\SetFigFont{12}{14.4}{\familydefault}{\mddefault}{\updefault}{\color[rgb]{0,0,0}$\e(k)$}%
}}}}
\put(2791,-1186){\makebox(0,0)[lb]{\smash{{\SetFigFont{12}{14.4}{\familydefault}{\mddefault}{\updefault}{\color[rgb]{0,0,0}$k_*$}%
}}}}
\put(2476,-151){\makebox(0,0)[lb]{\smash{{\SetFigFont{12}{14.4}{\familydefault}{\mddefault}{\updefault}{\color[rgb]{0,0,0}$\e(k)$}%
}}}}
\put(2476,-1186){\makebox(0,0)[lb]{\smash{{\SetFigFont{12}{14.4}{\familydefault}{\mddefault}{\updefault}{\color[rgb]{0,0,0}$k_1$}%
}}}}
\put(3151,-1186){\makebox(0,0)[lb]{\smash{{\SetFigFont{12}{14.4}{\familydefault}{\mddefault}{\updefault}{\color[rgb]{0,0,0}$k_2$}%
}}}}
\put(3646,-601){\makebox(0,0)[lb]{\smash{{\SetFigFont{12}{14.4}{\familydefault}{\mddefault}{\updefault}{\color[rgb]{0,0,0}$\l_+$}%
}}}}
\put(3646,-421){\makebox(0,0)[lb]{\smash{{\SetFigFont{12}{14.4}{\familydefault}{\mddefault}{\updefault}{\color[rgb]{0,0,0}$\m$}%
}}}}
\put(631,-1186){\makebox(0,0)[lb]{\smash{{\SetFigFont{12}{14.4}{\familydefault}{\mddefault}{\updefault}{\color[rgb]{0,0,0}$k_*$}%
}}}}
\put(1261,-1186){\makebox(0,0)[lb]{\smash{{\SetFigFont{12}{14.4}{\familydefault}{\mddefault}{\updefault}{\color[rgb]{0,0,0}$k$}%
}}}}
\put(3511,-1186){\makebox(0,0)[lb]{\smash{{\SetFigFont{12}{14.4}{\familydefault}{\mddefault}{\updefault}{\color[rgb]{0,0,0}$k$}%
}}}}
\end{picture}%

%% file: FigBl2.pdf_t
\begin{picture}(0,0)%
\includegraphics{FigBl2.pdf}%
\end{picture}%
\setlength{\unitlength}{4144sp}%
\begingroup\makeatletter\ifx\SetFigFont\undefined%
\gdef\SetFigFont#1#2#3#4#5{%
  \reset@font\fontsize{#1}{#2pt}%
  \fontfamily{#3}\fontseries{#4}\fontshape{#5}%
  \selectfont}%
\fi\endgroup%
\begin{picture}(5973,1665)(121,-1243)
\put(2161,239){\makebox(0,0)[lb]{\smash{{\SetFigFont{12}{14.4}{\familydefault}{\mddefault}{\updefault}{\color[rgb]{0,0,0}$\kappa$}%
}}}}
\put(136,-1141){\makebox(0,0)[lb]{\smash{{\SetFigFont{12}{14.4}{\familydefault}{\mddefault}{\updefault}{\color[rgb]{0,0,0}$-\pi$}%
}}}}
\put(2071,-1141){\makebox(0,0)[lb]{\smash{{\SetFigFont{12}{14.4}{\familydefault}{\mddefault}{\updefault}{\color[rgb]{0,0,0}$\pi$}%
}}}}
\put(541,-421){\makebox(0,0)[lb]{\smash{{\SetFigFont{12}{14.4}{\familydefault}{\mddefault}{\updefault}{\color[rgb]{0,0,0}$I_1$}%
}}}}
\put(1126,-421){\makebox(0,0)[lb]{\smash{{\SetFigFont{12}{14.4}{\familydefault}{\mddefault}{\updefault}{\color[rgb]{0,0,0}$I_2$}%
}}}}
\put(5941,239){\makebox(0,0)[lb]{\smash{{\SetFigFont{12}{14.4}{\familydefault}{\mddefault}{\updefault}{\color[rgb]{0,0,0}$\l$}%
}}}}
\put(4951,-241){\makebox(0,0)[lb]{\smash{{\SetFigFont{12}{14.4}{\familydefault}{\mddefault}{\updefault}{\color[rgb]{0,0,0}$\g_2$}%
}}}}
\put(4501,209){\makebox(0,0)[lb]{\smash{{\SetFigFont{12}{14.4}{\familydefault}{\mddefault}{\updefault}{\color[rgb]{0,0,0}$\g_1$}%
}}}}
\put(4456,-196){\makebox(0,0)[lb]{\smash{{\SetFigFont{12}{14.4}{\familydefault}{\mddefault}{\updefault}{\color[rgb]{0,0,0}$\g_\l$}%
}}}}
\put(541,-736){\makebox(0,0)[lb]{\smash{{\SetFigFont{12}{14.4}{\familydefault}{\mddefault}{\updefault}{\color[rgb]{0,0,0}$A_1$}%
}}}}
\put(1126,-736){\makebox(0,0)[lb]{\smash{{\SetFigFont{12}{14.4}{\familydefault}{\mddefault}{\updefault}{\color[rgb]{0,0,0}$A_2$}%
}}}}
\put(1801,-736){\makebox(0,0)[lb]{\smash{{\SetFigFont{12}{14.4}{\familydefault}{\mddefault}{\updefault}{\color[rgb]{0,0,0}$A_1$}%
}}}}
\put(1801,-421){\makebox(0,0)[lb]{\smash{{\SetFigFont{12}{14.4}{\familydefault}{\mddefault}{\updefault}{\color[rgb]{0,0,0}$I_1$}%
}}}}
\put(1396,-61){\makebox(0,0)[lb]{\smash{{\SetFigFont{12}{14.4}{\familydefault}{\mddefault}{\updefault}{\color[rgb]{0,0,0}$\G_\l$}%
}}}}
\put(2341,-556){\makebox(0,0)[lb]{\smash{{\SetFigFont{12}{14.4}{\familydefault}{\mddefault}{\updefault}{\color[rgb]{0,0,0}$\e$}%
}}}}
\end{picture}%

%% file: FigHall2.pdf_t
\begin{picture}(0,0)%
\includegraphics{FigHall2.pdf}%
\end{picture}%
\setlength{\unitlength}{4144sp}%
\begingroup\makeatletter\ifx\SetFigFont\undefined%
\gdef\SetFigFont#1#2#3#4#5{%
  \reset@font\fontsize{#1}{#2pt}%
  \fontfamily{#3}\fontseries{#4}\fontshape{#5}%
  \selectfont}%
\fi\endgroup%
\begin{picture}(1858,1614)(-807,32)
\put(-89,1469){\makebox(0,0)[lb]{\smash{{\SetFigFont{12}{14.4}{\familydefault}{\mddefault}{\updefault}{\color[rgb]{0,0,0}$A$}%
}}}}
\put(-89,929){\makebox(0,0)[lb]{\smash{{\SetFigFont{12}{14.4}{\familydefault}{\mddefault}{\updefault}{\color[rgb]{0,0,0}$D$}%
}}}}
\put(1036,929){\makebox(0,0)[lb]{\smash{{\SetFigFont{12}{14.4}{\familydefault}{\mddefault}{\updefault}{\color[rgb]{0,0,0}$\kappa_++\Delta$}%
}}}}
\put(1036,1109){\makebox(0,0)[lb]{\smash{{\SetFigFont{12}{14.4}{\familydefault}{\mddefault}{\updefault}{\color[rgb]{0,0,0}$\kappa_++\delta$}%
}}}}
\put(1036,1289){\makebox(0,0)[lb]{\smash{{\SetFigFont{12}{14.4}{\familydefault}{\mddefault}{\updefault}{\color[rgb]{0,0,0}$\kappa_+$}%
}}}}
\put(-404,884){\makebox(0,0)[lb]{\smash{{\SetFigFont{12}{14.4}{\familydefault}{\mddefault}{\updefault}{\color[rgb]{0,0,0}$\mathcal{C}$}%
}}}}
\end{picture}%

%% file: be3.bbl
\begin{thebibliography}{}

\bibitem{ASV} Avila, J.C., Schulz-Baldes, H. and Villegas-Blas, C.: Topological invariants of edge states for periodic two-dimensional models. {\tt arXiv:1202.0537}

\bibitem{BHZ} Bernevig, B.A., Hughes, T.L. and Zhang, S.-C.: Quantum spin Hall effect and topological phase transition in HgTe quantum wells. Science {\bf 314}, 1757-1761 (2006).

\bibitem{BGO} Br\"aunlich, G., Graf, G.M. and Ortelli, G.: Equivalence of topological and scattering approaches to quantum pumping. Commun. Math. Phys. {\bf 295}, 243-259 (2010).

\bibitem{EG} Essin, A.M., Gurarie, V.: Bulk-boundary correspondence of topological insulators from their Green's functions. Phys. Rev. B  {\bf 84}, 125132 (2011).
\bibitem{FrK} Fr\"ohlich, J. and Kerler, T.: Universality in quantum Hall systems. Nucl. Phys. B {\bf 354}, 369-417 (1991).

\bibitem{FrS} Fr\"ohlich, J. and Studer U.M.: Gauge invariance and current algebra in nonrelativistic many-body theory, Rev. Mod. Phys {\bf 65}, 733 (1993).

\bibitem{FrST} Fr\"ohlich, J., Studer U.M. and Thiran, E.: Quantum theory of large systems of non-relativistic matter, Les Houches Lectures 1994, Elsevier (1995); arXiv:cond-mat/9508062.

\bibitem{FZ} Fr\"ohlich, J. and Zee A.: Large scale physics of the quantum Hall fluid. Nucl. Phys. B {\bf 364}, 517-540 (1991).

\bibitem{FK} Fu, L. and Kane, C.L.: Time reversal polarization and a $Z_2$ adiabatic spin pump. Phys. Rev. B {\bf 74}, 195312 (2006).

\bibitem{FWNK} Fujita M., Wakabayashi K., Nakada K. and Kusakabe K.: Peculiar localized state at zigzag graphite edge. J. Phys. Soc. Jpn. {\bf 65}, 1920-1923 (1996).

\bibitem{H} Haldane, F.D.M: Model for a quantum Hall effect without Landau levels: Condensed-matter realization of the ``parity anomaly''. Phys. Rev. Lett. {\bf 61}, 2015-2018 (1988).

\bibitem{HK} Hasan, M.Z. and Kane C.L.: Topological insulators. Rev. Mod. Phys. {\bf 82}, 3045-3067 (2010).

\bibitem{Hat0} Hatsugai, Y.: Chern number and edge states in the integer quantum Hall effect. Phys. Rev. Lett. {\bf 71}, 3697 (1993).

\bibitem{Hat} Hatsugai, Y. and Ryu, S.: Topological origin of zero-energy edge states in particle-hole symmetric systems. Phys. Rev. Lett. {\bf 89}, 077002 (2002).

\bibitem{Hs} Hsieh, D., Qian, D., Wray, L., Xia, Y., Hor, Y.S., Cava, R.J. and Hasan, M.Z.: A topological Dirac insulator in a quantum spin Hall phase. Nature {\bf 452}, 970 (2008).

\bibitem{KM} Kane, C.L. and Mele, E.J: $Z_2$ Topological order and the quantum spin Hall effect. Phys. Rev. Lett. {\bf 95}, 146802 (2005).

\bibitem{Ka} Kato, T.: {\it Perturbation Theory for Linear Operators}. Springer-Verlag (1980).

\bibitem{Ko} Kohn, W.: Analytic properties of Bloch waves and Wannier functions. Phys. Rev. {\bf 115}, 809-821 (1959).

\bibitem{K} K\"onig, M., Wiedmann, S., Br\"une, C., Roth, A., Buhmann, H., Molenkamp, L.W., Qi, X.-L. and Zhang, S.-C.: Quantum spin Hall insulator state in HgTe quantum wells. Science {\bf 318}, 766 (2007). 

\bibitem{MB} Moore, J.E. and Balents, L.: Topological invariants of time-reversal-invariant band structures. Phys. Rev. B {\bf 75}, 121306(R) (2007).

\bibitem{Nak} Nakada, K., Fujita, M., Dresselhaus, G. and Dresselhaus, M.S.: Edge state in graphene ribbons: Nanometer size effect and edge shape dependence. Phys. Rev. B. {\bf 54}, 17954 (1996).

\bibitem{Na} Nakahara, M.: {\it Geometry, Topology and Physics}. Graduate Student Series in Physics, Institute of Physics Publishing (1990).

\bibitem{Pf} Pfeffer, W.F.: More on involutions of a circle. Amer. Math. Monthly {\bf 81}, 613 (1974).
		
\bibitem{P} Prodan, E.: Robustness of the spin-Chern number. Phys. Rev. B {\bf 80} , 125327 (2009).

\bibitem{QWZ} Qi, X.-L., Wu, Y.-S. and Zhang, S.-C.: Topological quantization of the spin Hall effect in two-dimensional paramagnetic semiconductors. Phys. Rev. B {\bf 74}, 085308 (2006).

\bibitem{RS} Reed, M. and Simon, B.: {\it Methods of Modern Mathematical Physics, III. Scattering Theory}. Academic Press (1979).

\bibitem{R} Roy, R.: $Z_2$ classification of quantum spin Hall systems: An approach using time-reversal invariance. Phys. Rev. B {\bf 79}, 195321 (2009).

\bibitem{SKR} Schulz-Baldes, H., Kellendonk, J. and Richter, T.: Simultaneous quantization of edge and bulk Hall conductivity. J. Phys. A: Math. Gen. {\bf 33} L27 (2000).

\bibitem{SWSH} Sheng, D.N., Weng, Z.Y., Sheng, L. and Haldane, F.D.M.: Quantum spin-Hall effect and topologically invariant Chern numbers. Phys. Rev. Lett. {\bf 97}, 036808 (2006).

\bibitem{T} Thouless, D.J.: Quantisation of particle transport. Phys. Rev. B {\bf 27}, 6083-6087 (1983).

\bibitem{W} Wen, X. G.: Chiral Luttinger liquid and the edge excitations in the fractional quantum Hall states. Phys. Rev. B {\bf 41}, 12838--12844 (1990).

\bibitem{Z} Zhang, S.-C.: The Chern-Simons-Landau-Ginzburg theory of the fractional quantum Hall effect. Int. J. Mod. Phys. B {\bf 6}, 25-58 (1992).
\end{thebibliography}
